\documentclass[fleqn,usenatbib]{mnras}

\usepackage{newtxtext,newtxmath}
\usepackage[T1]{fontenc}

\DeclareRobustCommand{\VAN}[3]{#2}
\let\VANthebibliography\thebibliography
\def\thebibliography{\DeclareRobustCommand{\VAN}[3]{##3}\VANthebibliography}

\usepackage{graphicx}	% Including figure files
\usepackage{amsmath}	% Advanced maths commands

\usepackage{orcidlink}

\usepackage{subcaption}
\usepackage{graphicx}
\usepackage{pdflscape} % for 'landscape' environment
\usepackage{multicol}
\usepackage{supertabular,booktabs}
\usepackage[normalem]{ulem}
\usepackage{amsmath}

\usepackage{placeins}
\usepackage{graphicx}
\usepackage{longtable}
\usepackage{threeparttablex}

\title[DIBs in MaNGA]{Diffuse interstellar bands in nearby galaxies with MaNGA: environmental trends with metallicity, extinction, ionization, and star-forming activity}

\author[K. Kouroumpatzakis et al.]{
K. Kouroumpatzakis,$^{1}$\orcidlink{0000-0002-1444-2016}\thanks{E-mail: kouroumpatzakis@asu.cas.cz}
S. Jiménez,$^{1}$\orcidlink{0000-0003-2808-3146}
B. Deshev,$^{2,1}$\orcidlink{0000-0002-7898-5490}
R. Wünsch,$^{1}$\orcidlink{0000-0003-1848-8967}
P. Jáchym,$^{1}$\orcidlink{0000-0002-1640-5657}
J. Cami,$^{3,4}$\orcidlink{0000-0002-2666-9234}
\newauthor
J. Kubát,$^{5}$\orcidlink{0000-0003-4269-8278}
and A. Srbljanović,$^{1}$\orcidlink{0000-0003-0091-063X}
\\
$^{1}$Astronomical Institute, Czech Academy of Sciences, Boční II 1401, CZ-14131 Prague, Czech Republic\\
$^{2}$Tartu Observatory, University of Tartu, Observatooriumi 1, 61602 Tõravere, Estonia\\
$^{3}$Department of Physics and Astronomy, Western University, London, ON, Canada\\
$^{4}$Institute for Earth and Space Exploration, Western University, London, ON, Canada\\
$^{5}$Astronomical Institute, Czech Academy of Sciences, Fričova 298, CZ-251 65 Ondřejov, Czech Republic
}

\date{Accepted 2026 November 10. 
Received 2026 November 10; 
in original form 2026 July 31}

\pubyear{\the\year{}}

\begin{document}
\label{firstpage}
\pagerange{\pageref{firstpage}--\pageref{lastpage}}
\maketitle

\begin{abstract}
Diffuse interstellar bands (DIBs) are absorption features of uncertain origin observed in the optical and near-infrared spectra of stars and galaxies. 
Although their correlation with dust extinction is well established, the physical conditions governing their strength across different interstellar medium (ISM) environments remain poorly understood. 
We investigate the environmental dependence of the DIB~$\lambda5780$ using spatially resolved MaNGA spectroscopy for $837$ nearby galaxies spanning a broad range of stellar masses, star-formation rates, and ISM properties. 
Stellar and nebular parameters are derived via spectral fitting, while DIB absorption is measured from high-signal-to-noise spatially binned residual spectra. 
We detect the DIB~$\lambda5780$ in 159 galaxies and report the most distant spatially resolved DIB detection to date, at 232~Mpc.
The DIB~$\lambda5780$ equivalent width correlates positively with both stellar and gas-phase extinction. 
DIB~$\lambda5780$ is preferentially detected in regions of high extinction, near- or super-solar metallicity ($12+\log({\rm O/H})\gtrsim8.5$), and moderate specific star-formation rates ($-10.5\lesssim\log({\rm sSFR}/{\rm yr}^{-1})\lesssim-9$). 
By contrast, it falls below our detection threshold in metal-poor or strongly ionized environments.
It is rarely detected in regions with little or no recent star formation, with the few detections in such environments occurring primarily in regions driven by composite ionization.
Our results indicate that DIB~$\lambda5780$ is preferentially detected, within the sensitivity limits of the MaNGA data, in dust-rich, chemically evolved, and moderately ionized environments.
\end{abstract}


%%%%%%%%%%%%%%%%%%%%%%%%%%%%%%%%%%%%%%%%%%%%%%%%%%

%%%%%%%%%%%%%%%%% BODY OF PAPER %%%%%%%%%%%%%%%%%%

\section{Introduction}
\label{sec:intro}

Diffuse Interstellar Bands (DIBs) are a collection of over 600 absorption features observed primarily in the optical and near-infrared spectra of stars obscured by interstellar dust \citep[e.g.,][]{1994A&AS..106...39J,2000PASP..112..648G,2019ApJ...878..151F,2022ApJS..262....2H}. 
First identified in the 1920s \citep{1922LicOB..10..146H}, their precise molecular carriers remain unknown, though carbonaceous species such as polycyclic aromatic hydrocarbons (PAHs), fullerenes, and long carbon chains are considered prime candidates \citep[e.g.,][and references therein]{1995ARA&A..33...19H,2006ARA&A..44..367S,2006A&A...451..973C,2018PASP..130g1001K}. 
Extensive studies within our Galaxy have shown that the strengths of prominent DIBs like $\lambda$5780 correlates with interstellar reddening and gas-phase tracers, yet exhibit differential sensitivity to the local radiation field \citep[e.g.,][]{1997A&A...326..822C, 2011A&A...533A.129V, 2011ApJ...727...33F}. 
This suggests that different DIBs trace varying physical and chemical environments in the interstellar medium \citep[ISM; e.g.,][]{2022MNRAS.510.3546F}.

Large-scale spectroscopic surveys, such as the Sloan Digital Sky Survey (SDSS) and Gaia, have recently enabled population-wide mapping of DIBs across the Galactic disk. 
For instance, \citet{2015MNRAS.452.3629L} extracted DIB signals from over 500,000 extragalactic SDSS spectra, and \cite{2024A&A...691A.282M} used about 290,000 Large Sky Area Multi-Object Fiber Spectroscopic Telescope (LAMOST) spectra to create all-sky DIB maps.
Similarly, \citet{2023MNRAS.521.3727V} identified new DIB features in nearly 900,000 Galactic Archaeology with Hermes (GALAH) spectra. 
These efforts revealed the potential of DIBs as tracers of diffuse gas across large spatial scales, facilitating the use of stacking and machine learning techniques to recover weak signals in noisy data.

The first confirmed extragalactic DIB detection predates modern surveys and was reported by \citet{1966MNRAS.131..299H}, who identified the broad 4430~\AA\ absorption feature in early-type stars in both the Small and Large Magellanic Clouds. 
This demonstrated that DIB carriers persist in chemically distinct, low-metallicity environments beyond our Galaxy. 
Subsequent studies by \citet{2007A&A...470..941C} and \citet{2006ApJS..165..138W} confirmed these detections and found that DIB strengths in the Magellanic Clouds are often weaker than in comparable Galactic sightlines, suggesting environmental dependencies related to metallicity, dust, and radiation fields.

More recent work has extended DIB studies to increasingly distant galaxies. 
\citet{2005A&A...429..559S} observed $\lambda$5780 and $\lambda$6284 DIBs in the host galaxy of SN 2001el (NGC~1448), while \citet{2008A&A...480L..13C, 2008A&A...492L...5C,2011ApJ...726...39C} performed the first systematic surveys of DIBs in M31 and M33. 
Using integral field unit (IFU) spectroscopy, \citet{2018A&A...615A..33M} mapped DIB features in the Antennae galaxies (NGC~4038/4039), and \citet{2025A&A...697A.151V} provided DIB maps in the luminous infrared galaxy merger NGC~6240. These studies established the viability of spatially resolved DIB observations across a range of galactic environments.
At even greater distances, high-redshift investigations have probed the detectability of DIBs in damped Lyman-$\alpha$ systems and quasar absorption line galaxies at $z \simeq 0.5$ \citep{2008AJ....136..994L}, pushing the limits of DIB detection into the younger universe.

The detection of intrinsically weak spectral features in extragalactic observations is frequently limited by the signal-to-noise ratio (S/N) of individual spectra.
Combining spectra is therefore widely used to enhance sensitivity to faint signals that may remain undetectable in individual spatial elements or sources.
Spectral stacking has been extensively applied in extragalactic studies, for example to recover weak atomic and molecular gas emission, while spatially resolved IFU analyses commonly employ adaptive binning to combine neighbouring spectra while preserving as much spatial information as possible \citep[e.g.][]{2011MNRAS.414..888E,2019AJ....158..231W,2019AJ....158..160B,2020A&A...638A.126D,2025A&A...693A.150K}.
Such approaches necessarily trade spatial information for increased spectral sensitivity, making them particularly useful for studying faint features such as DIB absorption.

In this study, we build upon these advances by analyzing DIB absorption in galaxies from the Mapping Nearby Galaxies at Apache Point Observatory (MaNGA) survey, part of SDSS-IV \citep{2015AJ....149...77D,2016AJ....152...83L}.
We employ adaptive spatial binning together with stellar-continuum subtraction to increase the sensitivity to DIB~$\lambda5780$ while retaining spatially resolved information across a large and diverse sample of galaxies.
Given the intrinsically weak nature of DIB absorption and the limited S/N of the MaNGA spectra for detecting these faint features, our analysis is not focused on individual DIB detections but instead on statistically robust trends that emerge across a large number of sub-galactic regions.
Our goal is therefore to characterize how DIB detectability depends on the local environment rather than to construct a catalogue of individual DIB measurements.

This paper is organized as follows:
Section~\ref{sec:sample_data} describes the sample selection and the final set of galaxies analyzed in this study.
Section~\ref{sec:Analysis} outlines the analysis methodology, and Section~\ref{sec:results} presents the main results.
A discussion of the findings is provided in Section~\ref{sec:discussion}, and the conclusions are summarized in Section~\ref{sec:conclusions}.
Throughout this work, we use the term ``environment'' to refer to the local physical conditions of the ISM rather than to the large-scale galaxy environment such as local galaxy density, groups, or clusters.
We adopt a flat $\Lambda$CDM cosmology \citep{2020A&A...641A...6P} with $\Omega_m = 0.315$ and $h = 0.674$.
We also adopt the solar oxygen abundance value of 12~+log(O/H)=8.69 from \citet{2021A&A...653A.141A}.
Confidence intervals (CIs) and uncertainties refer to the 16\%–84\% percentile range, and histogram bin widths follow Scott's rule \citep{10.1093/biomet/66.3.605}.
Unless otherwise specified, the term DIBs refers to the DIB$\lambda~5780$ absorption feature.

    \section{Sample \& data}
    \label{sec:sample_data}
    
    MaNGA uses IFUs to obtain spatially resolved spectroscopy for thousands of nearby galaxies over the wavelength range 3600–10300~\AA, with a spectral resolution of $R \sim 2000$ \citep{2015ApJ...798....7B}. 
    The survey employs IFUs of varying sizes (19, 37, 61, 91, and 127 fibers), corresponding to fields of view between $12''$ and $32''$ in diameter. 
    These observations sample sub-galactic regions within each galaxy, with physical spatial scales that depend on the galaxy distance. 
    By probing a large number of systems across a broad range of properties, MaNGA provides a statistically powerful dataset for investigating the environmental dependence of DIBs.
    
    In this work, we construct a galaxy sample by randomly selecting 837 objects from the high-quality MaNGA DR17 sample \citep{2022ApJS..259...35A}. 
    The selection is restricted to galaxies with reliable MaNGA data quality flags, but no explicit cuts are applied based on galaxy properties such as stellar mass ($M_\star$), star-formation rate (SFR), metallicity, or morphology.
    The choice of 837 galaxies is primarily motivated by the time and computational resources required for the analysis.
    We therefore adopt a random subsample that is sufficiently large to cover the broad range of galaxy properties present in MaNGA while keeping the complete analysis computationally tractable.
    
\begin{figure}
        \centering
        \includegraphics[width=\linewidth]{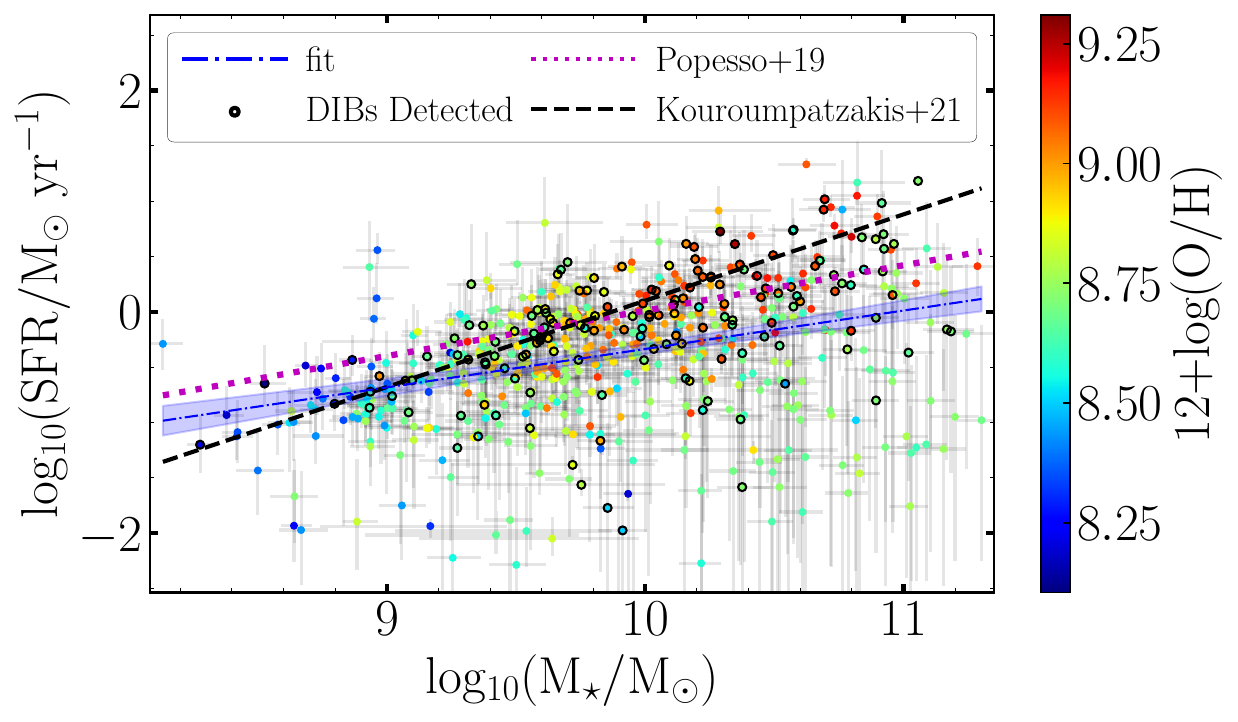}
        \caption{
        SFR over $M_\star$ of the selected sources.
        The points are color-coded by galaxy metallicity.
        Galaxies satisfying our DIB~$\lambda5780$ detection criterion are highlighted with black open circles.
        The blue dashed line shows the best linear regression fit to the full sample.
        The brown-dotted and black dashed lines represent the main-sequence relations from \protect\cite{2019MNRAS.483.3213P} and \protect\cite{2021MNRAS.506.3079K}, respectively, and are shown for reference.
        }
        \label{fig:MS_galaxy_sample}
    \end{figure}

    To characterize the global physical properties of the galaxies, we cross-match the MaNGA \texttt{drpall} catalog with the MPA-JHU database \citep{2003MNRAS.346.1055K,10.1111/j.1365-2966.2004.07881.x,Tremonti_2004} and the Heraklion Extragalactic Catalogue \citep[HECATE;][]{2021MNRAS.506.1896K,2026MNRAS.548ag522K}\footnote{\url{https://hecate.ia.forth.gr/}}.
    Both catalogs provide estimates of the SFR, $M_\star$, and gas-phase metallicity (expressed as $\log(\mathrm{O/H})+12$), but are based on different methodologies, including optical spectroscopy (MPA-JHU) and, in some cases, multiwavelength or photometric calibrations \citep[e.g., IR-based SFRs in HECATE;][]{2023A&A...673A..16K}.
    When multiple measurements are available, we adopt the MPA-JHU values as the primary reference, while using HECATE measurements to supplement missing entries and to assess the consistency of the derived properties.
    All quoted quantities correspond to luminosity-weighted measurements of the galaxies' integrated emission.
    These catalogue measurements are used only as global galaxy properties, primarily for the galaxy-level comparison shown in Figure~\ref{fig:MS_galaxy_sample} and for the representative integrated properties reported in Table~\ref{tab:List_table_1}.
    They are distinct from the spatially resolved measurements presented later in the paper, which are derived from the individual spatial bins of the MaNGA datacubes through our spectral analysis.
    
    The final sample spans a wide range of galaxy properties, including quiescent and actively star-forming systems ($10^{-12.3} \lesssim \mathrm{sSFR/yr^{-1}} \lesssim 10^{-8.4}$, where ${\rm specific{-}SFR \equiv SFR}/M_\star$), as well as metal-poor and metal-rich systems ($8.1 \lesssim 12 + \log(\mathrm{O/H}) \lesssim 9.3$). 
    Figure~\ref{fig:MS_galaxy_sample} shows the distribution of the selected galaxies in the SFR--$M_\star$ plane (i.e., the star-forming main sequence), color-coded by gas-phase metallicity. 
    The selected sample broadly traces the star-forming main sequence, but exhibits a somewhat flatter distribution compared to canonical relations \citep[e.g.,][]{2007A&A...468...33E,2014ApJS..214...15S,2019MNRAS.483.3213P,2021MNRAS.506.3079K}. 
    This reflects the inclusion of galaxies with low or suppressed star formation, extending the sample below the main sequence and ensuring coverage of both actively star-forming and more passive systems.
    This wide parameter coverage allows us to investigate DIB behavior across diverse interstellar medium conditions without imposing strong selection-driven biases.
    
\section{Analysis}
\label{sec:Analysis}

We focus our environmental analysis on DIB~$\lambda5780$, one of the strongest and most widely studied diffuse interstellar bands \citep[e.g.,][]{1995ARA&A..33...19H}. 
Beyond its strength and extensive use in the literature, this band offers practical advantages for MaNGA data. 
For instance, the wavelength-dependent continuum S/N decreases substantially toward the blue, making the broader DIB~$\lambda4430$ difficult to detect reliably despite its intrinsic strength, while its broad profile can also be heavily blended with stellar lines. 
Moreover, measuring DIB~$\lambda6284$ in MaNGA can be hindered by persistent telluric residuals near $\sim6280$~\AA, even after redshifting, as confirmed by recent telluric atlases and \texttt{Molecfit}-based analyses \citep{2015A&A...576A..77S}. 
Finally, DIB~$\lambda5780$ is widely used for environmental diagnostics and provides a robust reference for comparisons across Galactic and extragalactic studies. 
Therefore, focusing on this line maintains continuity with most previous DIB work while maximizing reliability within the constraints of MaNGA spectroscopy.

\subsection{Adaptive spatial binning}
\label{sec:PowerBin}

The analysis in this work is based on the fully reduced Data Reduction Pipeline (DRP) 3D data cubes\footnote{\url{https://www.sdss4.org/dr17/manga/manga-data/data-access/}}
provided by the MaNGA survey.
MaNGA datacubes are reconstructed from dithered fibre bundles, which introduces correlated noise between neighbouring spaxels through the cube reconstruction and resampling procedure\footnote{\url{https://www.sdss4.org/dr17/manga/manga-data/working-with-manga-data/}}.
Narrow absorption features such as DIBs, with ${\rm FWHM} \simeq 3~\AA$, are primarily constrained by the local spectral S/N and the consistency of their line profiles within each extracted spectrum.

To increase the sensitivity to weak spectral features while retaining spatially resolved information, we apply adaptive spatial binning using the \texttt{PowerBin} algorithm of \citet{2025MNRAS.544.1432C}.
Unlike the original Voronoi 2D-binning method of \citet{2003MNRAS.342..345C}, \texttt{PowerBin} constructs the spatial tessellation using Centroidal Power Diagrams.
Adaptive spatial binning is widely employed in IFU spectroscopy to increase the S/N of low-surface-brightness spectra while limiting the associated loss of spatial information \citep[e.g.,][]{2011MNRAS.414..888E,2019AJ....158..231W,2019AJ....158..160B}.
The signal ($S_{\rm cont}$) is defined as the median flux density over all spectral elements within the wavelength range 6000--6100~\AA, which is relatively free of strong stellar absorption and nebular emission features.
The corresponding noise ($N_{\rm cont}$) is taken as the standard deviation of the flux-density values within the same wavelength interval.
We therefore define the continuum-based signal-to-noise ratio as
\begin{equation}
    {\rm S/N}_{\rm cont} = \frac{S_{\rm cont}}{N_{\rm cont}} \quad .
\end{equation}

The values of ${\rm S/N}_{\rm cont}$ for the individual spaxels are supplied to \texttt{powerbin} as the capacity values used to construct the tessellation.
Here, ‘capacity’ refers to the scalar quantity assigned to each spatial element that PowerBin uses to define the target capacity of each bin.
For a bin containing multiple spaxels, these capacity values are summed during the bin-accretion step to determine whether the target capacity is reached.
No explicit correction for spatial covariance between neighbouring MaNGA spaxels is applied in this step.
We adopt a target ${\rm S/N}_{\rm cont}$ of approximately 400.
This target was selected empirically to increase the sensitivity to weak DIB absorption features while balancing the gain in spectral sensitivity against the loss of spatial resolution associated with larger bins.
The target ${\rm S/N}_{\rm cont}$ is therefore an additive capacity criterion used to construct the tessellation and should not be interpreted as the conventional S/N of the resulting spatial-bin spectrum.

After constructing the PowerBin spatial bins, we sum the spectra of the contributing spaxels at each spectral element to produce the final spatial-bin spectra.
The wavelength-dependent uncertainties of these spectra are estimated from the scatter among the contributing spaxel spectra, with the adopted $\sqrt{N_{\rm spax}}$ scaling corresponding to the assumption of independent contributions.
These uncertainties do not include an explicit correction for spatial covariance between neighbouring spaxels.
Because the reconstructed MaNGA datacubes contain correlated noise, the resulting uncertainties may therefore underestimate the true scatter in the combined spectra.

We independently assess the spectral S/N of the extracted spatial-bin spectra after the tessellation.  
For the final sample, the distribution of the resulting spectral S/N has a mode of approximately 60.  
We further test the adopted uncertainties using the final spatial-bin spectra by fitting a smooth continuum in the 6000--6100~\AA\ region and measuring the root-mean-square (RMS) of the residual fluctuations around this continuum after normalization by the adopted wavelength-dependent uncertainties.  
The resulting RMS distribution peaks at 0.97, close to the expected value of unity, but exhibits some bin-to-bin variations. 
We therefore retain the individual RMS values for further correction of the final measurement uncertainties, as described in Section \ref{sec:Sherpa_fits}.
This post-binning assessment provides an empirical characterization of the spectral quality and uncertainties of the spectra used in the subsequent DIB analysis.

\begin{figure}
    \centering
    \includegraphics[width=\linewidth]{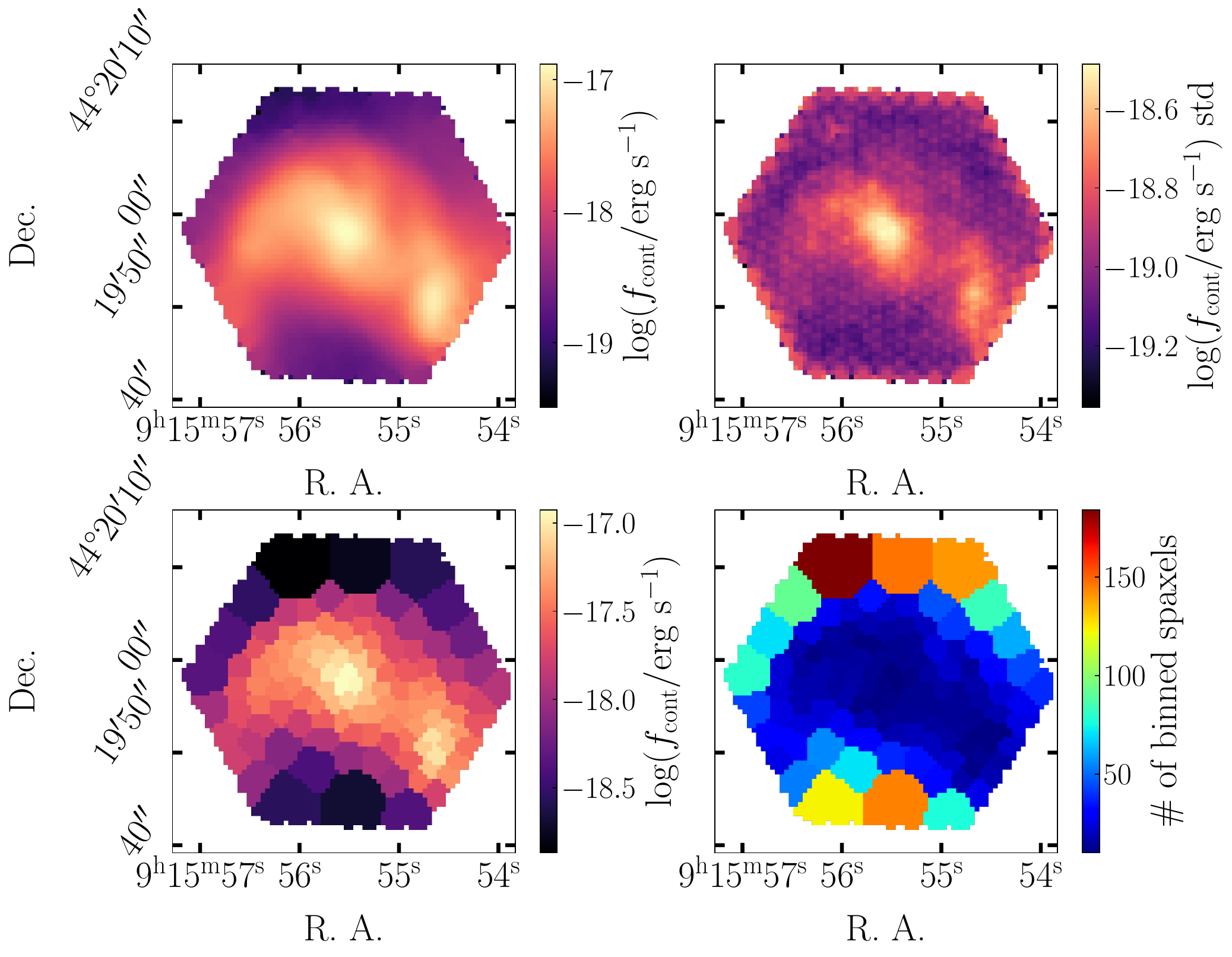}
    \caption{
        Application of the \texttt{PowerBin} spatial binning algorithm to the galaxy with MaNGA ID 1-585513.
        The top-left panel shows the median flux density in the red continuum, measured within the wavelength range 6000--6100~\AA.
        The top-right panel shows the corresponding standard deviation in the same wavelength range.
        The bottom-left panel shows the respective flux after applying the tessellation.
        The bottom-right panel shows the tessellation bins, where the color corresponds to the logarithm of the number of spaxels binned to achieve the target ${\rm S/N}_{\rm cont}$.
        }
    \label{fig:Vorronoi_Binning_1_585513}
\end{figure}

Figure~\ref{fig:Vorronoi_Binning_1_585513} illustrates the tessellation process for a merging galaxy system (UGC~4881 or KUG~0912+445A; MaNGA-ID~1-585513).
As in most cases, the lower surface brightness in the galaxy outskirts necessitates larger bins to approach the target ${\rm S/N}_{\rm cont}$.
Because the tessellation is performed on the reconstructed MaNGA datacube spaxels using this continuum-based S/N criterion, each final spatial bin contains at least one spaxel and therefore does not sample spatial scales smaller than the native spatial sampling of the MaNGA datacubes.

The mode of the distribution of the number of spaxels per bin for all analyzed sources is $\sim20$ spaxels.
Given the MaNGA spaxel scale of $0.5\arcsec$, this corresponds to spatial areas comparable in scale to the footprint of the $\sim2$--$3\arcsec$ PSF, although the exact correspondence depends on the shape of each bin and the observing conditions.
Although the data cubes are sampled on a fine grid of 0.5~\arcsec spaxels, the effective spatial resolution is set by the PSF, which extends over several spaxels.
Consequently, adjacent spaxels are not statistically independent, and the conventional $\sqrt{N_{\rm spax}}$ improvement in S/N expected for independent measurements is not exact for the reconstructed datacubes.
Apparent small-scale variations below the PSF scale in the native maps should therefore be interpreted with caution.
We consequently interpret the reported DIB detections as statistically significant line-of-sight absorption measurements within the extracted apertures rather than as sub-PSF spatial structures.

The spectra used in the following analysis correspond to the sum of all valid spaxel spectra within each spatial bin, performed at each spectral element.
The resulting spectra have different spectral S/N values depending on the local continuum quality and the number of contributing spaxels, and their quality is assessed explicitly after extraction.
In regions where the signal is weaker, typically at the outskirts of galaxies, the spatial bins encompass more spaxels.
As a result, these bins correspond to larger physical areas, effectively averaging over local variations within the sub-galactic region.
Because all environmental quantities are measured consistently within the same spatial bins, this spatial averaging is incorporated uniformly into the subsequent comparisons.

\subsection{Spectral fitting}
\label{sec:Spectral_fitting}

The visible spectra of galaxies and their sub-galactic regions include emission from multiple galactic components.
These components include stellar populations (SPs), the ISM, comprising ionized gas and dust, and, in some cases, emission from an active galactic nucleus (AGN), which can affect both the continuum and emission lines.
Spectral fitting is therefore required to model the stellar and nebular contributions and to identify absorption features associated with the ISM.
This approach has been adopted in several extragalactic DIB studies \citep[e.g.,][]{2018A&A...615A..33M,2025A&A...697A.151V}.

The spectral fitting in this work is performed with the widely used code Fitting Analysis using Differential evolution Optimization \citep[\texttt{FADO};][]{2017A&A...603A..63G}.
\texttt{FADO} is a full-spectrum fitting tool that simultaneously models both stellar continuum and nebular emission in galaxy spectra.
It employs a non-linear optimization scheme based on differential evolution and uses precomputed simple stellar population (SSP) libraries as input.
In this work, those from \citet{2003MNRAS.344.1000B} are adopted.
A distinctive feature of \texttt{FADO} is its self-consistent treatment of nebular continuum and line emission, allowing it to fit galaxies with strong star formation and emission-line activity.

Before fitting, all spectra are corrected for foreground Galactic extinction using the \cite{2001PASP..113.1449C} extinction law and are shifted to the rest frame using the redshift information from the MaNGA pipeline.
The spectra retain the native linear wavelength sampling of the MaNGA LINEAR datacubes, corresponding to 1~\AA\ per spectral pixel, and no interpolation or additional resampling is performed prior to the FADO fitting.
We do not explicitly account for the wavelength dependence of the MaNGA line-spread function.
We use the BC03 SSP library supplied with \texttt{FADO} (Base.BC03.L), and the synthetic spectra are computed with a wavelength step of 1~\AA.
We adopt a spectral resolution of 2.3~\AA\ FWHM for the input spectra through the FADO \texttt{-r} parameter, which internally convolves the SSP templates to this adopted resolution during the fitting.

In this work, the spectra are fitted with \texttt{FADO} over the spectral range $3500$ to $7000$~\AA, which encompasses the main diagnostic features while excluding wavelength regions with significantly increased noise outside this interval.
\texttt{FADO} does not require a dedicated normalization wavelength, since the continuum shape is fitted self-consistently across the entire modeled range.
This wavelength coverage ensures that the continuum around the DIBs is well constrained and that the fits incorporate the principal emission and absorption features required to characterize the stellar populations and ionized gas in each region.
Internal extinction is treated internally within \texttt{FADO}'s combined spectral synthesis framework, and no upper limit is imposed on the allowed attenuation.

The DIB regions and, for instance, the \ion{Na}{I} D doublet are intentionally not masked, because \texttt{FADO} is used to determine the stellar spectral component, including the stellar absorption features present in the observed spectra.
The DIBs and the interstellar component of \ion{Na}{I} D are not included in the \texttt{FADO} model.
We tested the effect of retaining these regions by repeating the fit for representative spatial bins with DIB~$\lambda5780$, DIB~$\lambda5797$, and the \ion{Na}{I} D doublet masked.
The differences between the resulting stellar components are approximately two orders of magnitude smaller than the local continuum level for the \ion{Na}{I} D doublet, and even smaller for the DIB regions.
For the representative spatial bins tested, this demonstrates that including these regions does not affect the fitted stellar component used for the DIB measurements.

The primary goal of the spectral fitting is to remove stellar spectral features that could otherwise affect the DIB measurements.
For each spatial-bin spectrum, the best-fitting stellar component obtained with \texttt{FADO} is therefore subtracted before the DIB fitting.
This step is particularly important in the DIB region, where stellar absorption features can overlap with or otherwise modify the apparent DIB absorption, as illustrated by the examples shown below.

We note that the residual spectra may exhibit fluctuations at other wavelengths, as expected given the finite S/N and limitations of stellar population modeling.  
However, features arising purely from noise or local continuum mismatches do not simultaneously satisfy the quantitative detection criteria adopted for DIB identification (see also Appendix~\ref{sec:Appendix_A}).  
Residual features introduced by imperfect stellar population modeling represent a potential source of systematic uncertainty.  
Nevertheless, the control-window analysis presented in Appendix~\ref{sec:Appendix_A} shows that such residuals do not reproduce the environmental trends observed for the DIB detections, indicating that the reported correlations are not driven by stellar subtraction artifacts.  

\begin{figure*}
    \centering
    \begin{subfigure}{.5\textwidth}
        \centering
        \includegraphics[width=0.99\linewidth]{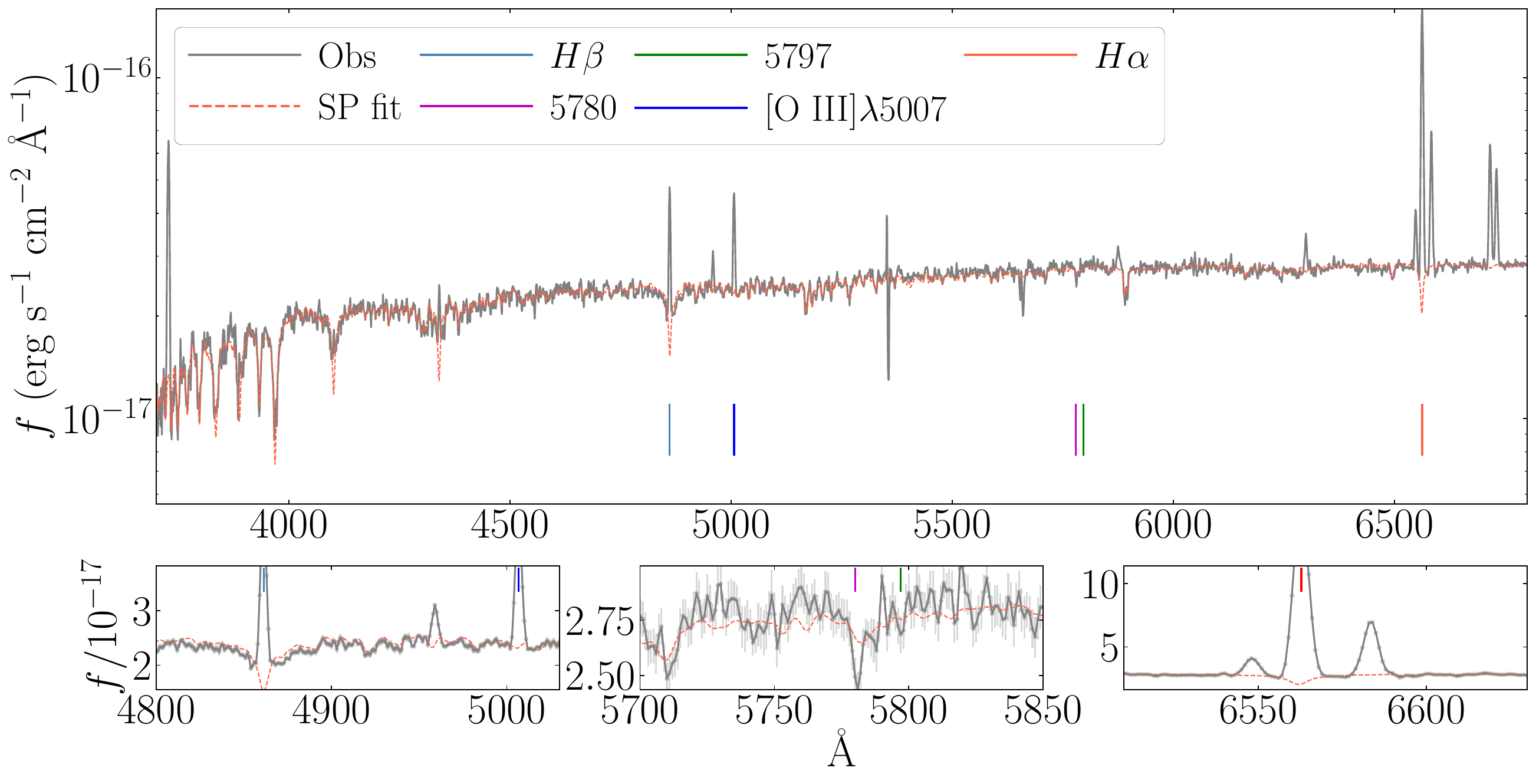}
    \end{subfigure}%
    \begin{subfigure}{.5\textwidth}
        \centering
    \includegraphics[width=0.99\linewidth]{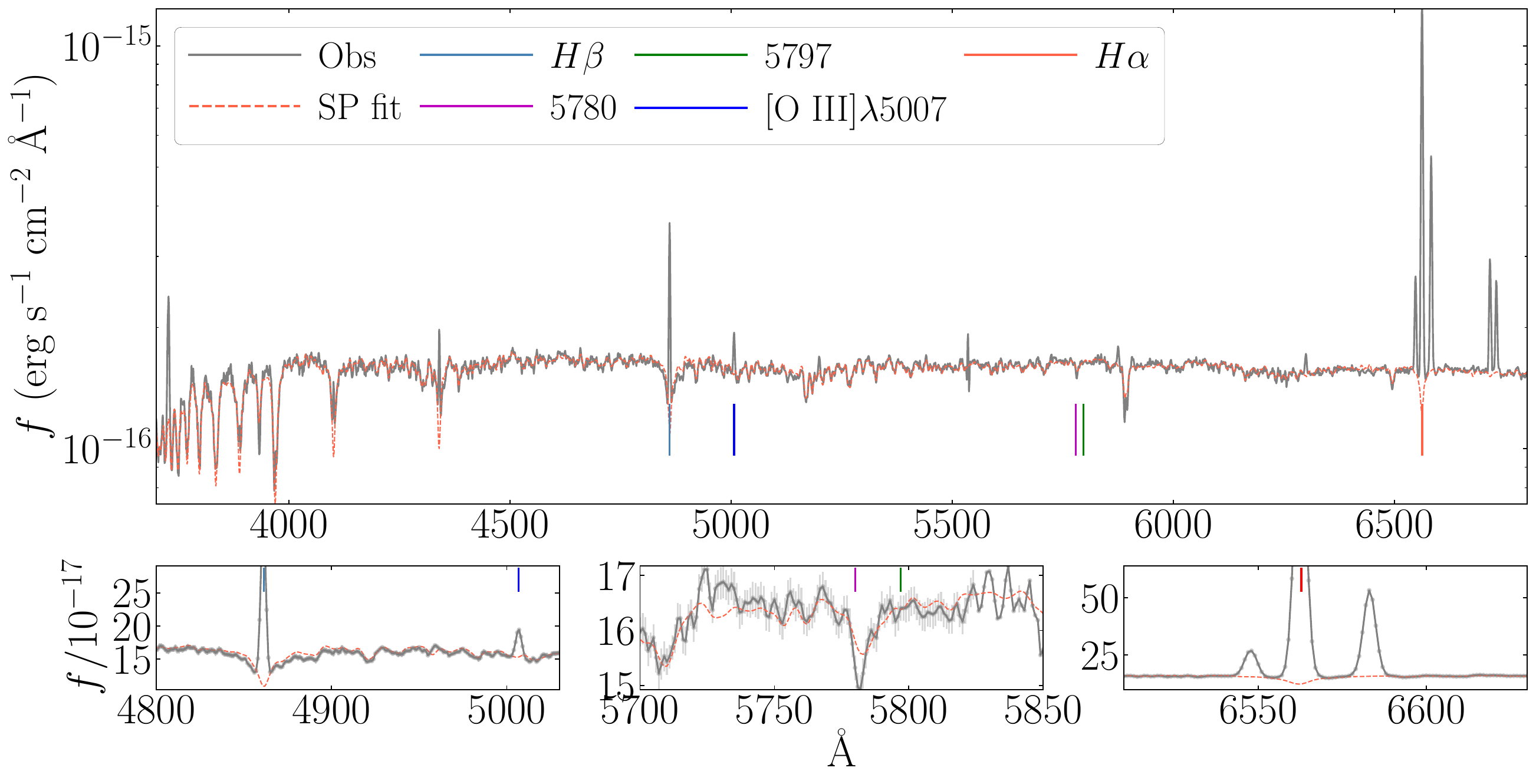}
        \end{subfigure}
    \caption{
    The \texttt{FADO} spectral fits of the rest-frame spectra corresponding to sub-galactic regions of the galaxies SDSS~J075049.38+405646.4 (MaNGA ID~1-43956; left)     and NGC~5383 (MaNGA ID~1-575771; right). 
    The gray line shows the observed rest-frame spectrum, and the red dashed line shows the best-fit stellar continuum. 
    The light blue, blue, red, magenta, and green vertical lines indicate the positions of the H$\beta$, [\ion{O}{III}]$\lambda 5007$, H$\alpha$, and the DIBs~$\lambda \lambda$5780, 5797 features, respectively. 
    The top panel shows the full extent of the optical spectrum, while the three lower panels provide zoomed-in views near the H$\beta$, DIBs, and H$\alpha$ regions, respectively.
    }
    \label{fig:FADO_fit}
\end{figure*}

As an example, Figure~\ref{fig:FADO_fit} shows the rest-frame spectra of sub-galactic regions of the galaxies SDSS~J075049.38+405646.4 and NGC~5383. 
The best-fit stellar component obtained with \texttt{FADO} is overplotted on the observed spectrum. 
This example illustrates that the absorption feature corresponding to DIB~$\lambda5780$ is visible extending beyond the stellar component of the spectral fit. 
However, spectral fitting is essential to accurately estimate the EW of this feature, since nearby wavelengths contain absorption lines such as \ion{Cu}{I} and \ion{Fe}{I} attributed to the stellar continuum \citep[e.g.,][]{2024A&A...691A.282M}. 
Additionally, the fit reveals that the Balmer lines exhibit both absorption and emission components, and the simultaneous modeling of these features enables a robust determination of their strengths.

Telluric absorption was addressed at two levels.
First, MaNGA DRP spectra include standard-star-based corrections for atmospheric transmission, which remove the bulk of the O$_2$ and H$_2$O telluric bands across the full wavelength range \citep[][]{2016AJ....151....8Y}.
Second, we independently verified that no significant telluric residuals affect the DIB~$\lambda5780$ feature at the redshifts of our sample.
The prominent telluric absorption complex near $\sim6280$~\AA\ lies well outside the observed wavelength range of DIB~$\lambda5780$ for all galaxies in the sample \citep[see, e.g., results from the code \texttt{Molecfit};][]{2015A&A...576A..77S}.
For the most distant system in our sample ($z\simeq0.0614$), the redshifted DIB~$\lambda5780$ appears at $\sim6152$~\AA, still well separated from the 6280~\AA\ O$_2$/H$_2$O band.
This redshift represents the upper end of the distribution of the selected sample and is not the result of an imposed redshift selection criterion.
We also examined the very narrow telluric features occasionally present around $\sim5900$~\AA\ \citep[reported in high-resolution studies and identified through \texttt{Molecfit};][]{2021MNRAS.502.4392L}.
At MaNGA's resolution (${\rm FWHM}\simeq2.9$~\AA), these lines are unresolved and are expected to contribute negligibly to the DIB measurements.
Together, these checks indicate that telluric absorption does not bias our DIB measurements across the sample's full redshift range.

\subsection{Fitting of DIBs}
\label{sec:Sherpa_fits}

The EWs of the DIBs are measured through spectral fitting of the residual spectrum using the \texttt{Sherpa} modeling and fitting application \citep{2001SPIE.4477...76F, SciPyProceedings_51}.
The fitting is performed over a spectral window covering 5750 to 5830~\AA.
The residual spectrum is obtained by subtracting the best-fitting stellar component from the observed spatial-bin spectrum.
We do not divide the observed spectrum by the FADO stellar component, since our aim is to remove the stellar spectral structure rather than to construct a normalized transmission spectrum relative to the fitted stellar model.
After subtraction, we restore a constant local flux level, defined as the median observed flux density in the adjacent 5745--5765~\AA\ and 5805--5825~\AA\ sidebands.
This restored level provides a positive and numerically stable baseline for the subsequent fitting and should not be confused with the continuum used for the final DIB equivalent-width (EW) measurements.
The residual spectrum is then modeled as the sum of a constant continuum and two Gaussian absorption components representing DIB~$\lambda5780$ and DIB~$\lambda5797$.
The local continuum is fitted simultaneously with the two Gaussian components, and the equivalent widths are subsequently measured relative to this fitted continuum.
We do not include a separate nebular-emission model in this narrow fitting step because the weak nebular features in this wavelength range do not overlap the DIB profiles and the contribution of [\ion{N}{II}]~$\lambda5755$ to the median flux level used to define the restored baseline is negligible.

The uncertainties used in the \texttt{Sherpa} fitting correspond to the wavelength-dependent uncertainty spectra associated with the continuum-subtracted spatial-bin spectra.
They are derived from the uncertainty information of the contributing MaNGA datacube spaxels following the procedure described in Section~\ref{sec:PowerBin} and are carried through the continuum-subtraction procedure used to generate the residual spectra analyzed with \texttt{Sherpa}.
The median spectral uncertainty replaces non-finite uncertainty values, and a minimum uncertainty floor corresponding to 5\% of the median uncertainty of the spectrum is imposed prior to the fitting procedure to prevent unrealistically small values from dominating the fit statistics.

The Gaussian components are initially assigned FWHM values of 2.8~\AA\ for DIB~$\lambda5780$ and 1.2~\AA\ for DIB~$\lambda5797$.
These values are used only to initialize the first optimization, during which the Gaussian line centers and widths are fixed.
The centers and widths are subsequently thawed, and the fit is repeated, allowing the final values to be determined by the data.
Tests using initial FWHM values spanning approximately 0.5--10~\AA\ showed that the final fitted widths are stable over the range encompassing the adopted initial values, indicating that the choice of initialization does not drive the final DIB solution.

Equivalent widths are computed by numerically integrating the fitted absorption profiles relative to the fitted continuum within fixed wavelength intervals around each DIB.
Uncertainties on the EWs are estimated through a Monte Carlo approach in which the observed spectrum is perturbed according to its wavelength-dependent uncertainties and the equivalent-width measurement is repeated multiple times while keeping the continuum fixed to the best-fit value.
This procedure is used only to estimate the statistical uncertainties on the measured EWs and is distinct from the spectral-perturbation tests described in Appendix~\ref{sec:Appendix_A}.
It is applied uniformly across the sample to ensure consistent DIB measurements in all selected regions.
The original observed spectrum, including its observational noise, is common to all Monte Carlo realizations, whereas the additional Gaussian perturbations are independently generated for each realization.
The EW uncertainty is estimated from the standard deviation of the resulting EW distribution, rather than from the deviation of the individual measurements from an underlying noise-free value.

Following the spectral fitting and EW measurements, we apply a bin-specific multiplicative correction to the uncertainties of the final measurements using the normalized RMS values measured in the 6000--6100~\AA\ continuum region (Section~\ref{sec:PowerBin}). 
Specifically, the uncertainties of each bin are multiplied by its corresponding RMS value, thereby bringing the RMS normalization to unity. Each RMS value is matched to the corresponding galaxy and spatial bin and used to rescale the uncertainties of the measured EWs and FWHMs, as well as those of the control bands.
This a posteriori correction adjusts the reported measurement uncertainties without modifying the wavelength-dependent spectral uncertainties used in the Sherpa fitting, the fitted spectral properties, or the measured EWs.

While both DIBs are fitted simultaneously to provide a consistent description of the local DIB spectral region, the quantitative analysis presented in this paper focuses exclusively on DIB~$\lambda5780$, as $\lambda5797$ is substantially weaker at the S/N and spectral resolution of the MaNGA data.
The $\lambda5797$ measurements are retained only as a consistency check and are presented in Appendix~\ref{sec:Appendix_A}.

\begin{figure}
    \centering
    \hbox{
    \includegraphics[width=0.49\linewidth]{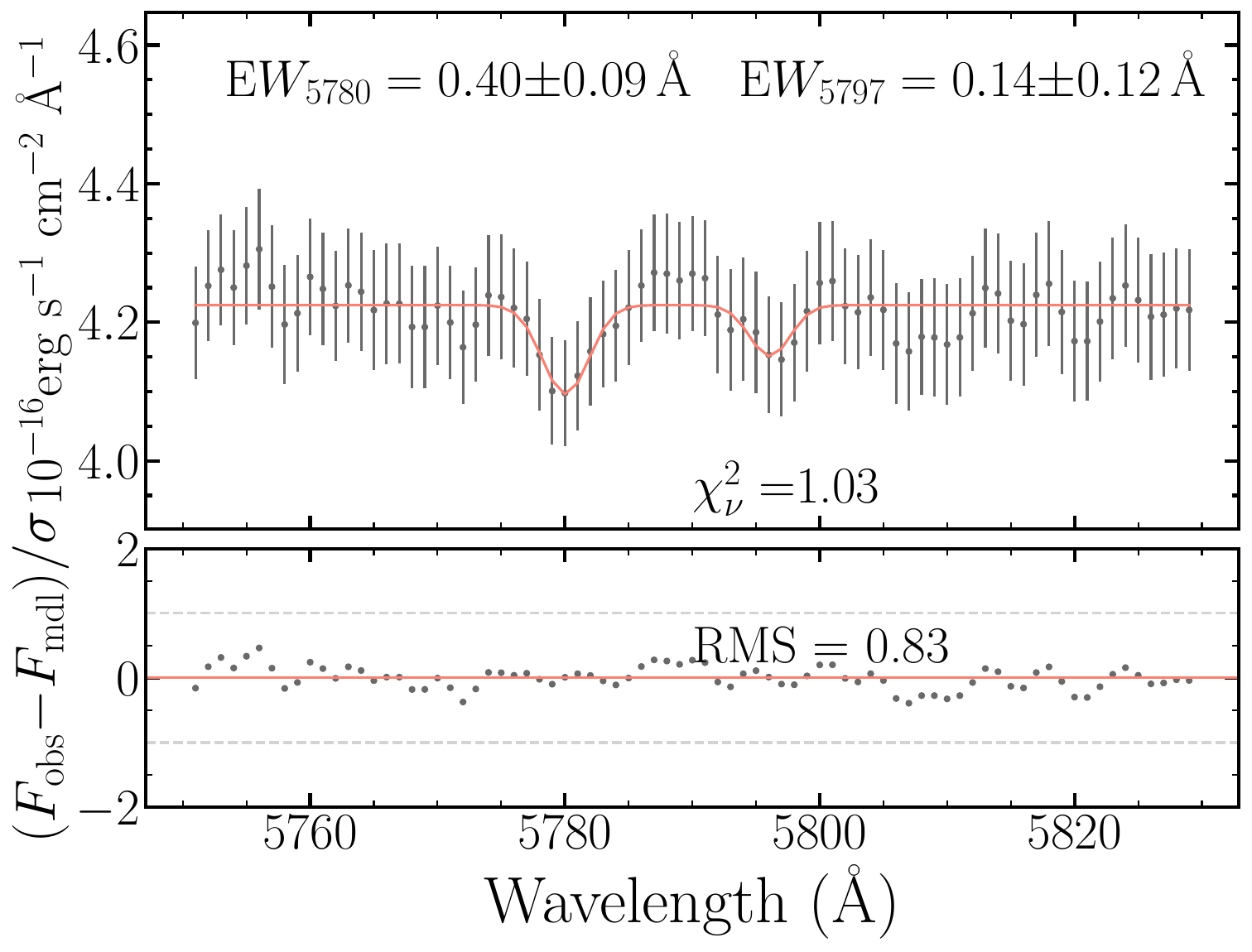}
    \includegraphics[width=0.49\linewidth]{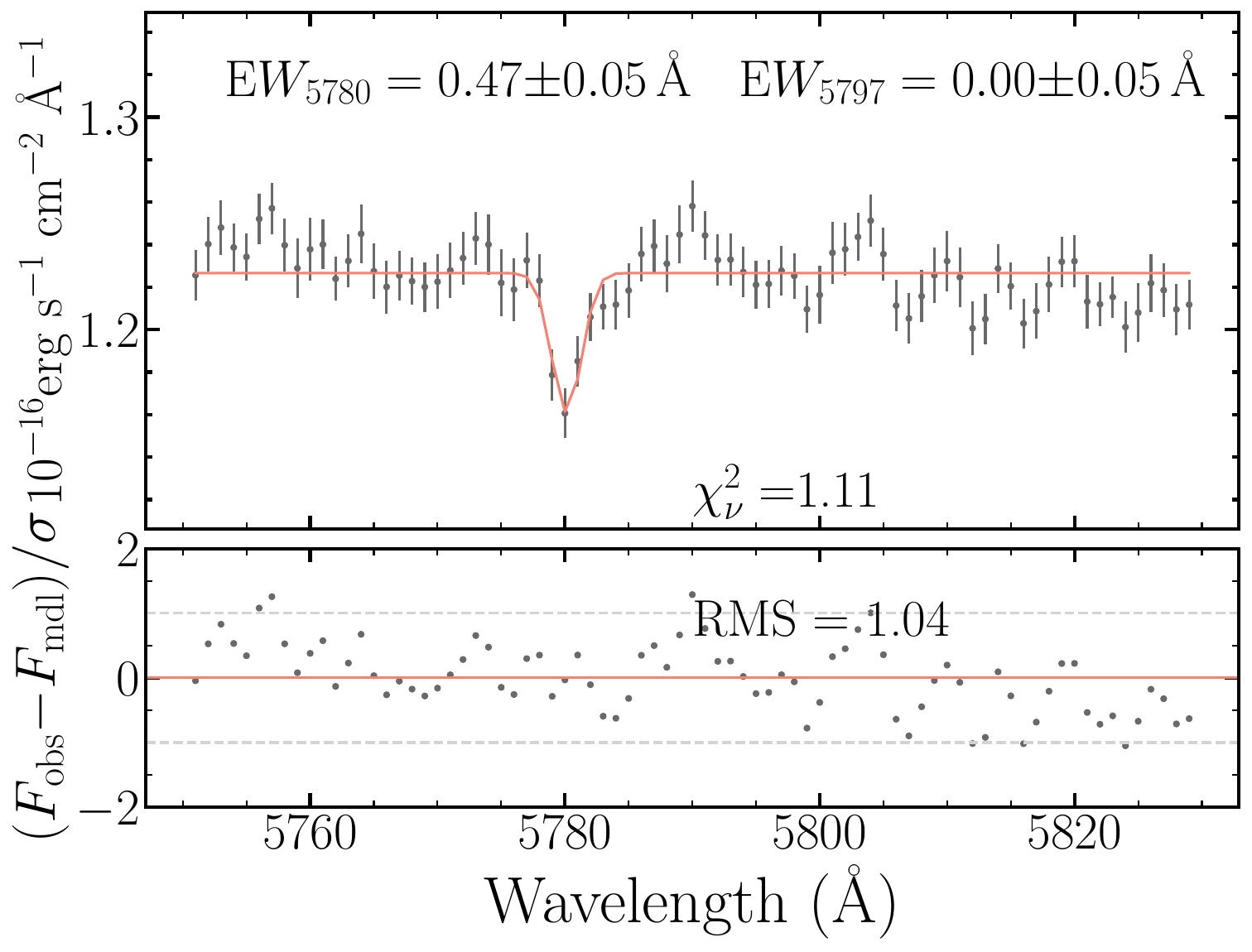}
    }
    \hbox{
    \includegraphics[width=0.49\linewidth]{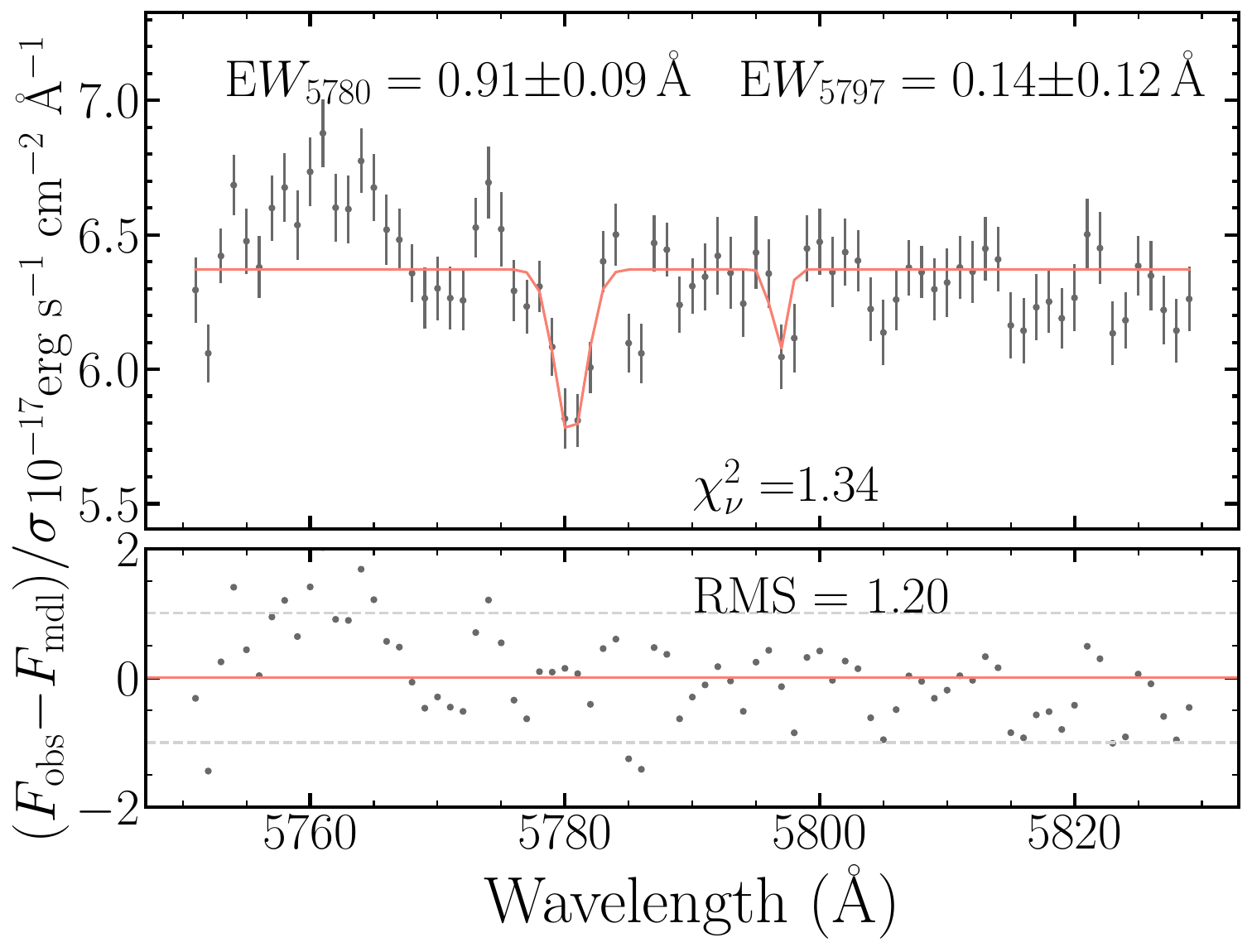}
    \includegraphics[width=0.49\linewidth]{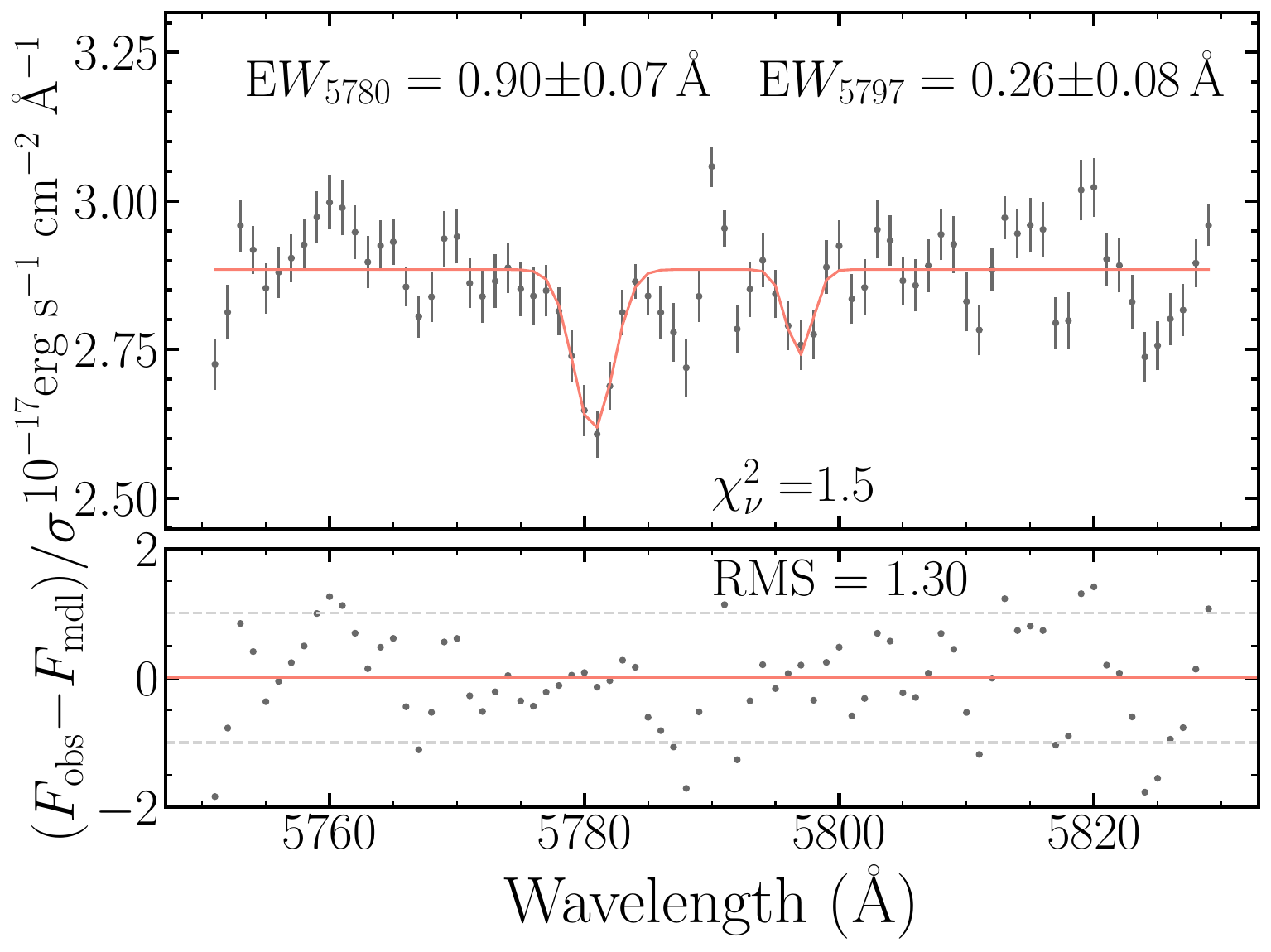}
    }
    \caption{
        Examples of the fitting of the DIB~$\lambda\lambda$5780,5797 absorption lines using \texttt{Sherpa}.
        The upper and lower panels correspond to spatially binned spectra extracted from the galaxies NGC~5383 and SDSS~J075049.38+405646.4, respectively.
        The plots in the right column correspond to the respective spectra, extracted from the same sub-galactic regions shown in Figure~\ref{fig:FADO_fit}.
        The gray error bars represent the residual spectrum and its wavelength-dependent uncertainties, and the red line indicates the best-fit \texttt{Sherpa} model.
        The small panels below each spectrum show the normalized residuals, $(F_{\rm obs}-F_{\rm model})/\sigma$, where $\sigma$ corresponds to the wavelength-dependent spectral uncertainties adopted in the fitting procedure.
    }
    \label{fig:Sherpa_fit}
\end{figure}

An example of the fitting process is shown in Figure~\ref{fig:Sherpa_fit}.
The fit corresponds to the same galaxies displayed in Figure~\ref{fig:FADO_fit}, following subtraction of the stellar continuum.
The quality of the fit is reflected in the small residuals (bottom panel) and the low reduced $\chi^2_\nu$.
Fits with high reduced $\chi^2_\nu$ values are excluded during the subsequent quality-selection stage of the analysis.

\section{Results}
\label{sec:results}

To ensure robust measurements, we restrict our analysis to measurements with $\rm S/N > 5$ in the respective parameter (e.g., SFR, $M_\star$, line fluxes).  
In addition to the adaptive spatial binning based on the continuum-quality criterion described above, we require that both spectral and DIB fits satisfy $\rm S/N > 5$ and $\chi^2_\nu < 2$.
The detection rate depends on the binning scale, since larger spatial bins may dilute the DIB signal by averaging over heterogeneous substructures.
It also depends strongly on the spectral quality.  
Sources with the most widespread DIB detections are typically bright in the red band in this sample, reflecting the importance of a strong continuum for detecting weak DIB absorption features.  

It is important to assess whether the reported DIB detections could be affected by spurious features arising from residual noise in the continuum-subtracted spectra.
To this end, we performed a series of control tests in which the same fitting procedure, detection thresholds, and quality cuts were applied to several wavelength windows where no interstellar absorption is expected.
These tests were carried out both far from and adjacent to the DIB spectral region to probe local continuum and residual systematics.
The resulting control detections occur at significantly lower rates than the DIB detections and do not exhibit the coherent dependencies on extinction, ionization, and other environmental properties observed for DIB~$\lambda5780$.
This demonstrates that the main environmental trends reported for the DIB detections are not dominated by noise-driven false positives.
A detailed description of these tests is provided in Appendix~\ref{sec:Appendix_A}.  

We detect the DIB~$\lambda5780$ in several galaxies located beyond 100~Mpc, with the most distant case being LEDA~1157026 (MaNGA~ID~1-179479), an edge-on galaxy at 232~Mpc and host of the SDSS-II SN~20130 supernova.  
To the best of our knowledge, this constitutes the most distant spatially resolved detection of a diffuse interstellar band to date, extending such studies well beyond the Local Group and nearby galaxy systems.  

\subsection{Description of the examined parameters}
\label{sec:Examined_parameters_description}

For each spatially binned region, we derive maps of stellar and gas-phase properties using spectral fitting and emission-line diagnostics, alongside the DIB-related quantities.
These maps (e.g., Figures~\ref{fig:Maps_1-575771}, \ref{fig:Maps_1-37213}, \ref{fig:Maps_1-135641}, \ref{fig:Maps_1-585513}, \ref{fig:Maps_1-585731}, \ref{fig:Maps_1-636015}, \ref{fig:Maps_1-179479}, \ref{fig:Maps_1-251307}, and \ref{fig:Maps_12-129612}) reveal substantial spatial variation in both SP characteristics and DIB absorption strength, allowing us to explore how DIBs respond to local galactic environments.
For instance, the galaxies presented in Appendix \ref{sec:Appendix_B} were selected to highlight sources with reliable DIB detections spanning a variety of intrinsic conditions.

Specifically, emission line fluxes and EWs, $M_\star$, SP age, stellar and gas-phase metallicities, and extinction
% , systemic velocity ($V$), and velocity dispersion ($\sigma$)
are obtained from the \texttt{FADO} spectral fits and subsequently reconstructed into spatial maps.
The SFR is derived from the extinction-corrected H$\alpha$ luminosity adopting the \citealt{2011ApJ...737...67M} calibration.
Gas-phase extinction is estimated from the Balmer decrement, assuming Case B recombination with $F_{\rm H\alpha}/F_{\rm H\beta} = 2.86$ \citep[][]{1989agna.book.....O}, and adopting the reddening law of \cite{2000ApJ...533..682C}.
FADO derives the stellar metallicity by fitting the observed continuum with a combination of SSPs spanning a wide range of ages and metallicities, while ensuring consistency with the nebular emission. 
Gas-phase metallicities are estimated adopting the method of \citealt{2004MNRAS.348L..59P}.

Additionally, we compute the [\ion{N}{II}]-based BPT \citep*{1981PASP...93....5B} classification \citep{1987ApJS...63..295V,1989agna.book.....O} for each spatially resolved spectrum to identify ionization sources such as AGN near galaxy centers or shocks, distinguishing them from pure \ion{H}{II} regions.
The classification was based on the theoretical \textit{extreme ionization line} of \cite{2001ApJ...556..121K} and the empirical line of \cite{2003MNRAS.346.1055K} which accounts for \textit{composite} sources.
Throughout this work, we use the [\ion{N}{II}]/\ion{H}{$\alpha$} BPT diagnostic as it offers the highest S/N and the most complete coverage across our sample, ensuring homogeneous classification.

Line ratios in the composite and low-ionization (nuclear) emission-line region (LI(N)ER)-like regimes may arise from a mixture of processes, including AGN activity, shocks, diffuse ionized gas, and evolved SPs such as post-AGB stars \citep[e.g.,][]{2019AJ....158....2B,2023A&A...679A..76D,2025A&A...693A..95D}. 
For this reason, we use the broader term “composite” rather than associating these regions with a specific mechanism.
Thus, for this work, BPT-based categories serve only as coarse empirical groupings of ionization conditions, rather than definitive mechanisms.
For instance, the BPT map presented in Figure~\ref{fig:Maps_1-37213} demonstrates that the galaxy IC~263 (MaNGA ID~1-37213) harbors an AGN whose ionizing radiation impacts the surrounding central regions, in agreement with other studies that have provided classification for this source \citep{2014ApJ...788...45T,2020ApJ...901..159C,2023A&A...674A..85A}.

We also examine the O32 index, defined here as $\mathrm{O32} \equiv f_{[{\rm O III}]\lambda 5007}/f_{[{\rm O II}]\lambda 3727}$, which is a widely used tracer of the ionization state of the gas and serves as a proxy for both the ionization parameter $U$ (the number of ionizing photons per hydrogen atom) and the effective temperature of the ionizing stars \citep[$T_{\rm eff}$; e.g.,][]{2016ApJ...816...23S,2025A&A...696A.133K}.
We also examine the $\mathrm{EW}_{\mathrm{H}\alpha}$ as an independent method of tracing recent star-forming activity.
The $\mathrm{EW}_{\mathrm{H}\alpha}$ is tightly correlated with sSFR, both on global galaxy scales \citep[e.g.,][]{2018MNRAS.477.3014B} and within the subgalactic regions studied here (Figures~\ref{fig:Corner_DIB_1-575771}–\ref{fig:Corner_DIB_1-179479}).
Furthermore, the studied subgalactic regions show a strong monotonic anti-correlation between SP age and either sSFR or $\mathrm{EW}_{\mathrm{H}\alpha}$, consistent with trends found for the integrated emission of galaxies \citep[e.g.,][]{2007MNRAS.381..263A}.

\subsection{Comparisons between DIB EWs and extinction}
\label{sec:DIBs_vs_EW}

\begin{figure*}
    \centering
    \includegraphics[width=\linewidth]{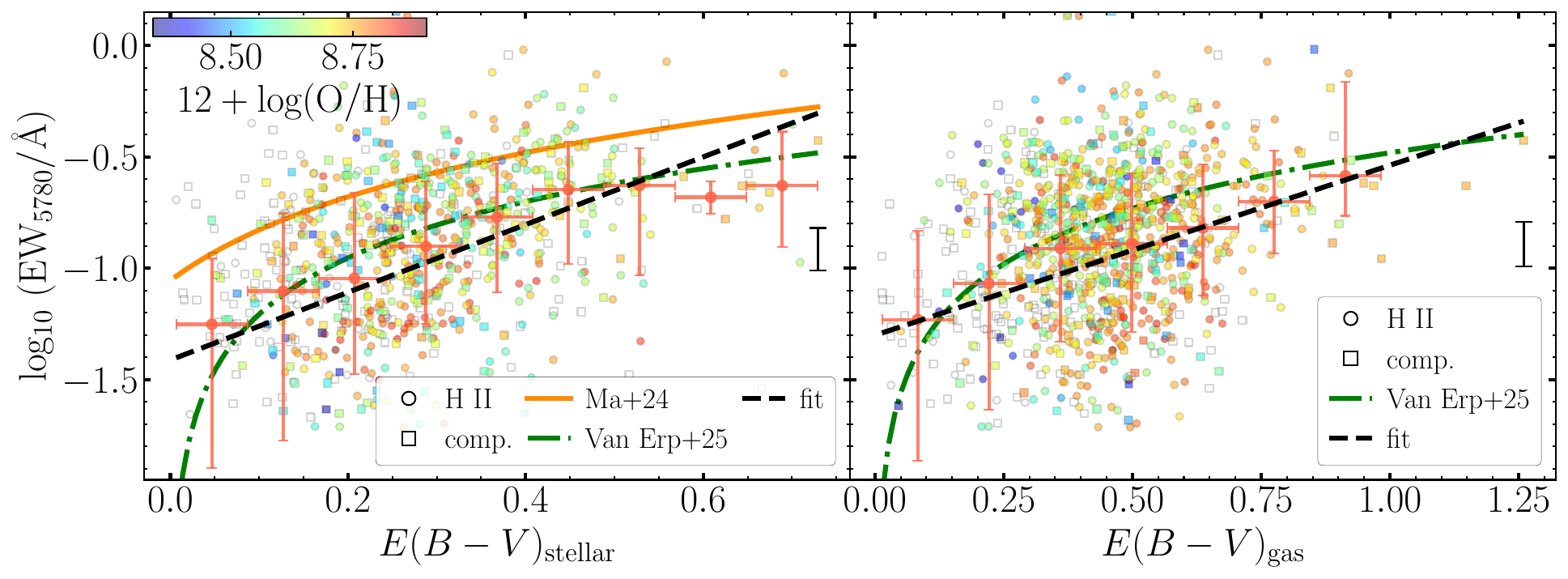}
    \caption{
    Relationships between the EWs of DIB~$\lambda5780$
    and the stellar and gas-phase extinction of sub-galactic regions in the sample, shown in the left and right panels, respectively.  
    Each point represents a spatially binned region classified via the BPT diagram as star-forming (circles) or composite/AGN (squares). 
    Points are color-coded according to the derived metallicity of each region.  
    Red symbols show the mode of the $\log_{10}(\mathrm{EW})$ distribution in bins of extinction with equal width, with vertical error bars indicating the CIs and horizontal error bars representing the bin extent in extinction.  
    The black error bar in each panel indicates the median uncertainty of the measurements.   
    Literature relations from \citet{2024A&A...691A.282M} and \citet{2025A&A...697A.151V} are shown as orange solid and green dashed lines, respectively.  
    The black dashed line shows a robust linear regression fit to the full sample.  
    The best-fit parameters are listed in Table~\ref{tab:fit_results}.  
        }
    \label{fig:DIB_vs_physical}
\end{figure*}

The correlation between DIB~$\lambda$5780 EW and extinction has been previously investigated in both Galactic and extragalactic environments.
Figure~\ref{fig:DIB_vs_physical} illustrates the comparison between the DIB~$\lambda$5780 EWs as a function of both gas-phase and stellar extinction, incorporating all analyzed sources.
Our measurements are consistent with this established trend, exhibiting a positive correlation with both stellar and gas-phase extinction.
A fitted linear regression shows that the correlation is steeper with stellar extinction.
This difference likely reflects the known offset between stellar and nebular reddening, rather than a direct physical change in the DIB–extinction relation.
In fact, the ratio of stellar to gas-phase extinction of all the datapoints analyzed by this work is  $ E(B-V)_{\rm stellar} \simeq 0.5^{+0.17}_{-0.10} \times E(B-V)_{\rm gas}$, consistent with previous studies \citep[e.g.,][]{2000ApJ...533..682C}.
The results of the fits are provided in Table \ref{tab:fit_results}.

\begin{table}
    \centering
    \caption{The best fit parameters of the comparison between DIB EW and gas-phase or stellar extinction in the form ${\rm log (EW_{DIB}/}$\AA)~$= \alpha + \beta \times E(B-V)$.}
    \begin{tabular}{c|cc}
         & $\alpha$ & $\beta$\\
         \hline
         & \multicolumn{2}{c}{stellar}\\
        DIB~$\lambda5780$ & $-1.42(\pm 0.06)$ & $1.54(\pm 0.15)$\\
         \hline
        & \multicolumn{2}{c}{gas-phase}\\
        DIB~$\lambda5780$ & $-1.30(\pm 0.06)$ & $0.77(\pm 0.09)$ \\
        \hline
    \end{tabular}
    \label{tab:fit_results}
\end{table}

Our measurements are broadly consistent with previous empirical relations between DIB strength and dust extinction reported in the literature.  
In particular, the DIB~$\lambda5780$ relation is in good agreement with the trends presented by \citet{2025A&A...697A.151V} based on the analysis of galaxy NGC~6240.  
The comparison with \citet{2024A&A...691A.282M} should be interpreted with caution because the two studies probe different sightline geometries and use different procedures to isolate the interstellar absorption.
\citet{2024A&A...691A.282M} derive their DIB~$\lambda5780$ relation from individual Galactic stellar sightlines, whereas our measurements are obtained from spatially resolved regions within external galaxies.
In addition, \citet{2024A&A...691A.282M} use stellar templates constructed from low-extinction reference stars to isolate the interstellar absorption, while our analysis subtracts the best-fitting stellar component obtained with \texttt{FADO} before fitting the DIBs.
The differences between the two approaches, together with differences in the sightline populations and DIB profile definitions discussed by \citet{2024A&A...691A.282M}, may contribute to the offset between the relations.
Nevertheless, significant scatter persists in both relations.

\subsection{Galaxy-to-galaxy variations in the DIB--extinction relation}
\label{sec:dib_individual_relations}

\begin{figure}
    \centering
    \includegraphics[width=\linewidth]{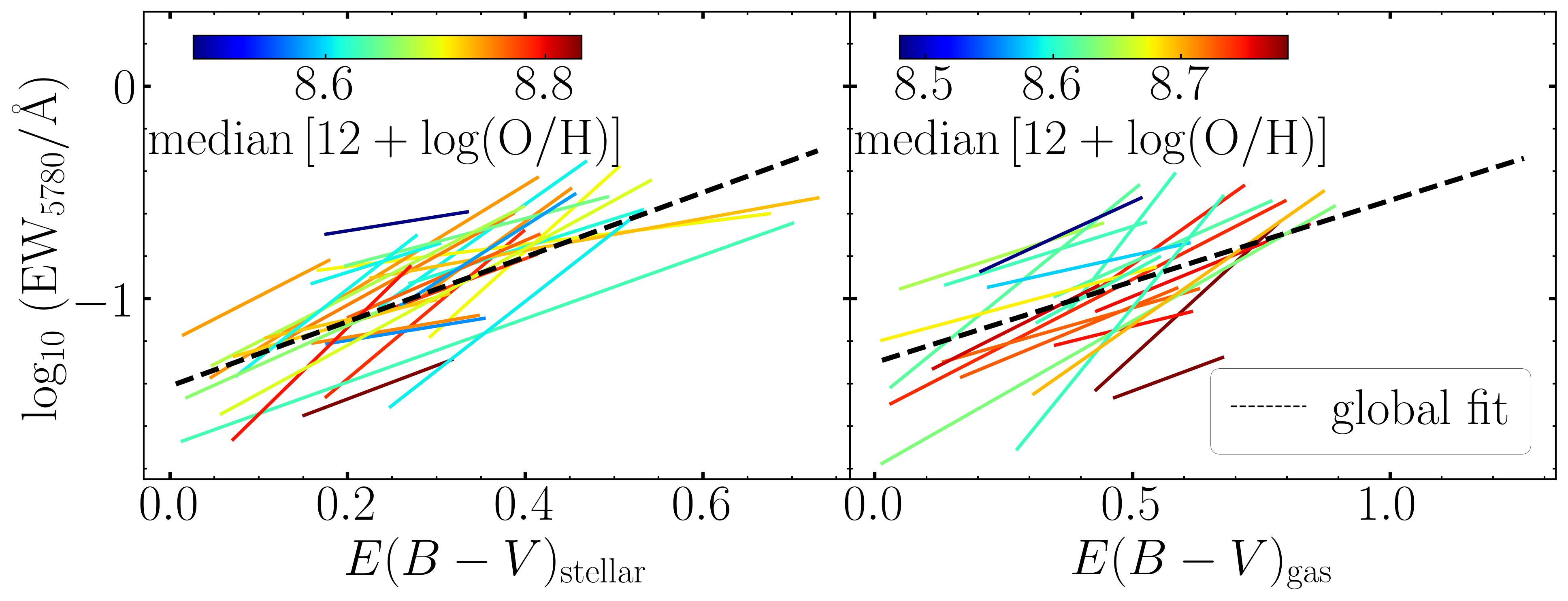}
    \caption{
    Linear regression fits between DIB~$\lambda5780$ EW and stellar (left panel) and gas-phase (right panel) extinction for individual galaxies.
    Only galaxies with at least 10 DIB~$\lambda5780$ detections are included.
    The stellar-extinction panel includes galaxies whose sub-galactic regions intrinsically span $\Delta E(B-V)_{\rm stellar}>0.1$~mag, while the gas-extinction panel includes those spanning $\Delta E(B-V)_{\rm gas}>0.2$~mag.
    Each colored line corresponds to the best-fitting relation for a single galaxy, while the black dashed line shows the global fit obtained from the full sample.
    The individual-galaxy relations are color-coded according to the median gas-phase metallicity, $12+\log(\mathrm{O/H})$, of each galaxy, with the color scale in each panel spanning the metallicity range of the galaxies shown.
    }
    \label{fig:DIB_vs_physical_reg_sep}
\end{figure}

The large scatter around the global DIB--extinction relations may partly reflect variations in the relation among individual galaxies.
Color-coding the individual regions in Fig.~\ref{fig:DIB_vs_physical} by their gas-phase metallicity suggests that metallicity may contribute to the observed scatter, particularly for the relation with gas-phase extinction.
To investigate the galaxy-to-galaxy variations more directly, Figure~\ref{fig:DIB_vs_physical_reg_sep} shows the best-fitting linear relations for individual galaxies with more than 10 DIB~$\lambda5780$ detections and whose sub-galactic regions intrinsically span $\Delta E(B-V)_{\rm stellar}>0.1$~mag or $\Delta E(B-V)_{\rm gas}>0.2$~mag for the respective relation.
Requiring a minimum number of DIB detections and a sufficient intrinsic extinction range reduces the influence of poorly constrained fits based on only a few regions or a narrow range in $E(B-V)$.
The individual relations exhibit substantial variations in both slope and normalization relative to the global relation.
Color-coding these relations by the median gas-phase metallicity of each galaxy suggests that metallicity may contribute to these variations, particularly for the gas-extinction relation, although the substantial scatter prevents a clear trend from being established.

\subsection{DIB detection and galaxy inclination}
\label{sec:inclination}

The detection of DIB~$\lambda5780$ may also depend on the inclination of the host galaxy, since more inclined systems can provide longer lines of sight through the interstellar medium.
We therefore investigated whether the axis ratio of the host galaxies differs between the DIB-detected and remaining galaxies in our sample.

For each galaxy, we adopt the minor-to-major axis ratio, $b/a$, provided by the HECATE catalogue \citep{2021MNRAS.506.1896K}.
The HECATE galaxy geometry is based on measurements from HyperLEDA \citep{2014A&A...570A..13M}, with the semimajor axes rescaled to match the $D_{25}$ isophotal values.
The DIB-detected galaxies have a histogram mode of $b/a=0.45$, compared with $b/a=0.75$ for the remaining galaxies.
The difference is also reflected in the median values, which are $b/a=0.590$ and $b/a=0.713$ for the DIB-detected and remaining galaxies, respectively.

The two galaxy-level distributions differ significantly according to a Kolmogorov--Smirnov test, with $D=0.248$ and $p=1.82\times10^{-7}$.
The DIB-detected galaxies therefore have systematically lower $b/a$ values than the remaining galaxies, corresponding to more inclined systems.
This trend is consistent with the possibility that longer lines of sight through the ISM in more inclined galaxies increase the observed DIB absorption.
However, the present analysis does not distinguish this geometrical effect from other possible effects associated with inclination, such as changes in the sampled dust column or DIB detectability.

\begin{figure}
    \centering
    \includegraphics[width=0.9\linewidth]{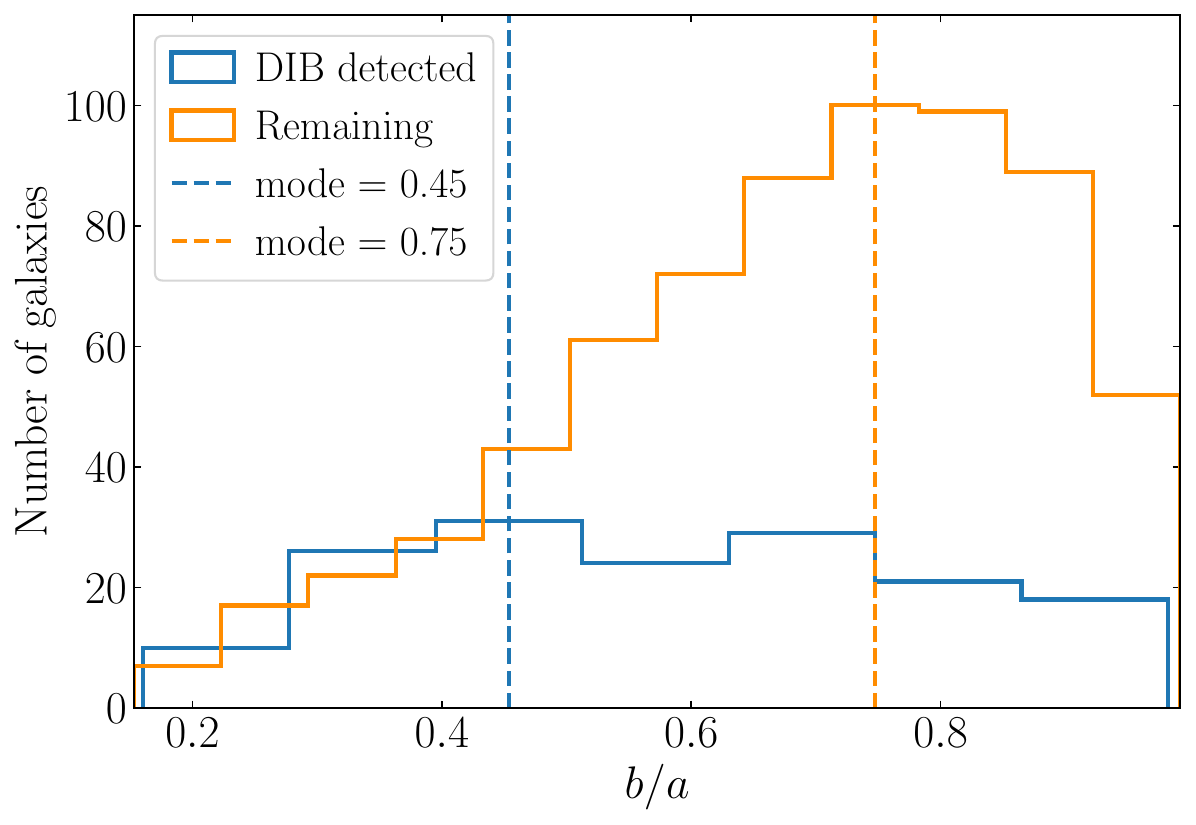}
    \caption{
    Distribution of the galaxy minor-to-major axis ratio, $b/a$, for the 159 galaxies satisfying the DIB~$\lambda5780$ detection criterion (blue histogram) and the remaining 678 galaxies in the sample (orange histogram).
    The vertical dashed lines indicate the corresponding modes of the two distributions, estimated using kernel-density estimation.
    }
    \label{fig:inclination}
\end{figure}

\subsection{Exploring environmental dependencies of DIBs}
\label{sec:DIBs_environment}

\begin{figure*}
    \centering
    \includegraphics[width=\linewidth]{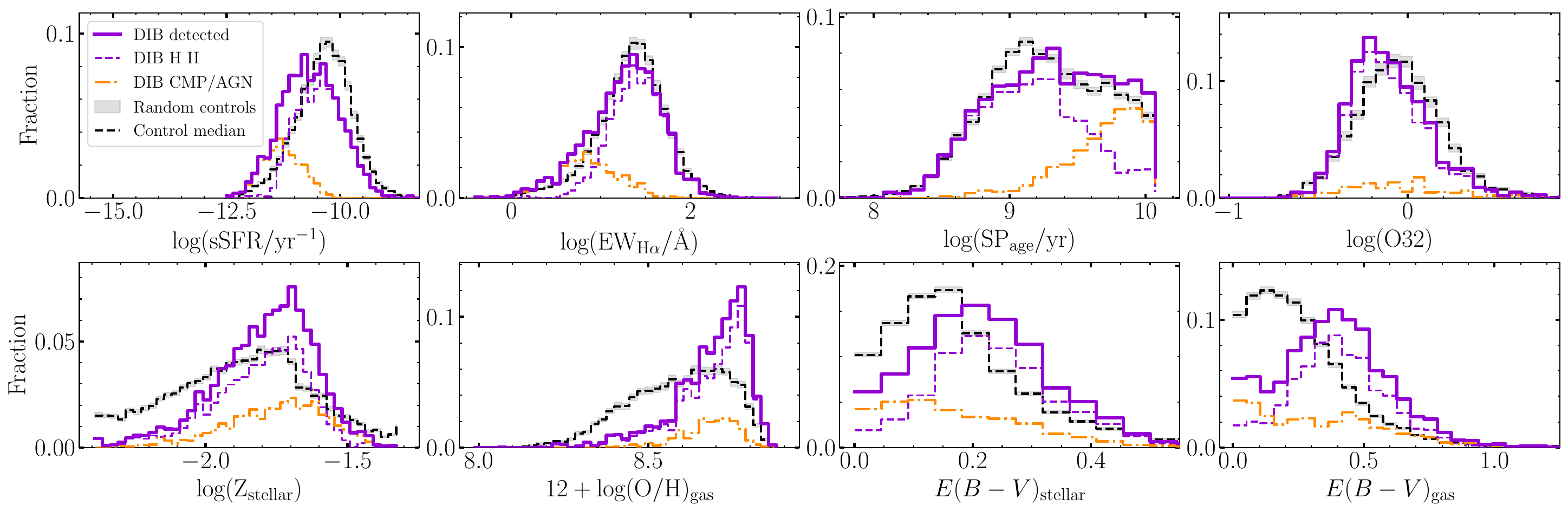}
    \caption{
    Distributions of several physical properties for spatially resolved regions with reliable DIB~$\lambda$5780 detections and for randomized control samples of non-DIB regions.
    The thick purple solid line shows the distribution of all DIB-detected regions, while the thin purple dashed and thick orange dash-dotted lines show the corresponding distributions for DIB detections in pure \ion{H}{II} and composite/AGN regions, respectively, as classified by the BPT diagram.
    The orange shaded region indicates the 16--84\% range of 5000 randomized control realizations, each containing the same number of spatial regions as the DIB sample, and the black dashed line shows the median control distribution.
    The DIB distributions are normalized to unit integral, while the H~II and composite/AGN DIB distributions are normalized to the total number of DIB-detected regions.
    From left to right and top to bottom, the panels show: sSFR, $\rm EW_{H\alpha}$, SP age, O32 index, stellar and gas-phase metallicities, and extinction.
    }
    \label{fig:Hist_DIB_All}
\end{figure*}

To assess the occurrence of the DIBs independently of their strength, we examine the conditions under which they are more frequently detected across galaxies and their sub-galactic regions.
Figure~\ref{fig:Hist_DIB_All} compares the distributions of several physical properties for the spatially resolved regions with reliable DIB~$\lambda$5780 detections with those obtained from randomized control samples of non-DIB regions.
Each control realization contains the same number of spatial regions as the DIB-detected sample, and the control distributions shown in the figure represent the median and the CIs obtained from 5000 randomized realizations.
This approach allows us to assess whether the distributions of the DIB-detected regions differ from those expected from random selections of regions from the non-DIB population, without imposing additional matching criteria on galaxy properties.
We also show the distributions of regions separated into purely star-forming and the rest (AGN and composite), as classified by the BPT diagram.

The DIB-detected regions show a clear shift toward lower recent star-forming activity compared with the randomized controls.
Since sSFR traces recent star formation per unit stellar mass, its distribution reflects the recent star-forming activity of each region.
The mode of the $\log(\mathrm{sSFR})$ distribution is lower by $\sim0.57$ dex in the DIB regions than in the randomized controls.
The DIB distribution also shows a slightly lower $\mathrm{EW}_{\rm H\alpha}$, although the difference between the modes is only $\sim0.03$ dex.
These results indicate that DIB detections are preferentially associated with regions with lower recent star-forming activity, while the difference in $\mathrm{EW}_{\rm H\alpha}$ is comparatively modest.
The H~II and composite/AGN subsamples contribute differently across the sSFR distribution, with the composite/AGN DIB detections contributing more strongly toward the low-sSFR regime.
The observed shift does not by itself establish whether the reduced star-forming activity is directly related to the survival of DIB carriers or instead reflects correlations with other physical properties of the regions.

The SP age distribution also shows a shift toward older stellar populations.
The mode of the DIB distribution is higher than that of the randomized controls by $\sim0.14$ dex.
The distributions nevertheless overlap substantially, indicating that the association with older stellar populations is weaker than the corresponding trends in sSFR, metallicity, or extinction.
The apparent peaks around mean ages of $\sim700$~Myr and $\sim2$~Gyr in the DIB distribution should therefore be interpreted as features of the distribution rather than as evidence for a distinct preferred stellar population age.
The AGN/composite-driven regions with DIB~$\lambda$5780 detections are also associated with even older SPs, consistent with their presence in low-sSFR regions.

The O32 distribution shows a clear shift toward lower ionization levels in the DIB-detected regions.
The mode of the $\log(\mathrm{O32})$ distribution is lower by $\sim0.12$ dex than that of the randomized controls.
The DIB distribution peaks at $\rm O32\simeq0.4$, while the control distribution is shifted toward higher ionization levels.
Thus, DIB detections are more commonly associated with regions characterized by relatively low ionization, although the two distributions overlap substantially.
This is consistent with DIBs being more readily detected in moderately ionized environments and less frequently detected in regions dominated by harder and stronger radiation fields, potentially due to changes in carrier abundance, ionization state, or detectability.

The most pronounced separation occurs in the distribution of gas-phase metallicity.
The mode of the gas-phase metallicity distribution is higher by $\sim0.08$ dex in the DIB-detected regions than in the randomized controls.
DIB~$\lambda5780$ is rarely detected in regions with $\rm 12 + \log(O/H) < 8.3$ within the sensitivity limits of our data, and the detection peak lies near $\rm 12 + \log(O/H) \simeq 8.8$, corresponding to super-solar metallicities.
The gas-phase metallicity therefore shows one of the clearest shifts between the DIB-detected and randomized control regions.
This trend is consistent with the idea that higher metal abundance in the ISM may be associated with conditions favorable for the formation or survival of large molecules, including potential DIB carriers.
The stellar metallicity distribution shows a smaller shift, with the mode higher by $\sim0.05$ dex in the DIB-detected regions.
The stronger difference in gas-phase metallicity suggests that the association of DIB detections with metallicity is particularly evident in the ISM abundance measurements.

As expected, the DIB detections are also shifted toward regions with higher extinction, in both stellar and gas-phase measurements.
The difference is particularly pronounced for gas-phase extinction, for which the mode of the DIB distribution is higher by $\sim0.28$ mag than that of the randomized controls.
The corresponding difference in the stellar-extinction modes is $\sim0.06$ mag.
The gas-phase extinction mostly traces dense, dusty star-forming regions, while the stellar extinction probes emission from older and more widespread stars that may have already cleared their dust cocoons.
The stronger shift in gas-phase extinction indicates that detectable DIB absorption is more strongly associated with regions having larger nebular extinction than with the stellar extinction measured from the stellar continuum.
This is consistent with previous observational results and supports the view that DIBs trace dust-enriched components of the ISM.

Overall, the randomized-control comparison shows that the clearest differences between DIB-detected and non-DIB regions occur in sSFR, gas-phase metallicity, and gas-phase extinction, while the differences in $\mathrm{EW}_{\rm H\alpha}$, O32, and stellar-population age are comparatively modest.
The H~II and composite/AGN DIB populations are shown separately in Figure~\ref{fig:Hist_DIB_All} to illustrate how these trends are distributed between the two ionization classes.
We emphasize that this analysis is intended to characterize associations in the local physical conditions of DIB-detected regions and does not by itself isolate independent causal drivers, since the investigated quantities are mutually correlated.

\section{Discussion}
\label{sec:discussion}

Building on the trends identified in Section~\ref{sec:DIBs_environment}, we interpret the environmental dependencies of DIB detectability across the MaNGA sample.  
The observed correlations with extinction, metallicity, and ionization reflect a combination of intrinsic environmental dependencies and detectability effects.  
In particular, DIBs are more frequently detected in regions associated with star formation than in the diffuse, widespread dust component of the ISM.  
We emphasize, however, that the absence of detections in certain environments should be interpreted in terms of detectability rather than a strict physical absence of carriers, given the sensitivity limits of the data.  

We find that DIBs are more frequently detected in regions with high metallicity ($\rm 12 + \log(O/H) > 8.5$), consistent with Magellanic Cloud studies showing that lower metallicities suppress DIB strengths due to the reduced abundance of carbon and dust grain precursors \citep{2006ApJS..165..138W,2007A&A...470..941C}.  
In our sample, DIB detections become increasingly rare in galaxies or sub-regions below $\rm 12 + \log(O/H) < 8.3$, consistent with DIBs being more readily detected in chemically mature ISM environments.  
Nevertheless, deeper or higher signal-to-noise observations may reveal DIBs even in lower-metallicity systems.  

DIB detections are preferentially associated with regions of modestly lower O32, indicating that detectable DIB absorption is less common under harder or more highly ionized conditions.
This is consistent with previous Galactic and extragalactic studies suggesting that hard radiation fields influence the detectability of DIB carriers \citep[e.g.,][]{2011A&A...533A.129V, 2015ApJS..216...33F}.  
While $\mathrm{EW}_{\mathrm{H}\alpha}$, a tracer of radiation intensity, shows only minor differences between regions with and without DIB detections, O32 reveals a clearer separation.  
This suggests that radiation hardness may be more closely associated with reduced DIB detectability, although this effect cannot be cleanly separated from variations in ionization parameter, since O32 is sensitive to both.  

DIB detections are more common in galaxies and regions with moderate star formation activity ($\rm -10.5 \lesssim \log(\mathrm{sSFR}/\mathrm{yr}^{-1}) \lesssim -9$), while they are less frequent in extreme starbursts and highly passive environments.  
In passive systems, only a few sporadic DIB detections are observed, primarily in regions classified as AGN or composite in the BPT diagram.  
These detections are associated with relatively mild ionization conditions, as indicated by their O32 distribution, suggesting that the presence of DIBs in such environments is linked to local conditions rather than global star formation activity.  
Collectively, these results are consistent with DIBs being more readily detected in evolved regions of the ISM that are enriched in metals and dust and exposed to moderate ionization levels.  

A growing body of observational evidence indicates that PAHs exhibit environmental dependencies similar to those found here for DIBs, including suppression in low-metallicity systems and in strongly ionized environments, as well as a positive correlation with dust extinction \citep[e.g.,][]{2005ApJ...628L..29E,2006A&A...446..877M,2007ApJ...656..770S,2022ApJ...931...38M}.  
In particular, the metallicity threshold for PAH emission is comparable to that inferred in our analysis, where DIB detections become rare below $12 + \log(\mathrm{O/H}) \simeq 8.3$.  
These similarities suggest that PAHs and DIB carriers may respond in a comparable way to the chemical enrichment and radiation environment of the ISM, although this does not necessarily imply a common origin \citep[e.g.,][]{2008ApJ...679..310G}.  
Recent JWST observations have further shown that PAH emission can persist down to metallicities of $12 + \log(\mathrm{O/H}) \simeq 7.7$ in some high-redshift ($z \gtrsim 2$) galaxies, albeit with rapidly declining strength \citep[e.g.,][]{2024A&A...690A..89S}.  

It should be emphasized that this comparison is phenomenological.  
While PAHs are plausible contributors to the population of DIB carriers, the two classes of features are not necessarily produced by identical molecules, and their correlations may instead reflect shared sensitivities to similar ISM conditions.  
Our results therefore suggest that the carriers of DIBs are widespread and influenced by common ISM processes, although their exact nature remains uncertain.  

Although only a fraction of galaxies ($\sim 1/5$) show significant DIB detections in the sample analyzed in this work, this is likely driven by the limited S/N of the stellar continuum in MaNGA spectra, the intrinsically faint nature of DIB absorption features, and additional observational constraints such as the moderate spectral resolution.  
Despite these limitations, our results show that DIB~$\lambda$5780 is detected across a broad range of galaxy types and environments, and are not restricted to extreme, dusty starburst systems.  
Most star-forming galaxies in the local Universe exhibit near-solar gas-phase metallicities \citep[e.g.,][]{2008ApJ...681.1183K,2013ApJ...765..140A,2021MNRAS.506.3079K}, with typical gas-phase dust extinction values peaking around $E(B{-}V) \simeq 0.3$ \citep[e.g.,][]{2016ApJ...818...13B,2021MNRAS.506.1896K,2023A&A...673A..16K}.  
These conditions suggest that the carbon-based macromolecules associated with DIB absorption are likely widespread in chemically enriched and dust-rich regions of the ISM in the local Universe.  

The detection of DIBs in common star-forming environments indicates that their carriers, likely large organic molecules, are not confined to rare or extreme ISM conditions.  
Both DIBs and PAHs tend to be weak in low-metallicity and strongly ionized environments, but are more readily detected in galaxies with moderate star formation, near-solar abundances, and well-developed dust reservoirs, which typify much of the local galaxy population.  
Carbon-rich molecules such as PAHs are considered important components of complex organic chemistry and have been proposed as contributors to the early organic inventory of planetary systems \citep[e.g.,][]{2002Natur.416..403M,2002SGeo...23..411B}.  
The detection of DIBs in spatially resolved regions of normal star-forming galaxies is therefore consistent with the presence of complex carbonaceous molecules in typical galactic ISM environments.  
In this context, DIBs serve as astrophysical probes of ISM conditions and as tracers of chemically enriched environments relevant to the study of complex organic chemistry in galaxies.  

\section{Conclusions}
\label{sec:conclusions}

We have conducted the most extensive spatially resolved extragalactic study to date of the DIB~$\lambda$5780 absorption feature, utilizing IFU spectroscopy from the SDSS-IV MaNGA survey.  
By analyzing subgalactic regions in 837 nearby galaxies spanning a wide range of star formation activity, extinction, metallicity, and ionization conditions, we identify key environmental dependencies that modulate DIB detectability.

Our main findings are:

\begin{itemize}
    \item We detect the DIB~$\lambda$5780 in the edge-on galaxy LEDA~1157026 at 232~Mpc, representing the most distant spatially resolved DIB measurement reported to date and demonstrating that such features can be traced well beyond the Local Group.

    \item The strongest and most consistent trend is a positive correlation between DIB~$\lambda$5780 EW and dust extinction.  
    The correlation is evident for both stellar and gas-phase extinction, although the fitted relations differ in slope owing to the known offset between stellar and nebular reddening.

    \item DIB detections predominantly occur in regions with high gas-phase metallicities ($\rm 12 + \log(O/H) \gtrsim 8.5$), consistent with DIBs being more readily detectable in chemically evolved ISM environments.

    \item DIBs are more commonly detected in regions with low-to-moderate ionization levels ($\rm O32 \lesssim 1$), while detections decline in high-ionization environments, indicating reduced detectability under strong radiation fields.

    \item DIBs are most frequently detected in regions with intermediate star formation activity ($\rm -10.5 \lesssim \log(\mathrm{sSFR}/yr^{-1}) \lesssim -9$).  
    In passive galaxies, detections are rare and typically confined to AGN/composite regions, as classified by BPT diagnostics, where ionization is likely not dominated by young stars.

    \item DIB detections are more frequently found in regions with relatively evolved stellar populations, consistent with their preferential occurrence in chemically mature ISM environments.

    \item The environmental dependencies of DIBs, particularly their suppression in low-metallicity and high-ionization regions, closely parallel those of PAHs.  
    These similarities suggest that DIB carriers may share comparable environmental dependencies with PAHs shaped by metallicity, dust abundance, and radiation field intensity.

    \item DIB detections are preferentially found in more inclined galaxies than the remaining sample, as indicated by their lower minor-to-major axis ratios.
    This association suggests that galaxy inclination and line-of-sight geometry may contribute to the detectability of  DIB absorption.
    
\end{itemize}

These results demonstrate that DIBs are not restricted to extreme dusty starburst or AGN systems but are instead a widespread, albeit subtle, feature of typical star-forming galaxies with near-solar metallicity and moderate ISM conditions.  
This work is consistent with the view that large carbonaceous molecules are common in the ISM of typical star-forming galaxies in the local Universe, although their abundance and distribution remain to be fully constrained.  

Future directions include detailed comparisons between DIB absorption and mid-IR PAH emission, dust-to-gas ratios, and UV extinction curves, as well as extending this methodology to additional DIB features (e.g., $\lambda$4430 and $\lambda$6284) and higher-redshift systems with upcoming IFU surveys.

\section*{Acknowledgements}

We sincerely thank the anonymous referee for their careful reading of the manuscript and constructive comments, which helped us improve the clarity and robustness of our analysis.
This work was supported by the institutional project RVO:67985815 and by the INTER-COST LUC24023 project of the INTER-EXCELLENCE II programme, funded by the Czech Ministry of Education, Youth and Sports.
This article/publication is based upon work from COST Action CA21126 -- Carbon molecular nanostructures in space (NanoSpace), supported by COST (European Cooperation in Science and Technology).
B.D. acknowledges funding from the HTM (grant TK202), the EU Horizon Europe programme through EXCOSM (grant No. 101159513), and the Estonian Research Council (grant PSG1200).
This research made use of the SAOImage DS9 software provided by the Smithsonian Astrophysical Observatory, the NASA/IPAC Extragalactic Database (NED), which is operated by the Jet Propulsion Laboratory, California Institute of Technology, under contract with NASA, and the NASA/IPAC Infrared Science Archive (IRSA), which is funded by NASA and operated by the California Institute of Technology.
This research made use of data from the Sloan Digital Sky Survey IV Mapping Nearby Galaxies at Apache Point Observatory (SDSS-IV MaNGA) survey. Funding for the Sloan Digital Sky Survey IV has been provided by the Alfred P. Sloan Foundation and the Participating Institutions. SDSS acknowledges support and resources from the Center for High Performance Computing at the University of Utah. The SDSS website is \url{https://www.sdss.org/}.

\section*{Data Availability}

The data underlying this article are publicly available. The MaNGA DR17 data products analysed in this work are available through the Sloan Digital Sky Survey (SDSS) Science Archive Server at \url{https://www.sdss.org/dr17/manga/}. Additional galaxy properties were obtained from the public MPA--JHU catalogue and the Heraklion Extragalactic Catalogue (HECATE). The derived DIB measurements and analysis tables generated during this study are available from the corresponding author upon reasonable request.



\bibliographystyle{mnras}
\bibliography{citations} % if your bibtex file is called example.bib

%\begin{thebibliography}{99}
%\end{thebibliography}

%%%%%%%%%%%%%%%%%%%%%%%%%%%%%%%%%%%%%%%%%%%%%%%%%%

%%%%%%%%%%%%%%%%% APPENDICES %%%%%%%%%%%%%%%%%%%%%

\appendix

% \section{Some extra material}

% If you want to present additional material which would interrupt the flow of the main paper,
% it can be placed in an Appendix which appears after the list of references.

\clearpage
% \onecolumn
% \begin{appendix}
\appendix
% \FloatBarrier

\setcounter{figure}{0} % Restart figure numbering
% \clearpage

\section{Assessment of Measurement Reliability}
\label{sec:Appendix_A}

\setcounter{figure}{0} % Restart figure numbering
\renewcommand{\thefigure}{A\arabic{figure}}% Figure counter representation
\renewcommand{\theHfigure}{A\arabic{figure}}% Hyperref figure hyperlink hook

\noindent

\subsection{Control-Window Analysis of Noise-Induced Detections}

To assess the contribution of spurious detections arising from residual noise in the continuum-subtracted spectra, we applied the same absorption-line fitting procedure, significance thresholds, and quality cuts used for the DIB analysis (Sections~\ref{sec:PowerBin}, \ref{sec:Spectral_fitting}, and \ref{sec:Sherpa_fits}) to several control wavelength windows where no interstellar absorption is expected.
The control windows were selected both far from ($\lambda = 5050\,\text{\AA}$) and closer to the DIB spectral region ($\lambda\lambda = 5600, 5750 ~ \text{\AA}$) to probe the false-positive rate under identical continuum, noise, and fitting conditions.

\begin{figure}
    \centering
    \includegraphics[width=0.99\linewidth]{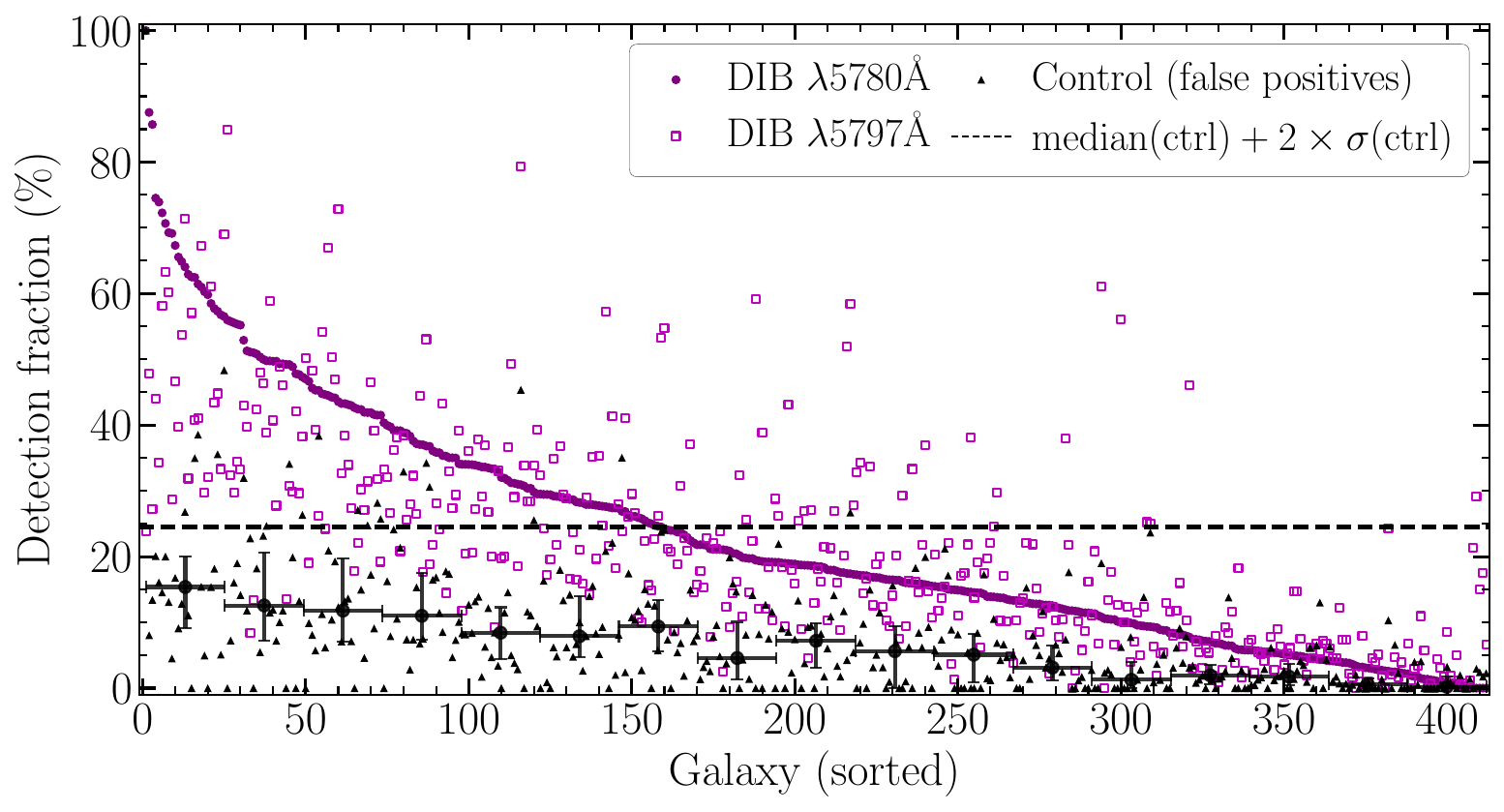}
    \caption{
    Detection fraction of absorption features as a function of their sorted number based on the detection fraction for all galaxies in the sample.
    Galaxies are sorted along the x-axis by decreasing detection fraction of the DIB $\lambda5780$, as listed in Table~\ref{tab:List_table_1}.
    Filled purple and open magenta circles show the detection fractions of DIB $\lambda5780$ and DIB $\lambda5797$, respectively. 
    Triangles indicate the detection fractions of the average of the control absorption windows at $\lambda5050$, $\lambda5600$, and $\lambda5750$.
    The black dashed horizontal lines mark the detection threshold, which is defined as $\mathrm{median}(\mathrm{ctrl}) + 2 \times \sigma_{\mathrm{ctrl}} = 24.5 \%$.
    The black errorbars show the modes and 67\% C.I.'s of the control absorption window measurements of the sources grouped in 17 bins across the examined sample.
    }
    \label{fig:Noise_fractions}
\end{figure}

\begin{figure*}
    \centering
    \includegraphics[width=\linewidth]{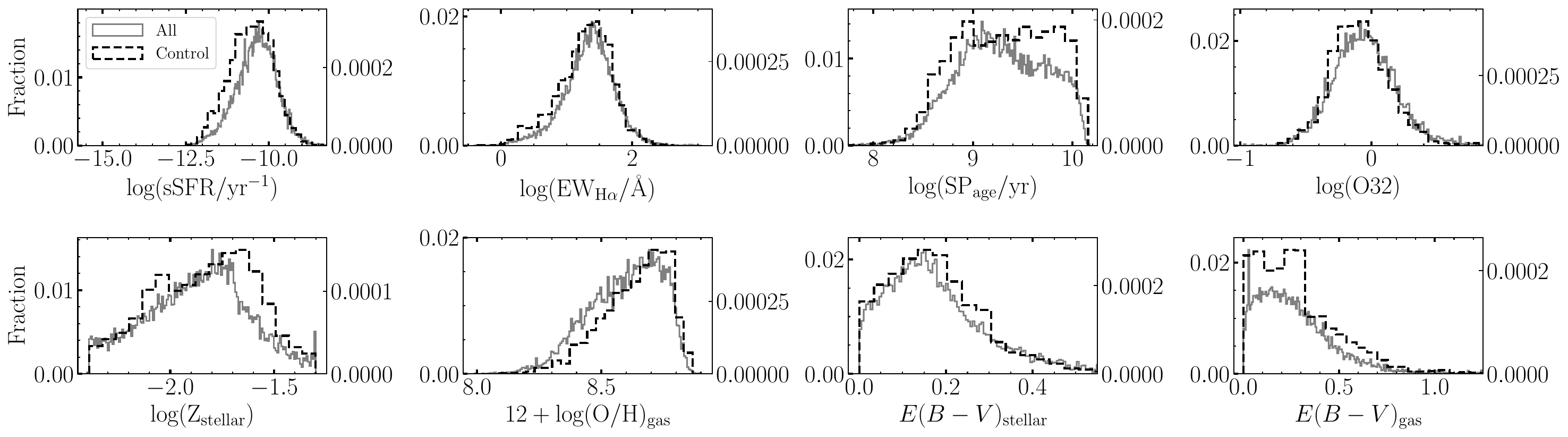}
    \caption{
    Distributions of environmental and physical parameters of the control wavelength-window detections in the continuum-subtracted residual spectra, shown for comparison with Figure~\ref{fig:Hist_DIB_All}.
    The gray histograms represent the parent distribution of all analyzed spatial bins across the full galaxy sample that satisfy the adopted quality cuts.
    The black histograms correspond to detections of the average of the control absorption windows at $\lambda5050$, $\lambda5600$, and $\lambda5750$, fitted using the same methodology, significance thresholds, and quality criteria.
    The left y-axis corresponds to the parent (gray) histogram, while the right y-axis corresponds to the colored detection-fraction distributions.
    }
    \label{fig:Noise_distributions}
\end{figure*}

The resulting control detections occur at significantly lower rates than the DIB detections and do not exhibit coherent dependencies on extinction or ionization conditions (Figures \ref{fig:Noise_fractions} and \ref{fig:Noise_distributions}).
However, the distribution of the control features is mildly skewed toward high metallicity environments.
This behavior is expected, as residual spectral structure after SP fitting is known to increase with metallicity due to enhanced line blanketing, SP complexity, and template mismatch in metal-rich systems \citep[e.g.,][]{2003MNRAS.344.1000B,2005MNRAS.362..799M,2020MNRAS.491.2025C}.
Such effects can give rise to rare spurious absorption-like features in galaxy spectra that mimic interstellar absorption features.

In contrast, the DIB $\lambda5780$ detections exhibit additional and independent selectivity toward higher extinction and lower ionization conditions, as well as coherent galaxy-scale trends.
These multi-parameter dependencies are not reproduced by the control features, demonstrating that the mild metallicity-dependent residual structure alone cannot account for the environmental behavior observed for the DIB detections.

Figure~\ref{fig:Noise_fractions} further shows that the detection fractions of the control features remain on average below $\sim 10\%$, substantially lower than those of the DIBs.
The median control detection fraction is $4.9\%$.
We define a conservative detection threshold based on the upper envelope of the control distribution as $\mathrm{median}(\mathrm{ctrl}) + 2 \times \sigma_{\mathrm{ctrl}} = 24.5\%$, where $\sigma_{\mathrm{ctrl}}$ is the standard deviation of the control detection fractions across the sample.
This threshold selects 159 galaxies above it.
The binned control-window measurements remain systematically below the adopted 24.5\% threshold across the full sample, indicating that the threshold is not driven by a small number of extreme outliers but lies well above the typical contamination level inferred from the control windows.

For these galaxies, the DIB detection rates clearly exceed the false positive level defined by the control windows.
Below this threshold, the observed DIB detection fractions approach the level expected from rare stochastic residuals and are therefore not interpreted as robust individual galaxy detections, although they remain statistically meaningful in the ensemble analysis.
Moreover, while the DIB~$\lambda5797$ detections broadly follow the same galaxy-to-galaxy trend as DIB~$\lambda5780$, they exhibit increased scatter due to their lower S/N.
In contrast, the control features remain distributed around their median level with no systematic dependence on galaxy ordering.
These tests demonstrate that the environmental trends reported in this work are not driven by residual noise fluctuations.

\subsection{Monte Carlo assessment of DIB measurement robustness}

To further assess the robustness of the DIB measurements against statistical fluctuations in the continuum-subtracted spectra, we performed Monte Carlo (MC) perturbation tests on a representative subset of galaxies spanning the full range of DIB detection fractions listed in Table~\ref{tab:List_table_1}.

For each spatial-bin spectrum, the continuum-subtracted residual spectrum was perturbed through $N_{\rm MC}=200$ independent Gaussian noise realizations. In each realization, random noise was added to every spectral pixel, with the width of the Gaussian distribution determined by the propagated spectral uncertainty adopted in the \texttt{Sherpa} fitting procedure (Section~\ref{sec:Sherpa_fits}). The full DIB fitting procedure was then repeated using the same fitting configuration and wavelength intervals adopted for the nominal measurements. From the resulting ensemble of realizations, we estimated the empirical MC uncertainty of the measured equivalent widths as the standard deviation of the recovered EW distributions:
\begin{equation}
\sigma_{\rm MC}({\rm EW}) =
\sqrt{
\frac{1}{N-1}
\sum_{i=1}^{N}
\left(
{\rm EW}_i - \langle {\rm EW} \rangle
\right)^2
}.
\end{equation}

\begin{figure}
    \centering
    \includegraphics[width=0.99\columnwidth]{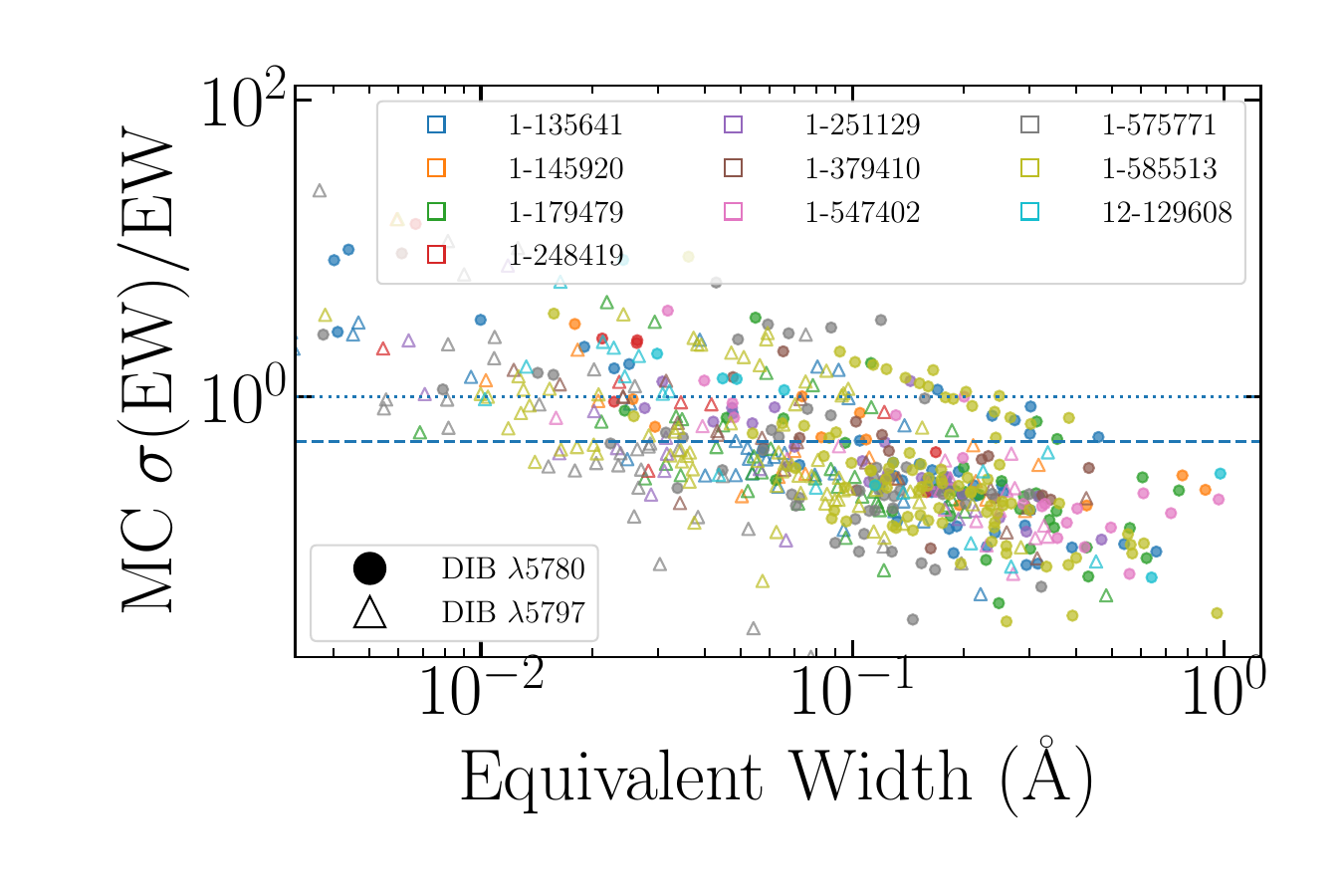}
    \caption{
    Fractional MC uncertainties of the DIB equivalent-width measurements for a representative subset of galaxies spanning the range of DIB detection fractions listed in Table~\ref{tab:List_table_1}.
    Filled circles show DIB~$\lambda5780$ measurements, while open triangles show DIB~$\lambda5797$ measurements.
    Different colors represent different galaxies.
    Dashed and dotted horizontal lines indicate fractional uncertainties of $\sigma_{\rm MC}({\rm EW})/{\rm EW}=0.5$ and $1.0$, respectively.
    }
    \label{fig:MC_uncertainties}
\end{figure}

Figure~\ref{fig:MC_uncertainties} shows the resulting fractional MC uncertainties,
$\sigma_{\rm MC}({\rm EW})/{\rm EW}$,
for both DIB~$\lambda5780$ and DIB~$\lambda5797$ measurements.
As expected, the weaker DIB~$\lambda5797$ generally exhibits systematically larger fractional uncertainties than DIB~$\lambda5780$, particularly at low equivalent widths, where the measurements are more sensitive to statistical fluctuations.
Conversely, stronger DIB detections display substantially smaller fractional MC uncertainties, indicating significantly more stable and robust fits under spectral perturbations.

The MC analysis shows a continuous transition between robust and increasingly stochastic measurements rather than a distinct separation between ``real'' and ``spurious'' detections.
The majority of the stronger DIB~$\lambda5780$ measurements remain stable under repeated perturbations, while DIB~$\lambda5797$ measurements become progressively more sensitive to statistical fluctuations at low EW values, consistent with its intrinsically weaker absorption strength.

These tests demonstrate that DIB~$\lambda5797$ is substantially more sensitive to statistical fluctuations than DIB~$\lambda5780$, consistent with its intrinsically weaker absorption strength.
We therefore restrict the quantitative environmental analysis and scientific conclusions of this work to DIB~$\lambda5780$.
The $\lambda5797$ measurements are retained only as a consistency check of the spectral fitting procedure.

\renewcommand{\thefigure}{B\arabic{figure}}% Figure counter representation
\renewcommand{\theHfigure}{B\arabic{figure}}% Hyperref figure hyperlink hook

\section{Visualization of Measurements in Individual Galaxies}
\label{sec:Appendix_B}

This Appendix presents representative examples of spatially resolved maps for several key physical properties of individual galaxies analyzed in this study.
The accompanying figures illustrate the spatial correspondence between these properties and the measured DIB equivalent widths.
Finally, we provide the list of galaxies with detected DIBs included in the analysis.

\begin{figure*}
    \centering
    \includegraphics[width=0.95\linewidth]{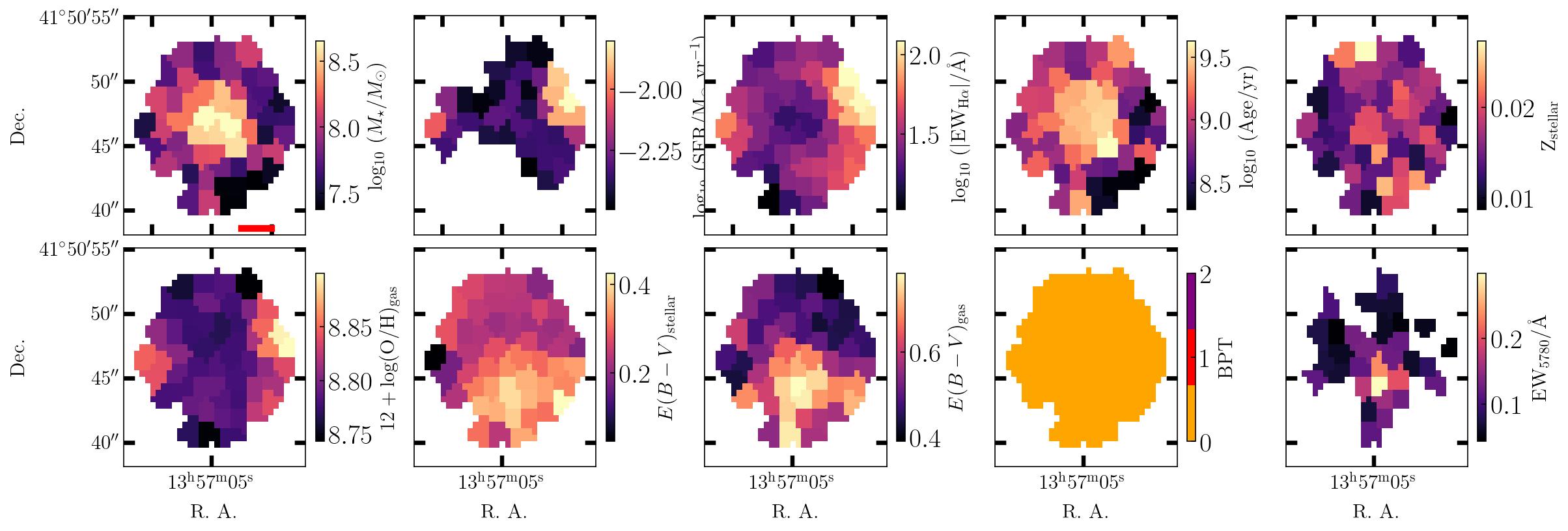}
    \caption{
    Spatially resolved maps of physical and spectral properties for NGC~5383 (MaNGA ID~1-575771).
    Each panel corresponds to a different parameter measured in individual spatial bins, derived from spectral and DIB fitting.
    Top row (left to right):
    $M_\star$, SFR, $\mathrm{EW}_{\mathrm{H}\alpha}$, SP age, and stellar metallicity.
    Bottom row:
    Gas-phase metallicity, extinction,
     BPT classification where 0 corresponds to purely star-forming regions, 1 to composite, and 2 to AGN/composite, and $\mathrm{EW}_{5780}$.
    The red line in the upper-left panel marks the projected length corresponding to 1~kpc at the distance of the galaxy.
    }
    \label{fig:Maps_1-575771}
\end{figure*}

\begin{figure*}
    \centering
    \includegraphics[width=0.95\linewidth]{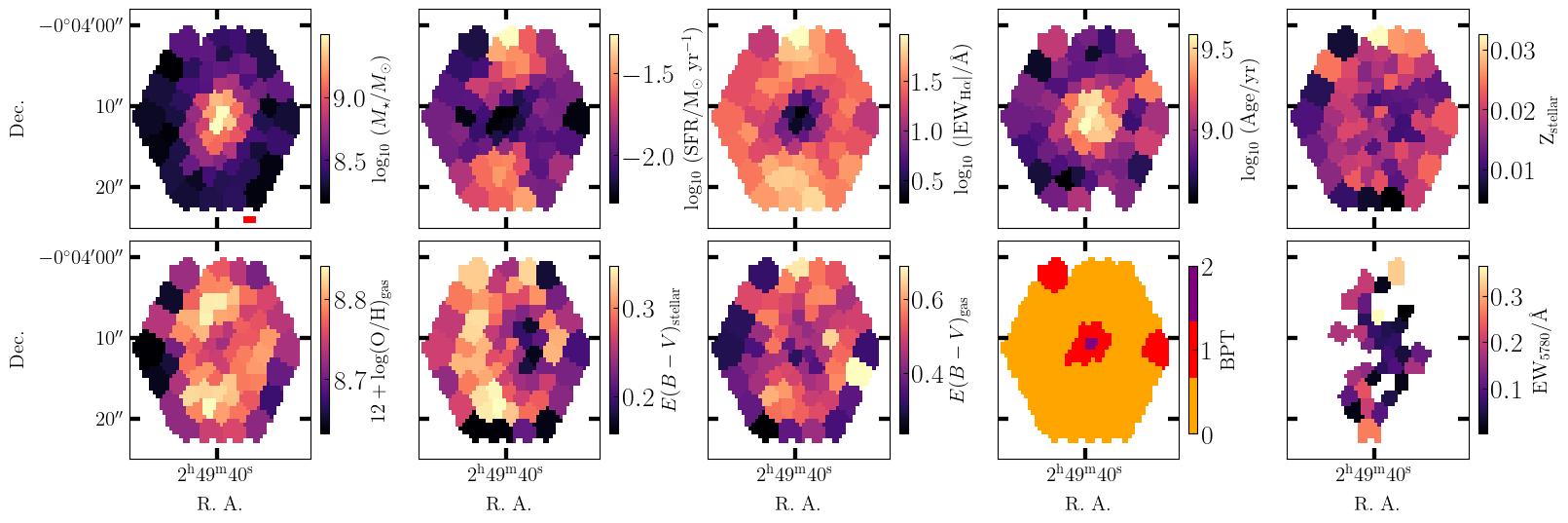}
    \caption{
    Spatially resolved maps of physical and spectral properties derived for the galaxy IC 263 (MaNGA ID~1-37213).
    The panels display the same parameters as shown in Figure~\ref{fig:Maps_1-575771}.
    The BPT classification map reveals the presence of an AGN, which produces a harder ionizing radiation field in the central region of the galaxy.
    }
    \label{fig:Maps_1-37213}
\end{figure*}

\begin{figure*}
    \centering
    \includegraphics[width=\linewidth]{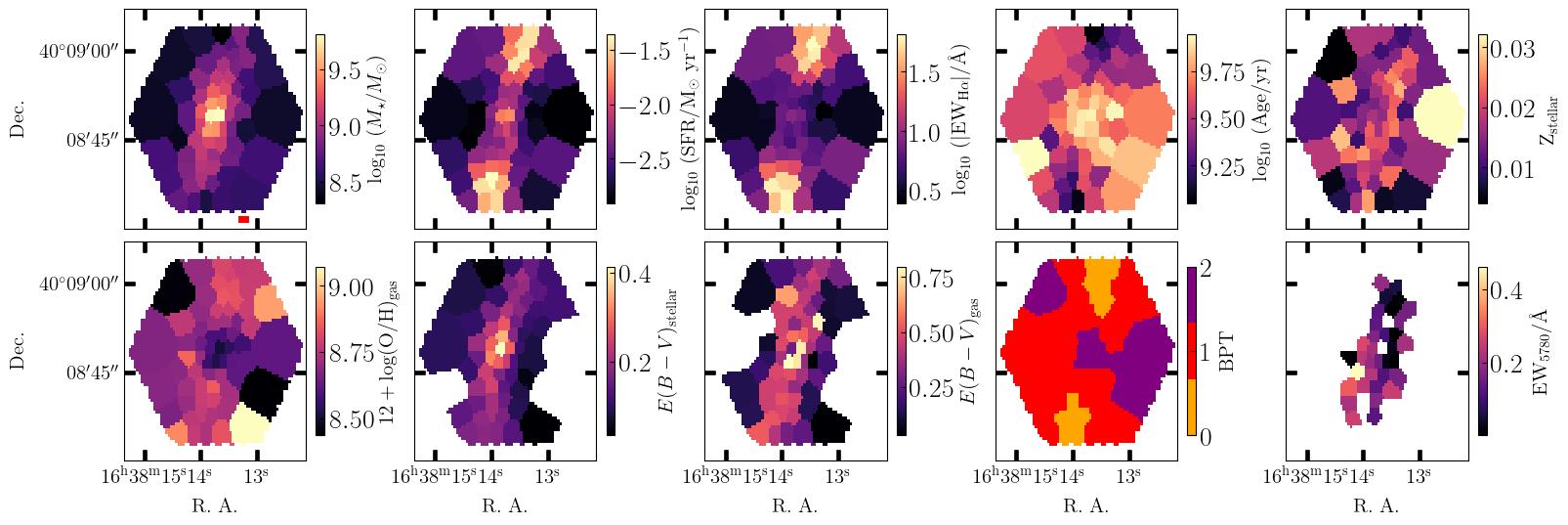}
    \caption{
    Spatially resolved maps of physical and spectral properties derived for the galaxy Z~224-81 (MaNGA ID 1-135641).
    The panels display the same parameters as shown in Figure~\ref{fig:Maps_1-575771}.
    % The BPT classification map reveals the presence of an AGN.
    }
    \label{fig:Maps_1-135641}
\end{figure*}

\begin{figure*}
    \centering
    \includegraphics[width=\linewidth]{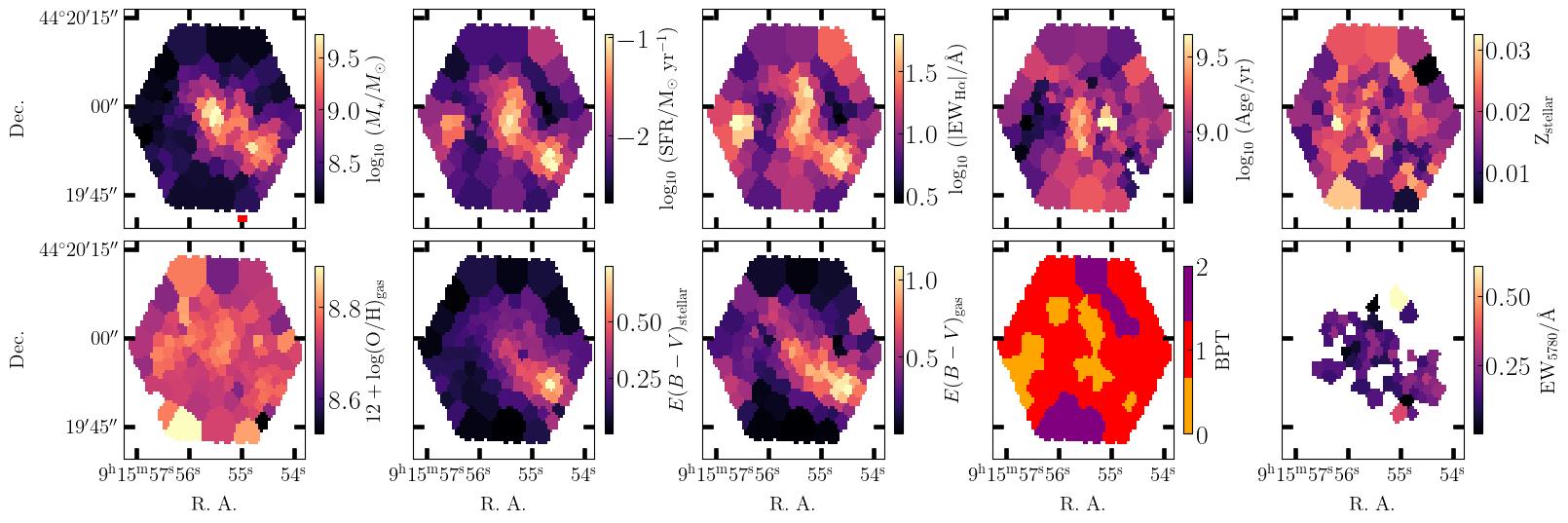}
    \caption{
    Spatially resolved maps of physical and spectral properties derived for the galaxy KUG~0912+445A (MaNGA ID 1-585513).
    The panels display the same parameters as shown in Figure~\ref{fig:Maps_1-575771}.
    This source also shows signs of composite/AGN activity, in agreement with the literature classification of the source \citep[e.g.][]{2011A&A...534A..31G}.
    }
    \label{fig:Maps_1-585513}
\end{figure*}

\begin{figure*}
    \centering
    \includegraphics[width=\linewidth]{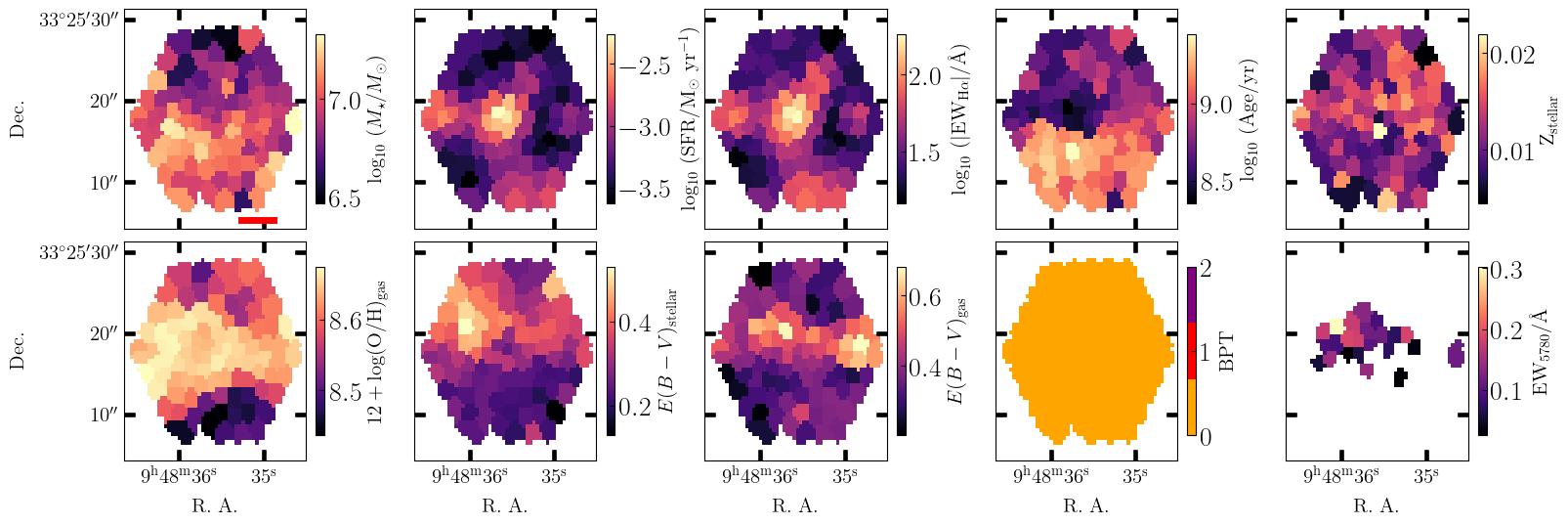}
    \caption{
    Spatially resolved maps of physical and spectral properties derived for the galaxy NGC~3003 (MaNGA ID~1-585731).
    The panels display the same parameters as shown in Figure~\ref{fig:Maps_1-575771}.
    % The galaxy shows no signs of AGN activity. 
    }
    \label{fig:Maps_1-585731}
\end{figure*}

\begin{figure*}
    \centering
    \includegraphics[width=\linewidth]{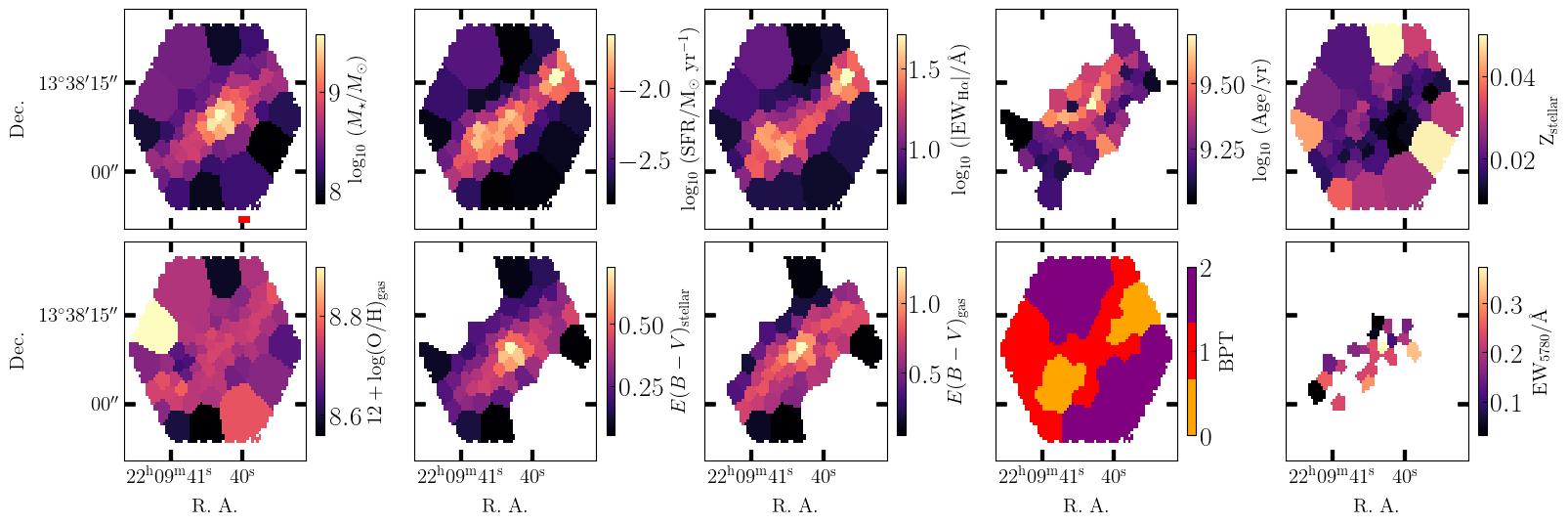}
    \caption{
    Spatially resolved maps of physical and spectral properties derived for the galaxy NGC~3003 (MaNGA ID~1-636015).
    The panels display the same parameters as shown in Figure~\ref{fig:Maps_1-575771}.
    % The galaxy shows no signs of AGN activity. 
    }
    \label{fig:Maps_1-636015}
\end{figure*}

\begin{figure*}
    \centering
    \includegraphics[width=\linewidth]{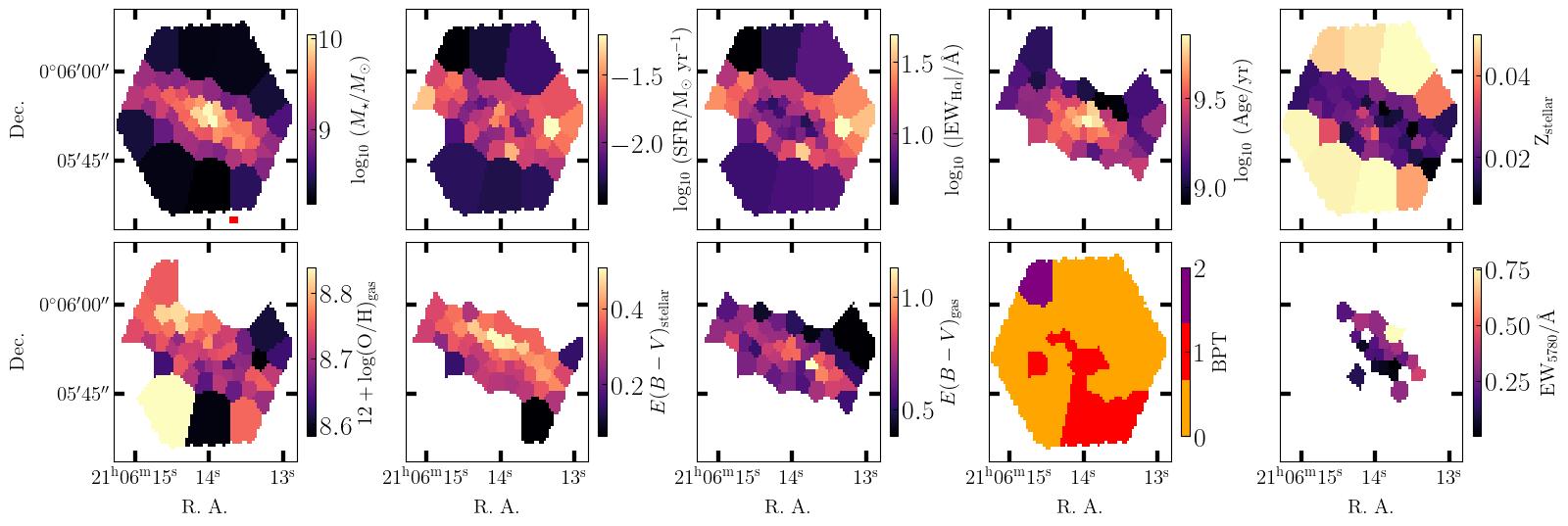}
    \caption{
    Spatially resolved maps of physical and spectral properties derived for the galaxy LEDA~1157026 (MaNGA ID~1-179479).
    The panels display the same parameters as shown in Figure~\ref{fig:Maps_1-575771}.
    This is the most distant galaxy where DIBs were detected in this work, lying at a distance of 232~Mpc. 
    }
    \label{fig:Maps_1-179479}
\end{figure*}

\begin{figure*}
    \centering
    \includegraphics[width=\linewidth]{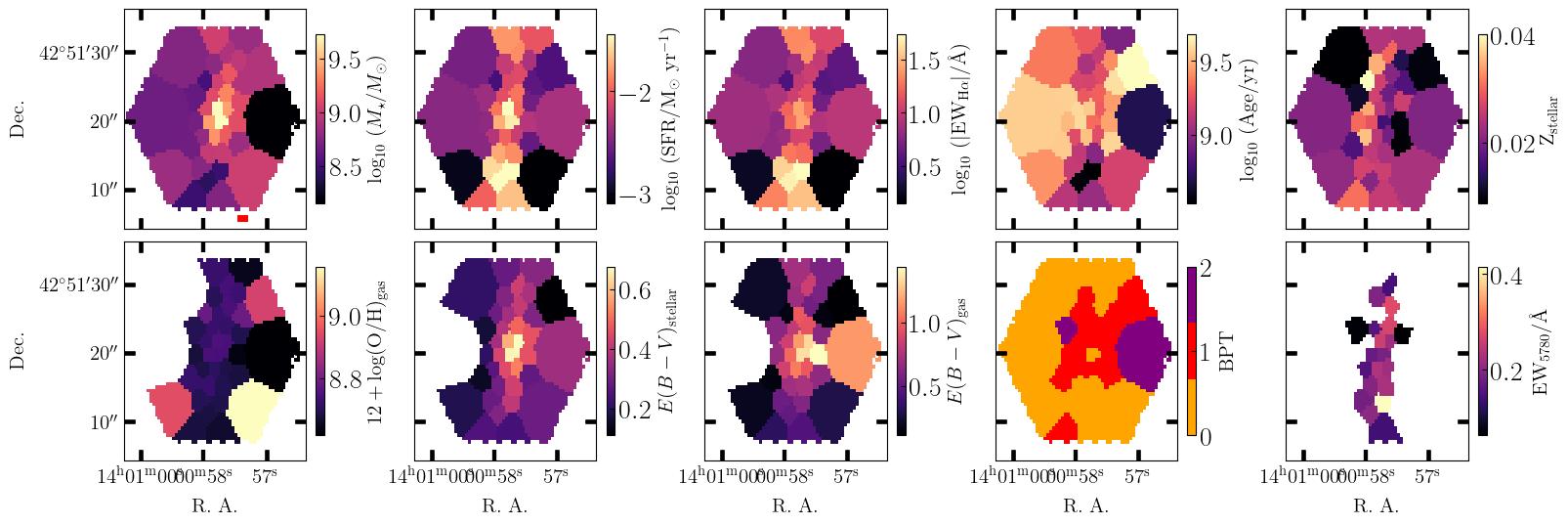}
    \caption{
    Spatially resolved maps of physical and spectral properties derived for the galaxy LEDA~2207577  (MaNGA ID~1-251307).
    The panels display the same parameters as shown in Figure~\ref{fig:Maps_1-575771}.
    }
    \label{fig:Maps_1-251307}
\end{figure*}

\begin{figure*}
    \centering
    \includegraphics[width=\linewidth]{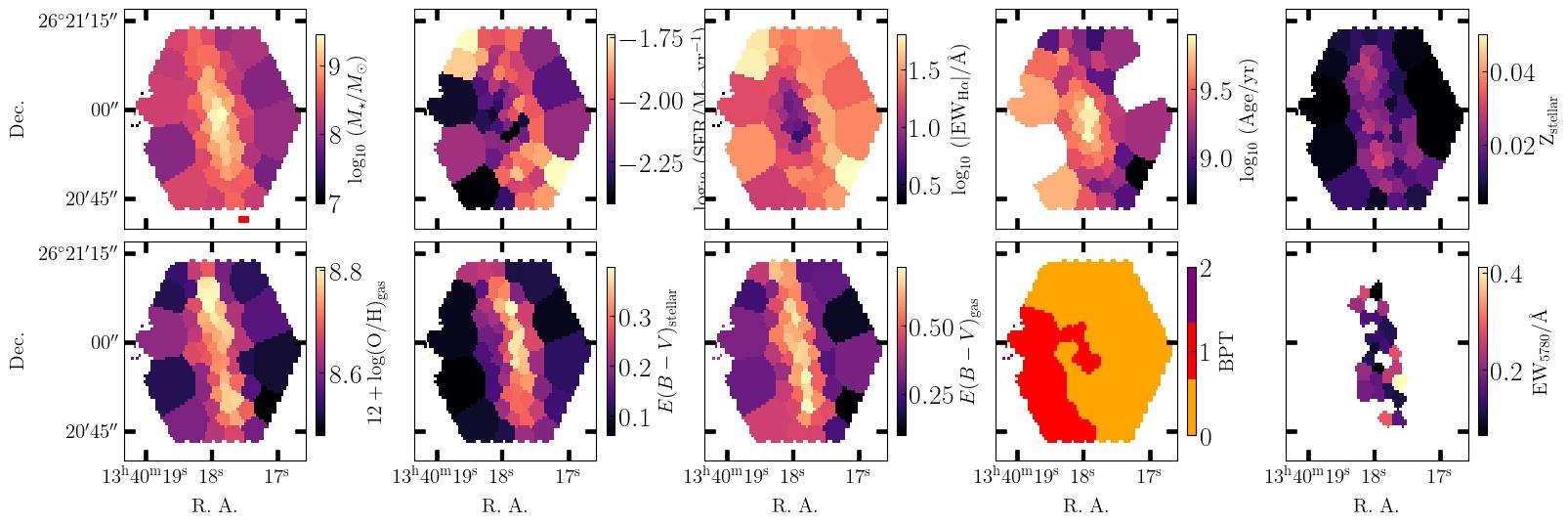}
    \caption{
    Spatially resolved maps of physical and spectral properties derived for the galaxy Z~161-116 (MaNGA ID~12-129612).
    The panels display the same parameters as shown in Figure~\ref{fig:Maps_1-575771}.
    }
    \label{fig:Maps_12-129612}
\end{figure*}

\begin{figure*}
    \centering
    \includegraphics[width=\linewidth]{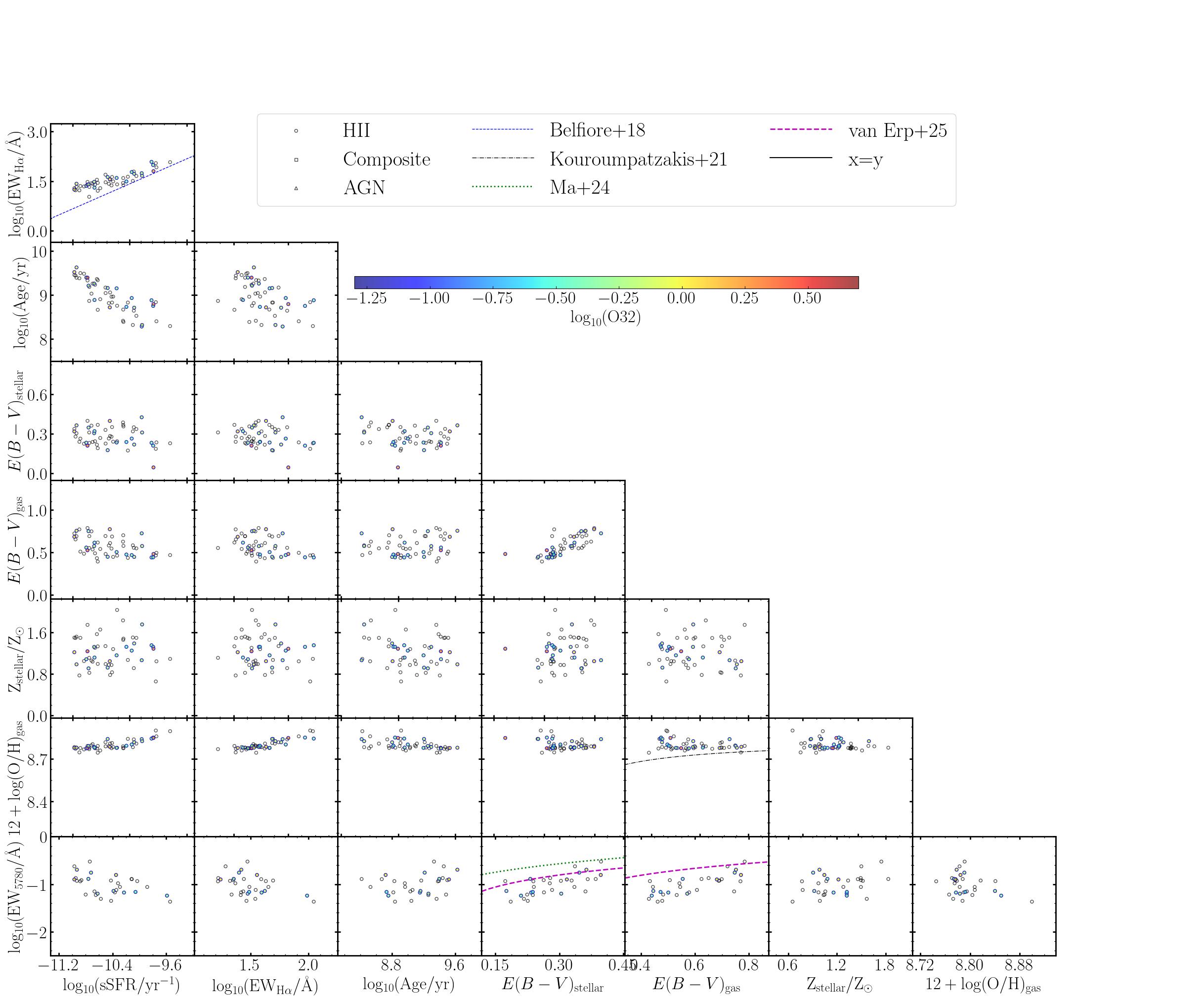}
    \caption{
    Corner plot showing the relationships between the equivalent width of DIBs~$\lambda\lambda$5780 and various physical properties of the sub-galactic regions in NGC~5383 (MaNGA ID~1-575771).
    Each point corresponds to a spatial bin with reliable DIB and physical parameter measurements (S/N~$>5$).
    The color scale indicates the O32 index, a proxy for the ionization parameter.
    The green dotted and purple dashed lines show the DIB~$\lambda$5780–$E(B{-}V)$ correlations from \citet{2024A&A...691A.282M} (our Galaxy) and \citet{2025A&A...697A.151V} (NGC~6240), respectively.
    The blue dashed line shows the relation between $\rm EW_{H\alpha}$ and sSFR \citep{2018MNRAS.477.3014B}, while the black dot-dashed line in the metallicity–extinction panel represents the relation for SDSS galaxies from \citet{2021MNRAS.506.3079K}.
    In the comparisons with the DIBs (bottom two rows), the red error bars correspond to the mode of the distribution within each of six equal-size bins.
    The comparisons reveal positive correlations between DIB strength and both gas-phase and stellar $E(B{-}V)$.
    For this particular galaxy, there is no clear correlation with sSFR, age, or metallicity.
    Super-solar metallicities are present throughout the galaxy, while star formation is ongoing, as indicated by elevated sSFR and $\rm EW_{H\alpha}$.
    An internal anti-correlation between sSFR and $\rm EW_{H\alpha}$ holds on sub-galactic scales, alongside a strong anti-correlation between sSFR and SP age.
    }
    \label{fig:Corner_DIB_1-575771}
\end{figure*}

\begin{figure*}
    \centering
    \includegraphics[width=\linewidth]{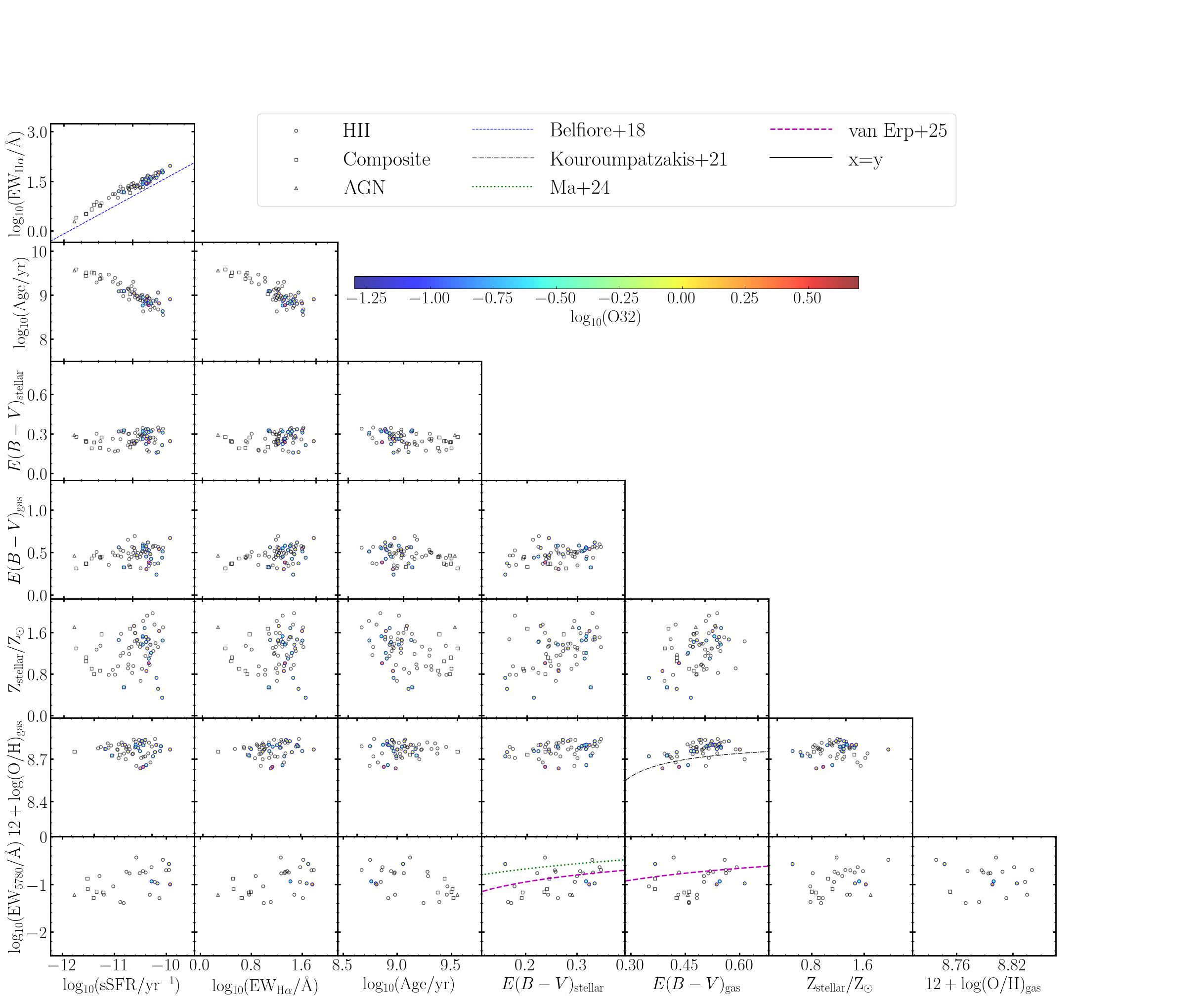}
    \caption{
    Corner plot of galaxy IC 263 (MaNGA 1-37213), similar to the one presented in Figure~\ref{fig:Corner_DIB_1-575771}.
    The plot reveals a lack of strong correlation between DIB strength and $E(B{-}V)$, which could be attributed to the small variance of extinction within the galaxy.
    The galaxy exhibits super-solar metallicities, low star-forming activity, and low O32 values.
    }
    \label{fig:Corner_DIB_1-37213}
\end{figure*}

\begin{figure*}
    \centering
    \includegraphics[width=\linewidth]{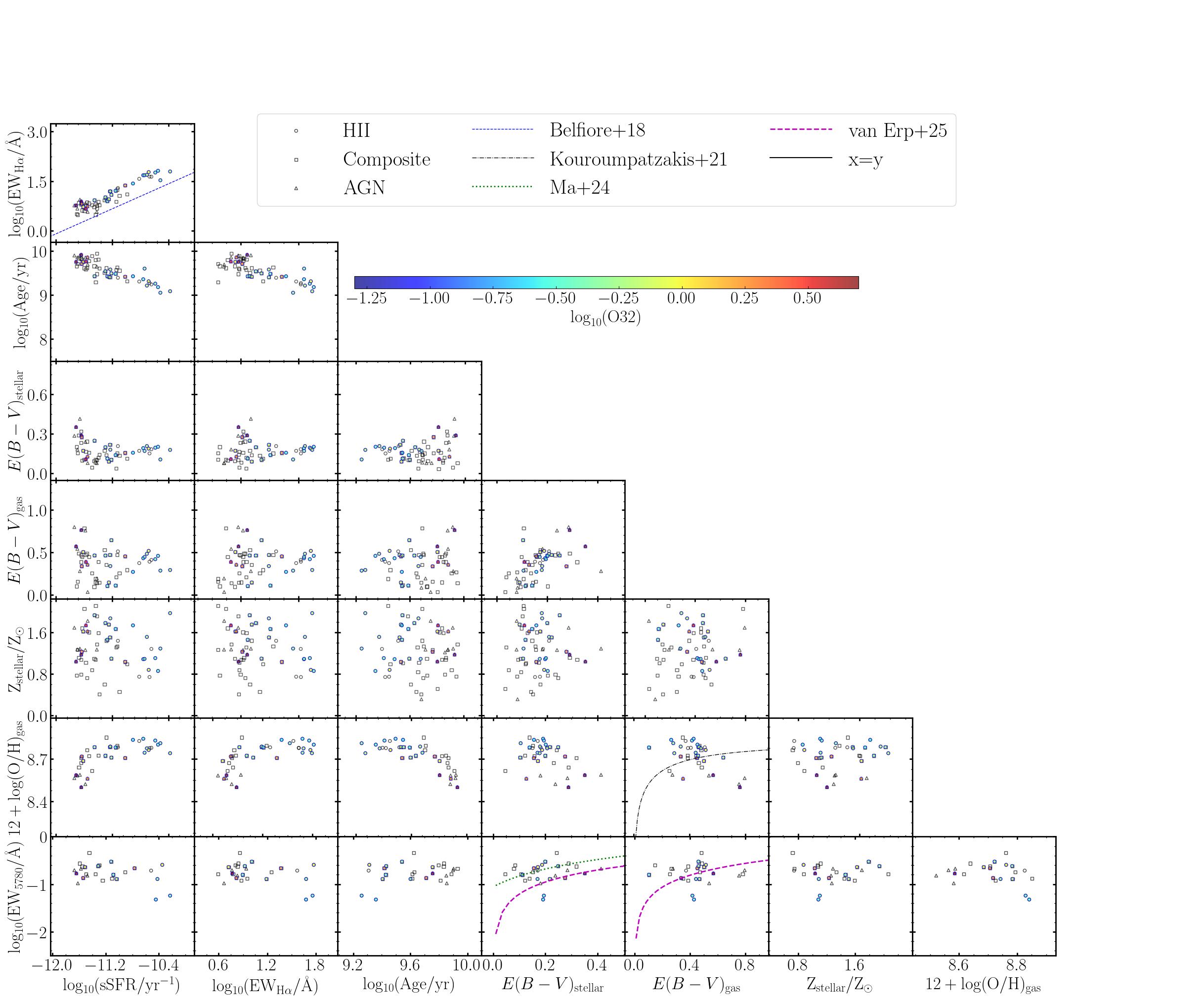}
    \caption{
    Corner plot of galaxy Z~224-81 (MaNGA 1-135641), similar to the one presented in Figure~\ref{fig:Corner_DIB_1-575771}.
    The plot reveals a positive correlation between DIB strength and $E(B{-}V)$.
    The galaxy exhibits near-solar metallicities and low star-forming activity.
    Some subgalactic regions exhibit elevated O32 values compared to overall low values within the galaxy.
    }
    \label{fig:Corner_DIB_1-135641}
\end{figure*}

\begin{figure*}
    \centering
    \includegraphics[width=\linewidth]{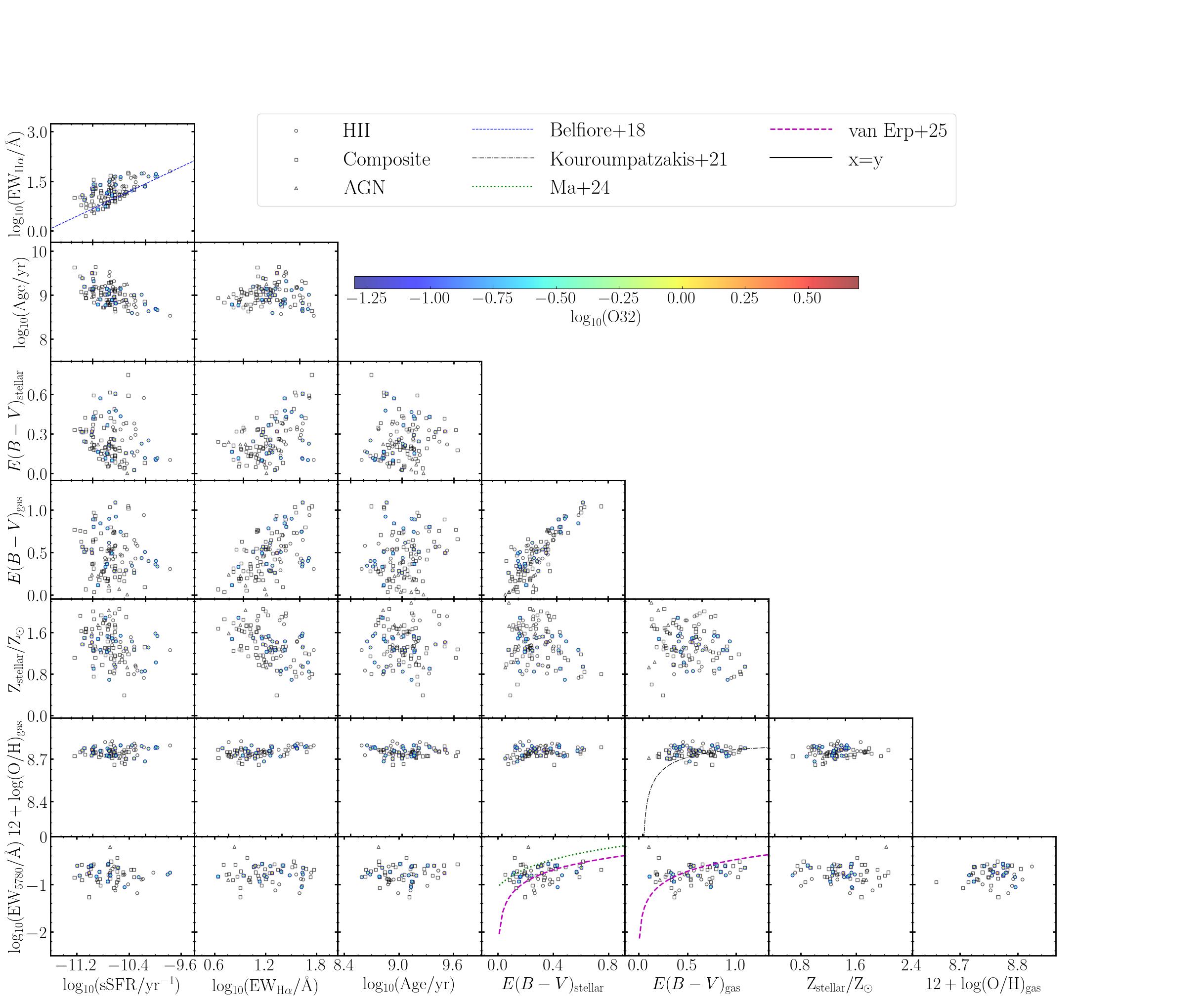}
    \caption{
    Corner plot of galaxy  KUG~0912+445A (MaNGA~ID 1-585513), similar to the one presented in Figure~\ref{fig:Corner_DIB_1-575771}.
    The comparisons reveal a weak correlation between DIB strength and stellar $E(B{-}V)$, but not with gas-phase extinction.
    In both cases, there is significant scatter.
    Moreover, the galaxy exhibits, on average, super-solar metallicities with moderate star-forming activity.
    }
    \label{fig:Corner_DIB_1-585513}
\end{figure*}

\begin{figure*}
    \centering
    \includegraphics[width=\linewidth]{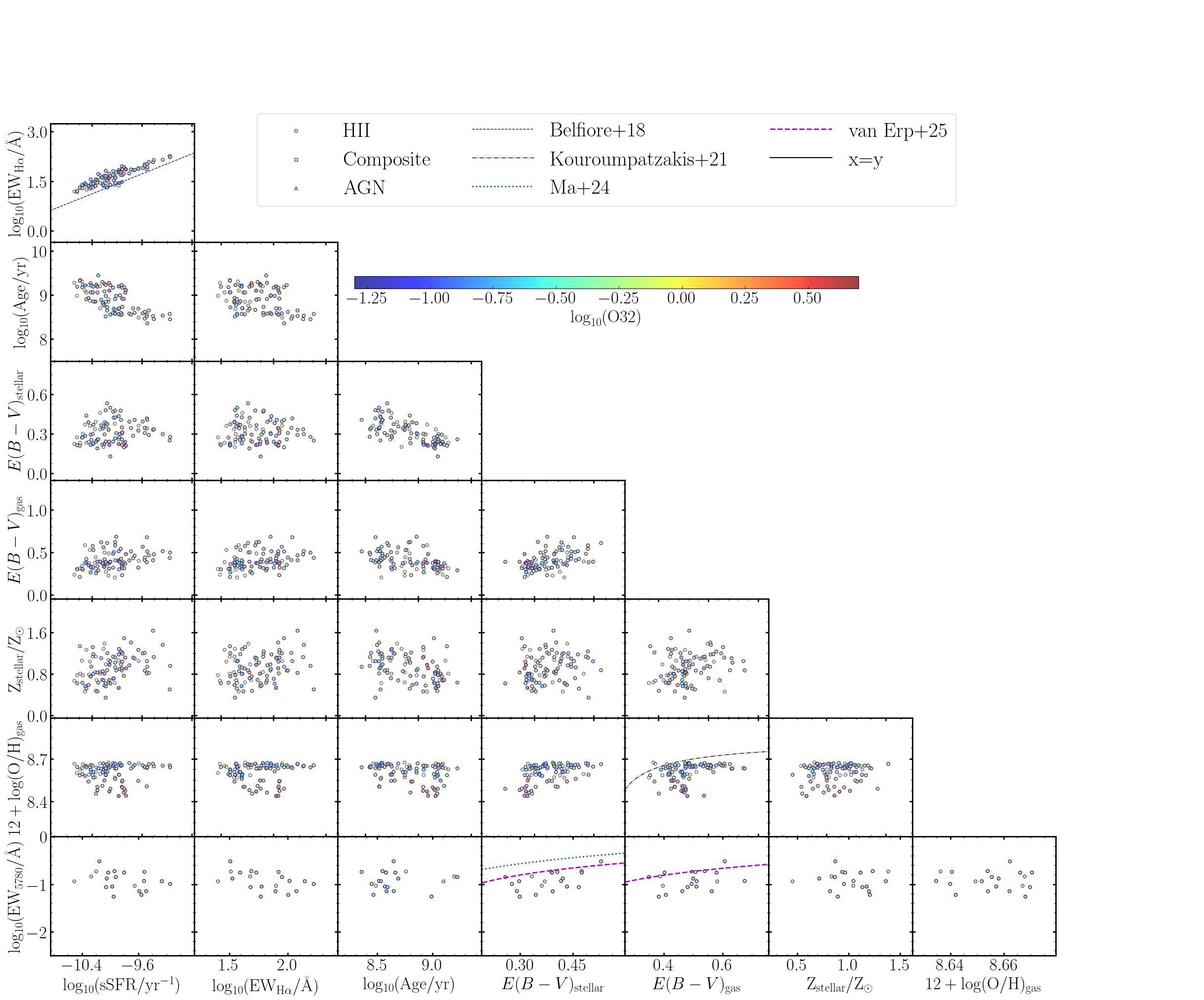}
    \caption{
    Corner plot of galaxy NGC~3003 (MaNGA 1-585731), similar to the one presented in Figure~\ref{fig:Corner_DIB_1-575771}.
    The plot suggests a positive statistical relation between DIB strength and $E(B-V)$, albeit with significant scatter.
    The galaxy exhibits sub-solar metallicities, high star-forming activity, but moderate O32 values.
    The DIB absorption feature is not observed in sub-galactic regions with intense radiation fields ($\rm O32 \gtrsim 1$), and these regions also strongly deviate from the galaxy-wide  \citet{2021MNRAS.506.3079K} metallicity--extinction relation. 
    }
    \label{fig:Corner_DIB_1-585731}
\end{figure*}

\begin{figure*}
    \centering
    \includegraphics[width=\linewidth]{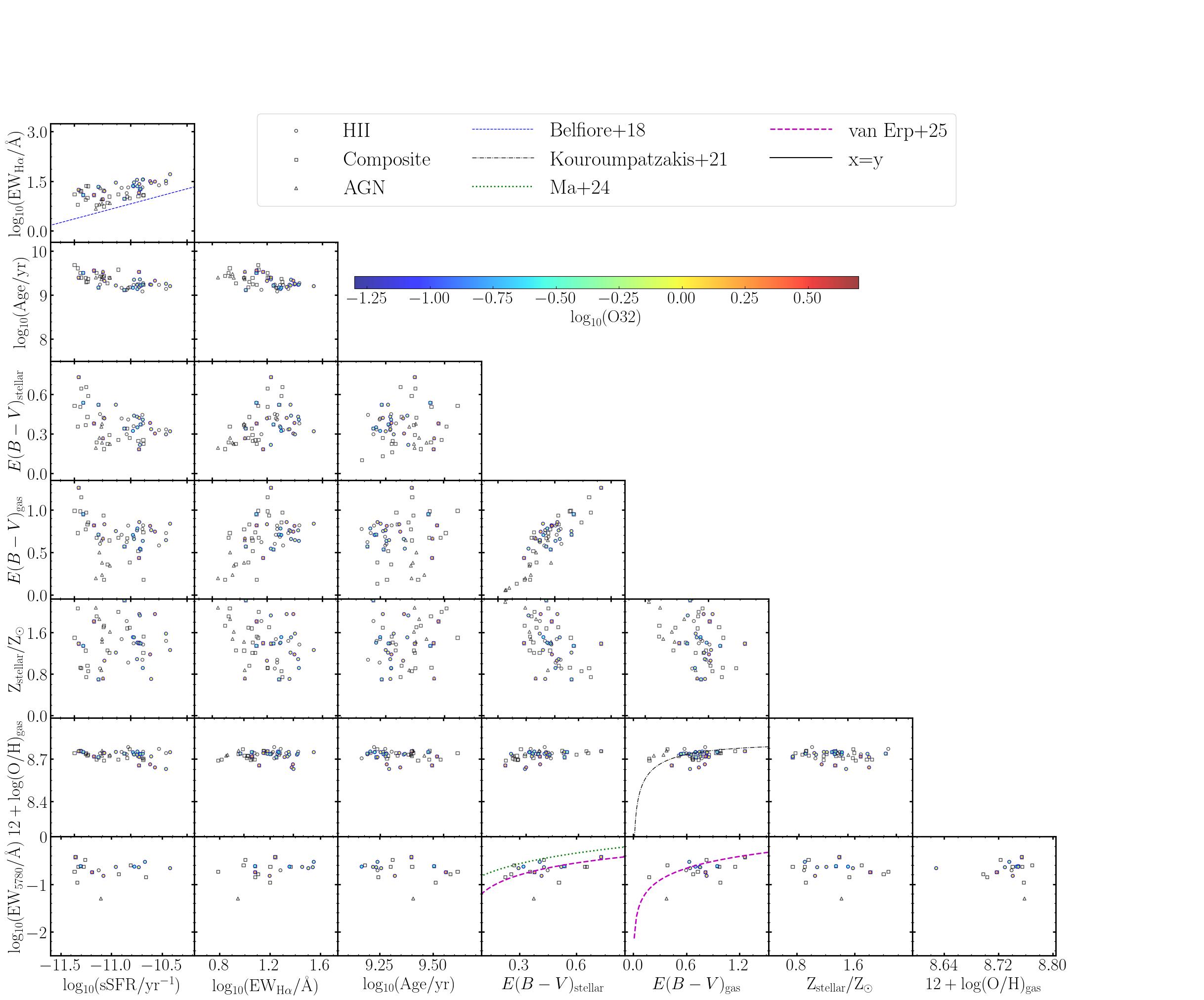}
    \caption{
    Corner plot of galaxy with MaNGA~ID 1-636015, similar to the one presented in Figure~\ref{fig:Corner_DIB_1-575771}.
    The plot reveals positive correlations between DIB~$\lambda$5780 EW and both stellar and gas-phase $E(B{-}V)$.
    The galaxy shows a narrow range of super-solar metallicities. 
    The sSFR and $\rm EW_{H\alpha}$ distributions indicate moderate to low star-forming activity.
    }
    \label{fig:Corner_DIB_1-636015}
\end{figure*}

\begin{figure*}
    \centering
    \includegraphics[width=\linewidth]{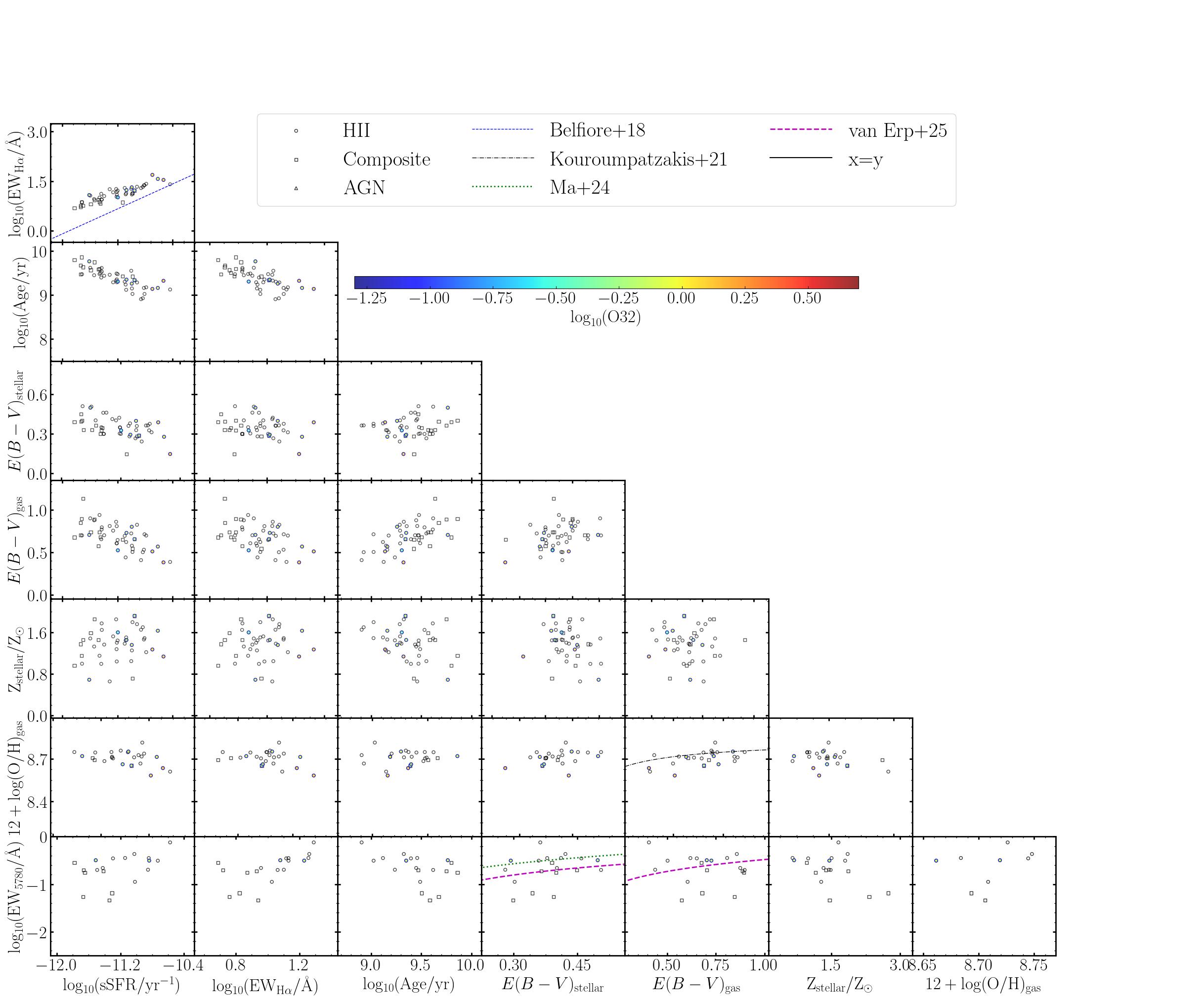}
    \caption{
    Corner plot of galaxy LEDA~1157026 (MaNGA~ID~1-179479).
    The DIBs EWs show a large scatter with respect to extinction, although the emission of most of the subgalactic regions are classified as composite by the BPT diagram.
    The galaxy exhibits near-solar metallicities but is quite passive, as indicated by its low sSFR and $\rm EW_{H\alpha}$. 
    }
    \label{fig:Corner_DIB_1-179479}
\end{figure*}

\begin{figure*}
    \centering
    \includegraphics[width=\linewidth]{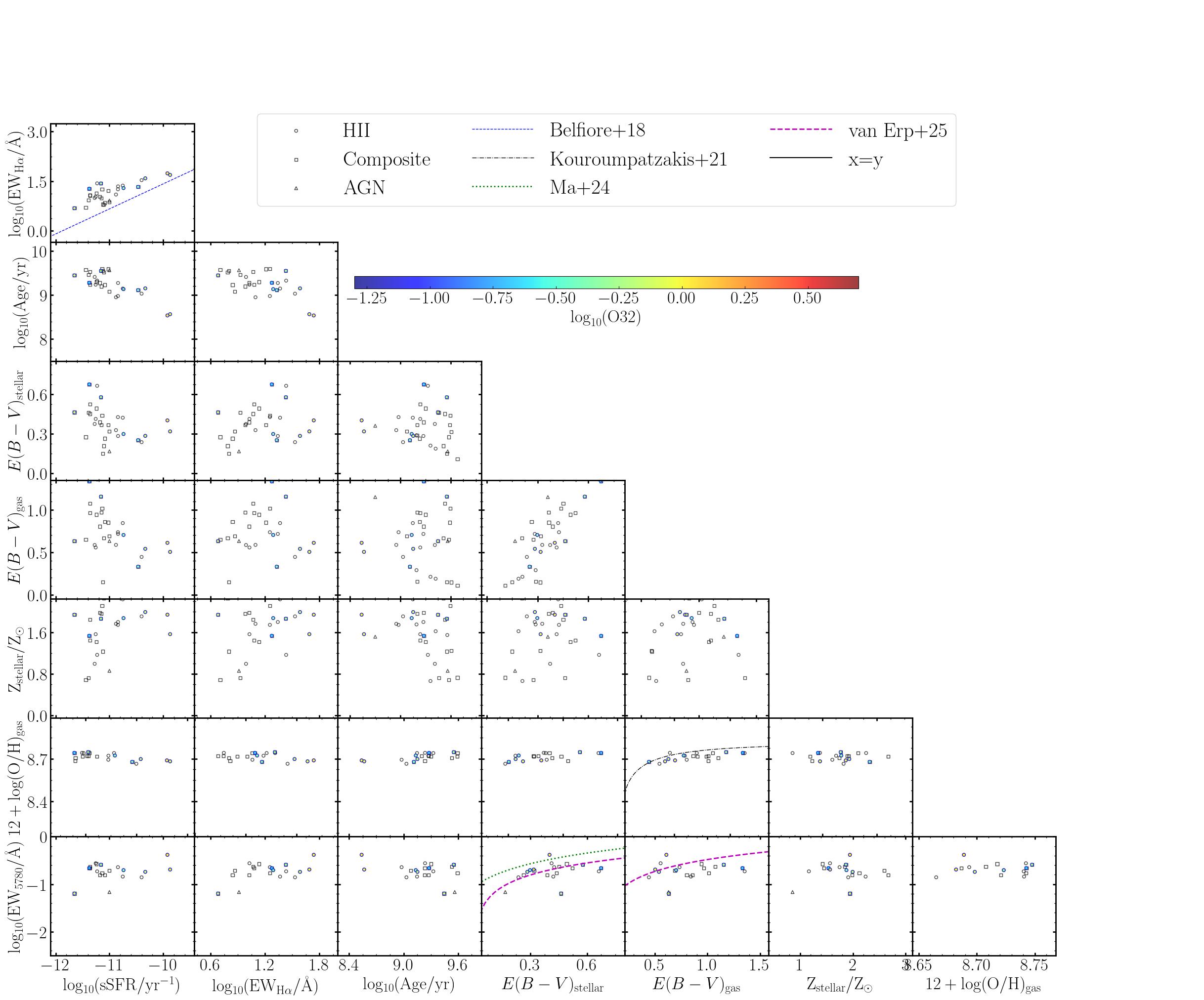}
    \caption{
    Corner plot of galaxy LEDA~2207577  (MaNGA~ID~1-251307).
    The DIB~$\lambda 5780$ EW shows strong correlation with extinction.
    The galaxy exhibits near-solar metallicities but is quite passive, as indicated by its low sSFR and $\rm EW_{H\alpha}$. 
    }
    \label{fig:Corner_DIB_1-251307}
\end{figure*}

\setcounter{table}{0} 
\renewcommand{\thetable}{A\arabic{table}}% Figure counter representation
\renewcommand{\theHtable}{A\arabic{table}}% Hyperref figure hyperlink hook

\onecolumn

\begin{ThreePartTable}
\begin{TableNotes}
\item[note 1] The sources are sorted by a combined metric proportional to $N_{\rm det} \times (N_{\rm det}/N_{\rm tot})$, which favors galaxies with both a high number of DIB~$\lambda5780$ detections and a high detection fraction.
\item[note 2] The column named $\lambda$5780 indicates the percentage of sub-galactic regions with reliable DIB~$\lambda$5780 measurements, relative to the total number of regions analyzed in each galaxy.
\end{TableNotes}

\begin{longtable}{c|llcccccc|c}
\caption{The selected sample analyzed by this work.}
\label{tab:List_table_1} \\

\hline
\# & MaNGA & Plate & R.A. & Dec. & D. & $M_{\star}$ & SFR & Metallicity & $\lambda 5780$\\
 & ID  & \# & (deg.) & (deg.) & (Mpc) & ${\rm log} (\frac{M_{\star}}{M_\odot})$ & $ {\rm log}(\frac{{\rm SFR}}{M_\odot/{\rm yr}})$ & ${\rm log} (\frac{O}{H})+12 $ & (\%)\\
\hline
\endfirsthead

\hline
\# & MaNGA & Plate & R.A. & Dec. & D. & $M_{\star}$ & SFR & Metallicity & $\lambda 5780$ \\
\hline
\endhead

\hline
\multicolumn{10}{r}{\textit{Continued on next page}}\\
\endfoot

\hline
\insertTableNotes
\endlastfoot

1 & 1-585513 & 8250-12704 & 138.9814 & 44.3328 & 181.4 & 11.0 & 0.2 & 8.7 & 50.4 \\
2 & 1-251129 & 8332-3702 & 207.8728 & 43.8064 & 151.1 & 10.2 & 0.6 & 9.0 & 69.3 \\
3 & 1-135641 & 8588-12704 & 249.5573 & 40.1468 & 137.5 & 10.8 & -1.6 & 8.7 & 51.3 \\
4 & 1-114325 & 7975-12703 & 324.0950 & 12.2363 & 130.6 & 10.7 & -0.4 & 8.6 & 48.1 \\
5 & 1-179479 & 7964-12704 & 316.5585 & 0.0978 & 232.4 & 11.3 & 0.4 & 9.1 & 49.3 \\
6 & 1-43864 & 9182-12702 & 118.2794 & 39.7444 & 126.5 & 10.6 & 0.1 & 9.0 & 42.4 \\
7 & 1-113418 & 7815-12704 & 319.3538 & 10.2316 & 196.9 & 11.0 & 0.2 & 8.8 & 57.3 \\
8 & 1-37213 & 8078-6103 & 42.4165 & -0.0699 & 127.6 & 10.7 & 0.3 & 8.7 & 44.6 \\
9 & 1-251307 & 8332-9102 & 210.2409 & 42.8556 & 146.6 & 10.7 & 0.5 & 9.1 & 58.5 \\
10 & 1-24357 & 7990-3703 & 262.0993 & 57.5454 & 129.1 & 9.6 & 0.8 & 8.8 & 55.6 \\
11 & 1-149650 & 8992-12701 & 171.3568 & 51.8839 & 225.4 & 10.5 & 0.1 & 8.7 & 69.1 \\
12 & 1-36832 & 8078-12701 & 40.8805 & 0.3068 & 120.4 & 11.1 & -0.8 & 8.7 & 31.9 \\
13 & 12-129612 & 7495-12702 & 205.0748 & 26.3496 & 128.4 & 10.6 & -0.0 & 8.6 & 33.4 \\
14 & 1-384396 & 9494-12704 & 126.6215 & 22.2612 & 128.7 & 10.3 & 0.2 & 9.0 & 41.9 \\
15 & 1-38381 & 8084-12705 & 51.0271 & -1.0579 & 113.6 & 10.5 & -0.3 & 8.7 & 45.3 \\
16 & 1-575771 & 8332-1902 & 209.2707 & 41.8463 & 36.3 & 10.9 & 0.7 & 8.8 & 67.3 \\
17 & 1-379291 & 8712-6101 & 118.7269 & 53.8463 & 158.4 & 10.8 & 0.7 & 9.2 & 39.7 \\
18 & 1-636015 & 7977-12704 & 332.4183 & 13.6358 & 123.7 & 11.1 & 1.2 & 8.7 & 43.6 \\
19 & 1-574504 & 9487-9102 & 123.8203 & 46.0753 & 187.0 & 10.9 & 1.0 & 8.7 & 43.1 \\
20 & 1-138102 & 8252-6102 & 144.5580 & 48.3883 & 116.4 & 10.3 & 0.0 & 9.1 & 39.1 \\
21 & 1-339041 & 8138-12704 & 116.6571 & 44.7905 & 139.5 & 11.2 & 0.2 & 8.8 & 29.0 \\
22 & 1-149572 & 8947-1901 & 171.3837 & 51.1804 & 122.5 & 10.2 & 0.6 & 9.1 & 64.1 \\
23 & 1-43865 & 8713-6104 & 118.3173 & 39.0501 & 186.5 & 10.9 & 0.7 & 8.7 & 38.3 \\
24 & 1-576436 & 9033-12705 & 224.0182 & 45.4053 & 164.5 & 10.9 & -0.1 & 8.8 & 28.2 \\
25 & 1-317822 & 9095-1901 & 242.8101 & 24.2250 & 147.5 & 10.6 & 0.7 & 9.2 & 62.5 \\
26 & 1-115128 & 7977-1902 & 332.4813 & 12.8181 & 162.8 & 10.3 & 0.4 & 9.1 & 62.9 \\
27 & 1-210971 & 8604-9102 & 246.4554 & 40.3452 & 131.4 & 10.3 & 0.8 & 8.8 & 32.9 \\
28 & 1-274689 & 8253-3703 & 157.3436 & 43.1706 & 123.1 & 10.0 & 0.2 & 9.1 & 47.7 \\
29 & 1-277272 & 8257-6101 & 165.2613 & 44.8882 & 133.2 & 10.3 & -0.1 & 8.7 & 37.0 \\
30 & 1-149556 & 8992-12702 & 172.3172 & 51.5232 & 149.1 & 10.6 & -1.3 & 8.7 & 34.0 \\
31 & 1-632610 & 9044-12701 & 231.1008 & 30.2774 & 141.2 & 11.0 & -0.4 & 8.7 & 40.0 \\
32 & 1-109394 & 8082-9102 & 50.1799 & -1.0023 & 164.2 & 10.6 & 0.3 & 8.7 & 37.6 \\
33 & 1-318334 & 9095-9102 & 243.0849 & 23.0020 & 146.3 & 10.6 & 0.3 & 9.1 & 31.0 \\
34 & 1-384095 & 9488-12701 & 126.5900 & 20.2705 & 114.0 & 10.1 & -0.3 & 8.9 & 36.1 \\
35 & 1-38160 & 8081-9102 & 49.8457 & 0.8235 & 152.3 & 10.4 & \dots & 8.6 & 46.7 \\
36 & 1-318725 & 9050-3704 & 245.9950 & 22.3934 & 168.4 & 10.7 & 1.0 & 9.2 & 44.2 \\
37 & 1-382712 & 9491-6101 & 119.1744 & 17.9912 & 188.2 & 10.6 & 0.7 & 8.6 & 35.0 \\
38 & 1-633219 & 9041-9102 & 235.9419 & 28.4152 & 148.5 & 10.5 & 0.3 & 9.2 & 29.5 \\
39 & 1-604085 & 8085-12704 & 52.6199 & 0.8115 & 139.5 & 10.5 & -0.0 & 8.7 & 33.8 \\
40 & 1-386605 & 8987-3701 & 136.2499 & 28.3477 & 223.2 & 10.0 & 0.8 & 9.1 & 42.6 \\
41 & 1-274519 & 8455-3701 & 157.1795 & 39.8389 & 132.5 & 9.9 & 0.4 & 9.1 & 42.0 \\
42 & 1-90190 & 8553-12702 & 233.5095 & 56.6857 & 53.8 & 11.0 & -1.2 & 8.6 & 30.4 \\
43 & 1-138140 & 8252-3704 & 145.3081 & 47.6886 & 214.4 & 10.6 & -0.1 & 8.8 & 87.6 \\
44 & 1-148780 & 8999-1901 & 163.6227 & 50.5775 & 112.2 & 9.4 & 0.2 & 8.9 & 55.4 \\
45 & 1-180039 & 8615-9102 & 321.4740 & 0.4161 & 143.9 & 10.5 & 0.1 & 8.7 & 28.2 \\
46 & 1-316920 & 9888-9101 & 235.3983 & 28.0496 & 144.6 & 10.2 & \dots & 8.7 & 38.7 \\
47 & 1-549071 & 9042-9102 & 235.0321 & 28.1523 & 141.7 & 10.5 & \dots & 8.8 & 27.8 \\
48 & 1-385547 & 9501-9102 & 130.3678 & 25.2345 & 132.0 & 10.4 & 0.1 & 9.1 & 27.3 \\
49 & 1-178315 & 9085-3702 & 259.5924 & 28.1442 & 189.7 & 10.1 & 0.1 & 9.0 & 85.7 \\
50 & 1-250997 & 8331-12704 & 205.2510 & 42.4314 & 126.2 & 10.8 & -0.3 & 8.8 & 25.5 \\
51 & 1-114082 & 7975-3701 & 324.1525 & 10.5067 & 183.6 & 9.8 & -0.2 & 8.9 & 62.5 \\
52 & 1-248213 & 9036-9102 & 240.5693 & 42.9169 & 110.2 & 10.7 & 0.4 & 9.1 & 25.9 \\
53 & 1-44013 & 8143-9101 & 119.4012 & 41.1146 & 189.4 & 10.4 & -0.6 & 8.6 & 45.3 \\
54 & 1-178586 & 7962-12703 & 261.2173 & 28.0783 & 218.9 & 11.2 & -0.2 & 8.7 & 25.8 \\
55 & 1-179386 & 7964-6101 & 316.5904 & -0.7712 & 223.4 & 10.9 & 0.4 & 8.8 & 33.6 \\
56 & 1-149507 & 8997-12705 & 172.1561 & 50.8274 & 146.3 & 10.6 & -0.2 & 8.7 & 45.6 \\
57 & 1-114956 & 7977-3704 & 332.7987 & 11.8007 & 122.1 & 10.4 & -0.9 & 8.7 & 39.2 \\
58 & 1-585731 & 8150-6103 & 147.1483 & 33.4216 & 21.9 & 10.0 & -0.2 & 8.6 & 29.5 \\
59 & 1-382465 & 9491-12703 & 118.8597 & 18.9755 & 197.8 & 10.1 & -0.4 & 8.9 & 50.8 \\
60 & 1-258760 & 8263-12702 & 186.0307 & 45.4345 & 107.2 & 10.2 & -0.8 & 8.7 & 33.5 \\
61 & 1-318463 & 9095-6104 & 243.5610 & 24.2810 & 204.6 & 10.2 & -0.1 & 9.1 & 74.5 \\
62 & 1-547415 & 7958-12703 & 259.0554 & 34.0554 & 173.5 & 9.7 & -1.1 & 8.9 & 49.2 \\
63 & 1-121486 & 8717-3704 & 117.5183 & 34.4793 & 131.2 & 10.8 & 1.0 & 9.2 & 35.8 \\
64 & 1-211103 & 8550-6103 & 247.6387 & 39.8307 & 110.7 & 10.5 & -1.3 & 8.8 & 28.7 \\
65 & 1-377686 & 8133-3701 & 112.0793 & 43.3021 & 199.8 & 10.2 & -0.6 & 8.8 & 61.5 \\
66 & 1-277271 & 8257-9101 & 165.4536 & 44.8138 & 114.1 & 10.0 & -0.3 & 9.1 & 34.0 \\
67 & 1-384952 & 9493-3704 & 129.1179 & 22.8474 & 174.6 & 9.7 & 0.2 & 8.9 & 35.0 \\
68 & 1-274706 & 8253-12701 & 157.0589 & 42.9681 & 205.9 & 10.2 & -0.4 & 9.0 & 43.3 \\
69 & 1-180172 & 7968-9101 & 322.3707 & -0.2196 & 228.8 & 11.2 & -0.2 & 8.7 & 37.0 \\
70 & 1-633000 & 9872-3701 & 233.2320 & 42.4383 & 88.5 & 10.1 & 0.0 & 9.0 & 42.4 \\
71 & 1-382860 & 9491-3702 & 118.9042 & 19.1550 & 187.8 & 9.6 & -0.1 & 8.9 & 73.9 \\
72 & 1-178557 & 9085-12703 & 260.3063 & 29.1859 & 216.3 & 10.7 & 0.8 & 9.2 & 56.5 \\
73 & 1-121784 & 9181-3701 & 119.6679 & 37.2651 & 183.7 & 9.9 & -0.3 & 8.8 & 64.9 \\
74 & 1-377378 & 8132-6102 & 111.1382 & 41.9506 & 119.9 & 10.0 & -0.6 & 8.7 & 57.7 \\
75 & 1-180163 & 7968-1901 & 322.4519 & -0.3136 & 135.5 & 10.6 & -1.8 & 8.6 & 61.0 \\
76 & 1-121837 & 9485-3702 & 120.0757 & 37.6093 & 189.5 & 10.2 & -0.6 & 8.9 & 55.8 \\
77 & 1-115320 & 7977-3703 & 333.0520 & 12.2052 & 124.4 & 10.0 & 0.2 & 9.1 & 35.0 \\
78 & 1-277328 & 8255-3701 & 164.4462 & 44.6546 & 170.0 & 9.6 & -0.2 & 8.7 & 72.3 \\
79 & 1-149610 & 8996-12701 & 171.1598 & 52.5334 & 123.1 & 10.4 & -0.3 & 9.1 & 29.8 \\
80 & 1-373878 & 9891-6102 & 228.4149 & 28.2445 & 209.2 & 9.9 & 0.4 & 8.9 & 56.8 \\
81 & 1-386078 & 9500-9102 & 133.5422 & 24.7767 & 197.3 & 10.0 & -0.3 & 8.7 & 41.5 \\
82 & 1-36945 & 8077-9101 & 41.6431 & -0.8435 & 195.9 & 10.5 & 0.0 & 8.7 & 49.8 \\
83 & 1-138164 & 8252-1902 & 146.0918 & 47.4599 & 116.9 & 9.4 & -1.1 & 8.5 & 100.0 \\
84 & 1-36849 & 8077-12702 & 40.9690 & 0.1526 & 123.4 & 10.1 & -0.3 & 8.9 & 30.4 \\
85 & 1-339010 & 8138-6101 & 115.9640 & 44.7243 & 141.9 & 10.3 & -0.2 & 8.6 & 47.0 \\
86 & 1-121871 & 9485-3703 & 120.7667 & 36.2645 & 202.0 & 10.1 & -0.2 & 9.0 & 49.2 \\
87 & 1-179556 & 8615-9101 & 319.9198 & 0.1209 & 150.7 & 10.7 & -0.2 & 8.7 & 25.2 \\
88 & 1-43956 & 8140-6104 & 117.7059 & 40.9463 & 189.1 & 10.0 & -0.0 & 8.8 & 55.2 \\
89 & 1-251043 & 8331-3704 & 206.0170 & 43.0740 & 122.9 & 10.4 & -0.4 & 8.6 & 47.8 \\
90 & 1-120941 & 8144-9101 & 114.6102 & 28.9631 & 70.7 & 9.8 & -1.2 & 9.0 & 28.7 \\
91 & 1-379410 & 8712-6103 & 120.2301 & 53.6706 & 185.0 & 9.8 & -0.0 & 8.8 & 49.8 \\
92 & 1-318728 & 9050-6102 & 245.8593 & 22.4857 & 170.8 & 9.4 & -0.9 & 8.6 & 60.3 \\
93 & 1-37751 & 9194-6104 & 47.2479 & -0.7933 & 174.1 & 9.6 & -0.1 & 8.9 & 40.4 \\
94 & 1-235270 & 8325-3703 & 212.5294 & 46.4396 & 118.7 & 9.9 & -0.2 & 8.6 & 41.5 \\
95 & 1-172948 & 8452-6102 & 156.1381 & 46.8312 & 194.0 & 10.9 & -0.5 & 8.7 & 24.7 \\
96 & 1-338828 & 8137-3704 & 115.6416 & 44.2159 & 191.0 & 10.1 & -0.1 & 9.0 & 43.0 \\
97 & 1-634056 & 8978-12701 & 248.7828 & 40.9914 & 136.0 & 10.8 & -0.0 & 8.6 & 27.7 \\
98 & 1-37126 & 8154-3701 & 44.7356 & -0.7661 & 198.9 & 10.4 & -1.0 & 8.6 & 65.6 \\
99 & 1-339138 & 8138-6103 & 117.3182 & 46.2050 & 173.8 & 10.3 & 0.1 & 9.0 & 49.3 \\
100 & 1-37727 & 8081-9101 & 47.7722 & -0.5465 & 127.5 & 10.5 & -0.1 & 8.7 & 26.1 \\
101 & 12-192149 & 7495-3702 & 205.0080 & 26.9789 & 124.5 & 9.8 & -0.5 & 9.0 & 49.7 \\
102 & 1-179561 & 8615-1902 & 319.7512 & -0.9640 & 87.6 & 10.0 & -0.0 & 9.0 & 36.8 \\
103 & 12-110756 & 7495-3703 & 205.2433 & 27.7263 & 128.9 & 9.6 & -0.6 & 8.8 & 48.9 \\
104 & 1-604835 & 8717-12701 & 117.2152 & 36.3904 & 128.9 & 10.3 & -0.1 & 8.7 & 27.9 \\
105 & 1-386695 & 8987-9102 & 137.9835 & 27.8993 & 217.4 & 10.1 & 0.1 & 9.1 & 44.1 \\
106 & 1-386452 & 8987-3703 & 136.2283 & 28.3843 & 121.3 & 10.6 & 0.2 & 9.1 & 37.2 \\
107 & 1-277289 & 8255-3704 & 165.2750 & 45.2933 & 169.4 & 9.8 & -0.8 & 8.6 & 59.8 \\
108 & 1-276547 & 8254-12704 & 163.4590 & 44.3644 & 223.3 & 10.2 & 0.4 & 9.1 & 29.2 \\
109 & 1-633256 & 9041-3702 & 236.4503 & 30.2024 & 142.8 & 10.2 & 0.2 & 9.1 & 27.4 \\
110 & 1-631680 & 8592-3703 & 224.3090 & 53.1520 & 131.4 & 10.1 & 0.1 & 9.1 & 29.2 \\
111 & 1-38392 & 8084-12704 & 50.6830 & -0.1796 & 172.2 & 10.1 & -0.3 & 8.9 & 29.4 \\
112 & 1-547926 & 8153-6102 & 39.6055 & -0.1884 & 120.7 & 10.0 & -0.1 & 9.0 & 26.9 \\
113 & 1-248410 & 8979-3701 & 241.2085 & 42.0358 & 112.9 & 9.7 & -0.6 & 8.9 & 70.7 \\
114 & 1-210989 & 8550-3701 & 246.6527 & 39.8587 & 142.1 & 10.3 & -0.1 & 8.7 & 44.8 \\
115 & 1-43957 & 8140-6102 & 117.8186 & 40.9521 & 184.2 & 9.6 & -0.2 & 8.9 & 31.2 \\
116 & 1-37769 & 8080-9101 & 47.3400 & -0.4079 & 205.1 & 9.8 & 0.2 & 8.9 & 27.7 \\
117 & 1-180193 & 8616-3704 & 322.2420 & -0.3424 & 137.2 & 10.3 & -0.1 & 8.8 & 35.8 \\
118 & 1-247771 & 9037-9102 & 235.9107 & 43.8248 & 170.3 & 9.9 & -0.4 & 8.8 & 34.1 \\
119 & 1-314332 & 9024-1902 & 224.4378 & 33.1656 & 135.9 & 9.7 & 0.4 & 8.7 & 33.3 \\
120 & 1-37335 & 9193-9102 & 46.9438 & -0.7876 & 174.1 & 9.4 & -0.1 & 8.8 & 33.9 \\
121 & 1-109073 & 8078-12704 & 42.1568 & -0.5543 & 199.3 & 10.1 & -0.0 & 8.9 & 25.0 \\
122 & 1-121735 & 8717-3703 & 118.3182 & 35.5726 & 209.6 & 9.9 & 0.4 & 9.0 & 43.3 \\
123 & 1-48545 & 8721-12702 & 134.5994 & 55.0549 & 200.9 & 10.1 & -0.0 & 9.0 & 24.6 \\
124 & 1-379241 & 8711-12703 & 119.3447 & 52.7127 & 184.6 & 9.7 & -0.7 & 8.7 & 41.6 \\
125 & 1-155558 & 8249-3702 & 137.0327 & 45.9210 & 121.0 & 10.0 & 0.1 & 8.9 & 28.3 \\
126 & 1-403684 & 8934-12703 & 196.7842 & 28.6333 & 179.0 & 9.7 & -0.0 & 8.8 & 29.1 \\
127 & 1-419240 & 8442-6101 & 200.2734 & 31.8313 & 284.0 & 10.7 & 0.9 & 9.1 & 27.9 \\
128 & 1-487452 & 8448-6103 & 166.6314 & 22.7776 & 101.1 & 10.4 & -0.6 & 8.6 & 39.2 \\
129 & 1-631278 & 8595-3703 & 221.4380 & 51.5808 & 134.1 & 10.9 & 0.2 & 8.8 & 38.9 \\
130 & 1-149471 & 8992-6103 & 173.0378 & 50.7655 & 222.8 & 10.6 & -1.3 & 8.6 & 31.6 \\
131 & 1-547402 & 7958-9101 & 258.4958 & 33.6071 & 176.1 & 9.7 & \dots & 8.4 & 33.6 \\
132 & 1-55612 & 10001-3703 & 134.5915 & 57.6850 & 185.0 & 10.0 & -0.2 & 8.5 & 35.3 \\
133 & 1-386322 & 9506-3703 & 134.2140 & 27.2701 & 83.3 & 9.9 & -2.0 & 8.5 & 27.5 \\
134 & 1-251119 & 8332-6103 & 207.6574 & 43.7641 & 222.8 & 10.6 & -0.6 & 8.7 & 32.0 \\
135 & 1-113219 & 7815-9102 & 317.3747 & 10.0519 & 186.6 & 9.4 & -0.3 & 8.7 & 30.7 \\
136 & 1-386108 & 9500-6101 & 132.8305 & 24.9279 & 204.5 & 10.0 & -0.0 & 8.9 & 41.1 \\
137 & 1-547295 & 7960-3703 & 259.1835 & 32.8063 & 185.1 & 9.8 & -0.2 & 8.8 & 56.0 \\
138 & 1-121789 & 9485-3701 & 119.5320 & 36.8313 & 178.8 & 10.1 & 0.3 & 9.1 & 52.9 \\
139 & 1-135239 & 9029-1902 & 246.4872 & 41.5221 & 195.2 & 10.0 & 0.1 & 9.2 & 51.1 \\
140 & 1-456616 & 9876-3703 & 194.9159 & 27.5765 & 74.9 & 9.0 & -0.6 & 9.0 & 26.3 \\
141 & 1-48157 & 10001-12701 & 133.3711 & 57.5984 & 178.7 & 9.7 & -0.0 & 8.7 & 27.9 \\
142 & 1-274440 & 8455-1902 & 155.7087 & 39.3689 & 118.0 & 9.5 & -0.6 & 8.9 & 36.8 \\
143 & 1-48450 & 8724-3704 & 132.2615 & 53.8385 & 137.4 & 9.4 & -0.8 & 9.0 & 51.0 \\
144 & 1-48136 & 10001-3702 & 132.9128 & 57.1074 & 115.5 & 10.0 & 0.0 & 9.1 & 28.8 \\
145 & 1-37744 & 8080-3701 & 47.5284 & -0.4680 & 167.0 & 9.4 & -0.5 & 8.8 & 47.3 \\
146 & 1-547107 & 7957-3704 & 256.5658 & 35.4172 & 167.2 & 9.4 & -0.5 & 9.0 & 41.9 \\
147 & 1-373381 & 9002-9102 & 223.7109 & 30.9580 & 112.3 & 9.3 & -0.9 & 8.7 & 27.4 \\
148 & 1-37062 & 8077-3702 & 41.8464 & 0.0588 & 111.7 & 10.2 & 0.1 & 9.1 & 26.9 \\
149 & 1-178233 & 7962-6102 & 259.5472 & 26.9411 & 198.7 & 9.7 & -0.1 & 8.8 & 27.6 \\
150 & 1-211027 & 8602-6102 & 247.2360 & 39.6096 & 159.6 & 9.3 & -0.4 & 8.9 & 29.5 \\
151 & 1-113405 & 7972-3704 & 316.8413 & 11.0664 & 194.0 & 10.1 & -0.2 & 8.4 & 34.0 \\
152 & 1-209108 & 8485-3704 & 235.5806 & 47.9435 & 169.2 & 9.6 & -0.4 & 8.8 & 31.2 \\
153 & 1-318513 & 9048-6101 & 244.0360 & 24.0795 & 145.7 & 9.9 & -0.7 & 8.5 & 29.5 \\
154 & 1-48586 & 8721-6101 & 135.1639 & 53.9832 & 173.8 & 9.5 & -0.2 & 8.9 & 35.4 \\
155 & 1-318435 & 9095-6101 & 243.1605 & 23.6070 & 143.8 & 9.5 & -1.9 & 8.7 & 50.0 \\
156 & 1-420770 & 8444-1902 & 202.5088 & 31.7767 & 108.6 & 9.4 & -0.8 & 8.9 & 34.0 \\
157 & 1-113663 & 8618-3703 & 318.8046 & 9.9131 & 143.4 & 9.7 & \dots & 8.7 & 24.5 \\
158 & 1-248419 & 8313-1902 & 241.6261 & 41.3433 & 156.2 & 9.4 & -0.6 & 8.9 & 26.1 \\
159 & 1-145920 & 8149-3703 & 119.3568 & 27.4404 & 120.5 & 9.6 & -0.8 & 8.8 & 30.9 \\

\end{longtable}
\end{ThreePartTable}

\twocolumn

%%%%%%%%%%%%%%%%%%%%%%%%%%%%%%%%%%%%%%%%%%%%%%%%%%

% Don't change these lines
\bsp	% typesetting comment
\label{lastpage}
\end{document}